\documentclass[a4paper,12pt]{report}
\usepackage[left=2.5cm,right=2cm,top=2cm,bottom=3.5cm]{geometry}
\usepackage{graphicx}
\usepackage{mathtools}
\usepackage{amsmath}
\usepackage[utf8]{inputenc}
\usepackage[english]{babel}
\usepackage{fancyhdr}
\usepackage{caption}
\usepackage{gensymb}
\usepackage{slashed}
\usepackage{amssymb}
\usepackage{bbold}
\usepackage{dirtytalk}
\usepackage{feynmp-auto}
\usepackage{hyperref}
\usepackage{tikz}
\usetikzlibrary{decorations.markings}
\graphicspath{{images/}}

\newtheorem{theorem}{Theorem}
\newtheorem{definition}{Definition}

\begin{document}
	
	\begin{titlepage}
		
		\centering
		\vspace*{7cm}
		{\LARGE\bf Gribov Ambiguity \par}
		\vspace*{2cm}
		{\Large Thitipat Sainapha\par}
		{\Large Department of Physics, Faculty of Science\par}
		{\Large Chulalongkorn University\par}
		\vfill
		
		{Submitted in partial fulfillment of the requirements for the degree of Bachelor of Science in Department of Physics, Faculty of Science, Chulalongkorn University\\ Academic Year }
	\end{titlepage}

\newpage








\pagenumbering{roman}

\chapter*{\centering{Abstract}}
\addcontentsline{toc}{chapter}{Abstract}
\hspace{0.25in} Gribov ambiguity is a problem that arises when we try to single out the physical gauge degree of freedom in non-Abelian gauge theory by imposing the covariant gauge constraint. Unfortunately, the solution of the gauge constraint is not unique, thus the redundant gauge degree of freedom, called Gribov copies, remains unfixed. One of the traditional methods to partially resolve the Gribov problem is to restrict the space of gauge orbits inside the bounded region known as the Gribov region. The meaning of \say{partially resolve} is that this procedure can solve only the positivity's problem of the Faddeev-Popov operator but the Gribov copies are still there. However, on the bright side, the restriction to the Gribov region leads to the modification of the gluon propagator. Additionally, the new form of the gluon propagator yields the violation of the reflection positivity which is considered as the important axiom of the Euclidean quantum field theory. This shows that the gluon field in the Gribov region is an unphysical particle or technically confined. In this review article, we will start by discussing the traditional Faddeev-Popov method and its consequence on the proof of the unitarity of the perturbative Yang-Mills theory. Next, we will discuss the blind spot of the Faddeev-Popov quantization and study the mathematical and physical origin of the Gribov problem. Then, the method of the Gribov restriction will be elaborated. After that, we demonstrate the modification of the gluon field after restricting inside the bounded Gribov region. Finally, we show that the new form of the gluon leads to the violation of the reflection positivity axiom.

\chapter*{\centering{Acknowledgments}}
\addcontentsline{toc}{chapter}{Acknowledgments}
\hspace{0.25in}
I would like to thank Rujikorn Dhanawittayapol, my supervisor, for allowing me to study what I want to. I want to thank my study group's members, especially, Chawakorn Maneerat, Saksilpa Srisukson and Karankorn Kritsarunont for very useful discussions. I also would like to thank my senior, Sirachak Panpanich, for good suggestions. I'm very grateful to give a special thanks to Kei-Ichi Kondo who give many helpful comments on my thesis. Additionally, I also would like to thank Paolo Bertozzini who gives comments on my axiomatic quantum field theory's part of the thesis. In fact, it is very impressive to thank the organizers of the YKIS2018b to let me participate in the conference which makes me found my research interest. Finally, I want to thank Minase Inori, Yura Hatsuki, Ariabl'eyeS, Yurica/Hanatan, $\not=$ME and other bands for very wonderful songs when I feel sick about life.

\newpage
\addcontentsline{toc}{chapter}{Table of Contents}
\tableofcontents

\chapter*{\centering{Conventions}}
\addcontentsline{toc}{chapter}{Conventions}
\begin{itemize}
    \item Einstein summation convention is used.
    \item Formally we work in the spacetime manifold $\mathcal{M}$ of $D=d+1$ dimensions where $d$ is the number of spatial dimensions.
    \item The metric signature is taken to be $(+, - , - , ... , -)$. However, mainly, we will work with Euclidean signature $(+,+,...,+)$.
    \item The natural unit ($c=\hbar=k_B=1$) is used.
    \item We sometimes write $A_\mu A_\mu\equiv (A_\mu)^2$ and $A_\mu A_\mu A_\nu A_\nu =(A_\mu A_\mu)^2$ so on and so forth.
    \item The Fourier transformation is taken to be 
    \begin{displaymath}
    \varphi(x)=\int\frac{d^Dp}{(2\pi)^D}\;\varphi(k)e^{-ik\cdot x},
    \end{displaymath}
    and
    \begin{displaymath}
    \varphi(k)=\int d^Dx\; \varphi(x) e^{ik\cdot x}.
    \end{displaymath}
\end{itemize}
\chapter{Introduction}
Quantum chromodynamics (QCD) is a gauge theory describing the strong interaction - one of four fundamental interactions of nature. In the absence of quark, QCD is described through the pure Yang-Mills (YM) theory which is a non-Abelian analogue of Maxwell's model of electromagnetism. Vacuum solution of the pure classical YM theory in (3+1)-dimensional Minkowski(or Euclidean) space(time), by construction, undergoes a conformal symmetry (this fact can be found, for example, in appendix C of \cite{Actor1979} and \cite{hep-th/0405286}), thus the existence of YM's mass gap is forbidden in principle. However, all known phenomena support the existence of the mass gap since the formation of colorless bound states of colorful particles will be not allowed to happen unless the mass gap do exist. This fact was organized to be one of the most difficult problems in mathematics. As mentioned above, the existence of the mass gap and the absence of free colorful particles are closely related to each other. The latter phenomenon, relate more to physicists, is known as (color) confinement which can not be understood by using the traditional YM construction of QCD. Color confinement can be classified further into quark and gluon confinement. For the former case, we have known for a long time that quarks, specifically in long-distance or infrared (IR) limit, only appears to be bound states called hadrons. There are many possible candidates for describing the quark confinement, for example, the dual conductivity model constructed by the condensate of magnetic monopole in the vacuum (See also \cite{Kondo:2018pwd,1409.1599,hep-ph/0310102,hep-lat/9710049} for further information). For gluon confinement case, we do not have any experimental evidence so far, however, we believe so much that gluon, generally colorful, cannot appear isolately and also forms a bound state called glueball \cite{Fritzsch:1975tx}.\\
\indent Clearly, because of conformal invariance at the classical level, the confinement effect is therefore believed to be a very pure quantum effect. Since YM theory is a gauge theory, its quantization is also quite difficult to be done properly. For instance, the most traditional way, so-called Faddeev-Popov (FP) quantization \cite{Faddeev1967}, is also not complete in the sense that the FP method requires the very ideal gauge fixing choice. One merit of using the FP quantization is the introduction of new attractively harmful degrees of freedom, which violates a so-called spin-statistics theorem \cite{Streater:1989vi}, known as ghost modes. This leaves quantum YM theory to be invariant under new global transformation proposed firstly by Becchi, Rouet, Stora and Tyutin \cite{Becchi1974,RCP25_1975__22__A10_0,Becchi1975,0812.0580}. The BRST transformation can be used to classify the unphysical state out from the overall degrees of freedom. This fact leads Kugo and Ojima to prove the unitarity of the quantum Yang-Mills theory and propose how unphysical spectra, involving ghosts, are confined \cite{Kugo1979,Kugo:1979gm}. Unfortunately, the confinement of (transverse) gluon field is not proven by this particular argument. \\
\indent In the later 1970s, Gribov pointed out the incompleteness of the FP method in almost all of gauge fixing choices, this problem is well-known in the name Gribov ambiguity \cite{Gribov1978}. Moreover, in numerical lattice calculations, the Gribov ambiguity also leads to a meaningless indefinite result $0/0$ known as the Neuberger problem \cite{Neuberger:1986xz} which can be solved by introducing the mass term into the massless YM theory. In particular, the failure of the FP quantization opened the new door of exploration in theoretical high energy physics. There are many ways to solve this ambiguity, for example, imposing rather a non-local gauge fixing condition than a local one e.g. lightcone gauge. However, the main resolution is the one was proposed by Gribov himself. He suggested the way to generalize further the FP method by restricting the gauge orbit inside the Gribov region, in fact, inside a more restrict region called fundamental modular region (FMR). This generalization gives a mass of non-trivial form to the gluon field. As a byproduct, a new look of gluon spectral function generally violates the fundamental axiom of quantum field theory (QFT) called reflection positivity, hence, gluon will become an unphysical asymptotic state - technically confined. This will complete the analytical proof of the violation of the reflection positivity axiom in the YM theory restricted inside the Gribov region as pointed out in several papers, for example, \cite{0912.5153,hep-lat/0703022,hep-lat/0406020,1506.09021,1902.08894}.  
\pagenumbering{arabic}
\section{Outline of This Review}
We will start the next section by discussing the difficulty of quantizing gauge theories and how the FP procedure avoids this difficulty. In the subsection, we will discuss the consequence of FP quantization and then study the argument to understand the confinement proposed by Kugo and Ojima. In chapter 2, we will begin to discuss the big blind spot of FP procedure in covariant gauge conditions such as Landau gauge fixing (will be mainly focused on in this review). Then, we discuss how the naive FP method fails even in numerical computations. After we understand how harmful the Gribov ambiguity is, we will give both mathematical and physical evidence of how Gribov ambiguity gets into the YM model. After that, we will review the possible resolution of the Gribov ambiguity by introducing the notion of the Gribov region and FMR. Now then we will study the semi-classical solution of the Gribov ambiguity which will illustrate how gluon mass of Gribov form gives rise. In the final chapter, we will discuss the consequence of Gribov mass term to prove the gluon confinement by starting with reviewing the axiom of QFT both Wightman axioms \cite{Wightman} and, Euclideanized version, Osterwalder-Schrader axioms \cite{Osterwalder:1973dx}. Finally, we will show that the Gribov-type massive gluon propagator violates the reflection positivity axiom leading to the sign of gluon confinement.
\chapter{Traditional Quantization}
For the Yang-Mills (YM) theory, the path integral quantization is much more convenient to perform than the canonical one. We firstly introduce a useful tool called (Euclidean) partition function as follows
\begin{equation}
    Z(J)\equiv \int \mathcal{D}\varphi \; e^{-S[\varphi]-\int d^{D}x\;J(x)\varphi(x)},
\end{equation}
where $S[\varphi]$ is an (Euclidean) action functional of any field operator $\varphi$ and $J$ represents a source of that operator.
As we have already mentioned, the YM theory is a non-abelian gauge theory so the relevant field is, of course, a YM gauge field or YM connection $A_\mu(x)\equiv A^a_\mu(x) T^a$ where $T^a$ is a generator of Lie group $G$. The partition function is expressed as
\begin{equation}
\label{2.2}
    Z(J)=\int \mathcal{D}A e^{-S[A]_{YM}-\int d^{D}xJ^a_\mu A_\mu^a},
\end{equation}
with the YM action
\begin{equation}
\label{2.3}
    S[A]_{YM}=\frac{1}{4}\int d^{D}x \;F^a_{\mu\nu}(x)F^a_{\mu\nu}(x) 
\end{equation}
$F_{\mu\nu}$, so-called YM field strength tensor or YM curvature, is defined to be $\equiv \partial_\mu A_\nu-\partial_\nu A_\mu -ig[A_\mu,A_\nu]$ with YM coupling constant $g$. In term of component of Lie algebra, we have
\begin{equation}
\label{2.4}
    F^a_{\mu\nu}=\partial_\mu A^a_\nu -\partial_\nu A^a_\mu-gf^{abc}A^b_\mu A^c_\nu.
\end{equation}
In the last term of expression \eqref{2.4}, we have used the Lie algebra among generators $[T^b,T^c]=if^{bca}T^a$. The totally anti-symmetric tensor $f^{abc}$ is known as the structure constant of the Lie group. Now let's restrict ourselves to consider only the quadratic part of the action functional to focus only on the kinetic part of the theory
\begin{equation}
\label{2.5}
    \begin{split}
         Z_{quad}(J)&=\displaystyle{\int \mathcal{D}A \exp{\left[-\int d^Dx\;\left(\frac{1}{4}(\partial_\mu A^a_\nu -\partial_\nu A^a_\mu)^2+A^a_\mu J^a_\mu\right)\right]}}\\
         &=\displaystyle{\int \mathcal{D}A \exp{\left[-\int d^Dx\;\left(\frac{1}{2}(\partial_\mu A^a_\nu\partial_\mu A^a_\nu -\partial_\mu A^a_\nu\partial_\nu A^a_\mu)+A^a_\mu J^a_\mu\right)\right]}}\\
         &=\displaystyle{\int \mathcal{D}A \exp{\left[-\int d^Dx\;\left(\frac{1}{2}A^a_\nu(\partial_\mu \partial_\nu -g_{\mu\nu}\partial_\rho \partial_\rho )A^a_\mu+A^a_\mu J^a_\mu\right)\right]}}\\
         &\equiv\displaystyle{\int \mathcal{D}A \exp{\left[-\int d^Dx\;\left(\frac{1}{2}A^a_\mu\square_{\mu\nu}A^a_\nu+A^a_\mu J^a_\mu\right)\right]}},
    \end{split}
\end{equation}
The partition function \eqref{2.5} is nothing but Gaussian integral easily to evaluate. Using the identity which will be derived in appendix \eqref{A.1}. We thus have
\begin{equation}
\label{2.6}
    Z_{quad}(J)=\frac{1}{\det\square_{\mu\nu}}e^{\frac{1}{2}\int d^{D}x\; J^a_\mu\square_{\mu\nu}^{-1}J^a_\nu}.
\end{equation}
The result obtained in \eqref{2.6} will be technically correct if an inverse of $\square_{\mu\nu}$ really do exist while, in fact, it does not. To make sense of this, let's compute the equation of motion by minimizing the action functional with respect to gauge field $A_\mu$. We simply obtain
\begin{equation}
    \square_{\mu\nu}A^a_{\nu}=J^a_\mu+\mathcal{O}(A^2).
\end{equation}
This means that operator $\square_{\mu\nu}$ demonstrates a map (at leading order) from a set of gauge fields to a set containing their source fields. Since, in principle, the gauge field $A_\mu$ belongs to an equivalence class $[A]\equiv \{A\}/\mathcal{G}$ (gauge field belongs to a set of all possible field $A_\mu$ modulo out by the gauge transformation) called gauge orbit. In English, this shows that $A_\mu$ is not typically unique in the sense that other gauge field of the form
\begin{equation}
\label{2.8}
    ^UA_\mu=UA_\mu U^{-1}-\frac{i}{g}(\partial_\mu U)U^{-1},
\end{equation}
for any element $U\in G$, it also describes the same physical situation as $A_\mu$ can do. Following this logic, we can deduce that $\square_{\mu\nu}$ is precisely not injective (one-to-one) but many to one implying that it is impossible to be bijective, hence, it is non-invertible at the first place. Consequently, we need to eliminate redundant gauge degrees of freedom, formally known as fixing the gauge, in a systematical way before performing a consistent quantization of the YM theory. (In mathematical language, it is known as choosing the one specific representative out from the gauge orbit.) Specifically, in this review, we will stick in the Landau gauge (or sometimes called Lorenz gauge), i.e. $\partial_\mu A^a_\mu=0$.
\section{Faddeev-Popov Quantization}
One way to fix the gauge in path integral quantization, we will follow the most traditional way called Faddeev-Popov (FP) quantization as already mentioned in an introduction. The main idea is we will impose the gauge condition, behave as the constraint, by inserting the delta function into path integrand. However, we cannot just naively add delta function into the partition function. So what should we do exactly? Let's recall the well-known identity of delta function
\begin{equation}
\label{2.9}
    \delta(f(x))=\sum_i\frac{\delta(x-x_i)}{|f'(x_i)|},
\end{equation}
where $x_i$ is an element of kernel of function $f$, $x_i\in \ker(f)$, i.e. $f(x_i)=0$. Taking the integration over $x$ in both sides of relation \eqref{2.9} and using the distribution properties of delta function $\int dx\; \delta(x-x_i)=1$. We then get
\begin{equation}
\label{2.10}
    \int dx\; \delta(f(x))=\sum_i\frac{1}{|f'(x_i)|}.
\end{equation}
To obtain the form of \say{1} we are seeking for, we need to evaluate the summation in the left-hand side of \eqref{2.10}. However, this summation is precisely not too trivial to evaluate easily. To be honest, one way to evaluate this summation is to not have the summation at the beginning. In other words, what we really want to say is to simplify this we have to require that there exists only one value of $x_i$ satisfying the condition $f(x_i)=0$. Therefore, the calculation of this summation is no longer necessary. Thus, after imposing the condition we have discussed, we find out the form of \say{1} as
\begin{equation}
\label{2.11}
    1=|f'(x_i)|\int dx\; \delta(f(x)).
\end{equation}
In the YM theory, we will generalize the identity \eqref{2.11} into the path integral version as following expression
\begin{equation}
\label{2.12}
    1=\Delta_{FP}\int \mathcal{D}\alpha\; \delta(G(^\alpha A))
\end{equation}
where $G(^\alpha A)$ is a gauge constraint we are talking about, $\alpha^b$ denotes an infinitesimal parameter of particular gauge transformation $U\approx 1-i\alpha^b T^b+\mathcal{O}(\alpha^2)$ and $\Delta_{FP}$ is called Faddeev-Popov determinant defined to be
\begin{equation}
\label{2.13}
    \Delta_{FP}\equiv \left|\frac{1}{g}\det M_{ab}\right|\equiv \left|\det\left(\frac{\delta G^a(^\alpha A)}{\delta \alpha^b}\right)\right|_{G(^\alpha A)=0},
\end{equation}
factor $g$ in denominator is just a convention. Determinant has been used since the path integration measure is nothing but the product of integration measure so $|f'(x_i)|$ becomes the product of eigenvalues of FP matrix, hence, it becomes the determinant of that particular matrix as expressed in \eqref{2.13}. Another way, probably a simpler way, to make sense of this determinant is to think it as a Jacobian associated to the transformation from the integral measure $\int \mathcal{D}G(^\alpha A)$ into other form of measure $\int\mathcal{D}\alpha$.\\
\indent Before we continue the work, let us emphasize very important point, which follows from the condition we have imposed to ignore the summation symbol in \eqref{2.10}, that the expression \eqref{2.12} will be true if and only if there is only one gauge field $A_\mu$ that satisfies the gauge fixing condition $G(A)=0$. Technically speaking, we demand strongly that the gauge orbit crosses or intersects the gauge fixing constraint surface only once! Unfortunately, in general situations, this condition is too ideal as we will discuss in detail in section 3.

\indent Now we can insert that non-trivial $1$ of the form \eqref{2.12} into the path integral \eqref{2.2} (from now on we will turn off the source without loss of generality)
\begin{equation}
\label{2.14}
    Z(0)=\int \mathcal{D}A \int \mathcal{D}\alpha \; \delta(G(^\alpha A))e^{-S_{YM}}\Delta_{FP}= \int \mathcal{D}\alpha\int \mathcal{D}A \; \delta(G(^\alpha A))e^{-S_{YM}}\Delta_{FP}.
\end{equation}
Note that we need to be careful about the position FP determinant in the path integral since FP determinant is independent of the gauge transformation, which comes from the fact that the functional derivative $\frac{\delta G}{\delta \alpha}$ does not depend on $\alpha$ (infinitesimal gauge transformation is effectively linear in group parameter), but nothing guarantees that FP determinant is independent of the gauge field especially in non-Abelian gauge theories. One might ask immediately how can we change the order of path integral measure as we have done in the last step in \eqref{2.14}. To answer that, we will claim first that $\mathcal{D}A$ measure is a so-called Haar measure with respect to gauge group $G$ which means it is completely invariant under the gauge transformation. Intuitively speaking, the path integration measure of gauge field $A_\mu$ does an integration over all possible configuration of gauge fields implying that this integral is performed in the sense that it has already involved all contributions from gauge transformation. Thus, we can swap the measure without making any harmful karma.\\
\indent In the next steps, we will use a dirty trick a bit to simplify a partition function. Firstly, we will perform a gauge transformation from the gauge field $A_\mu$ into $^\alpha A_\mu$. This transformation changes nothing since $\mathcal{D}A$ is a Haar measure and $S[A]_{YM}$ is invariant under gauge transformation. Then $\alpha$ behaves just like dummy indices so we can change the dummy variable back into the original one, $^\alpha A_\mu\rightarrow A_\mu$, without hesitation. With these little steps, the path integral (2.14) is therefore dependent on an infinitesimal parameter $\alpha$ no longer.
\begin{equation}
    Z(0)=\left( \int \mathcal{D}\alpha\right)\int \mathcal{D}A \; \delta(G(A))e^{-S_{YM}}\Delta_{FP}=V_{\mathrm{group}}\int \mathcal{D}A \; \delta(G(A))e^{-S_{YM}}\Delta_{FP},
\end{equation}
where the integration over all possible gauge parameter $\alpha$ is nothing more and nothing less but a volume $V_{group}$ of gauge group manifold (a gauge group is a Lie group so it is also a manifold) which is generally infinite. However, even though it is not well-defined in principle, we will treat this infinity as a normalizing factor and throw it out from the partition function. Hence, it will not affect the physical phenomena at all.\\
\indent Let's specify the gauge fixing condition, we choose the class of gauge condition
\begin{equation}
\label{2.16}
    G^a(A)=\partial_\mu A_\mu^a(x)-\omega^a(x),
\end{equation}
where $\omega^a(x)$ is an arbitrary scalar field contributing nothing in the physical theory and also the FP determinant. Since the expression \eqref{2.16} is the local expression, i.e. describing the only specific value of $\omega^a$, thus, at the end of the calculation, we prefer the averaging over this particular variable to obtain the result describing global information. To average this variable, we will use the Gaussian weight to parametrize the distribution due to the fact that $\omega^a$ is arbitrary. On the other hand, it can be treated as a random parameter represented by normal (Gaussian) distribution. Finally, after we integrating over $\omega^a$, we need to divide out by the normalization constant which can be dropped out from the path integration as usual. To conclude, the partition function becomes
\begin{equation}
\label{2.17}
\begin{split}
   Z(0)&=   (...)\displaystyle{\int \mathcal{D}A \int\mathcal{D}\omega\delta(\partial_\mu A^a_\mu-\omega^a)\exp{\left(-\frac{1}{2\xi}\int d^{D}x(\omega^a)^2\right)}e^{-S_{YM}}\Delta_{FP}} \\
     & =(...)\displaystyle{\int \mathcal{D}A \exp{\left[-\left(S_{YM}+\frac{1}{2\xi}\int d^{D}x\;(\partial_\mu A^a_\mu)^2\right)\right]}\Delta_{FP}}.
\end{split}
\end{equation}
In this sense, this step can be effectively realized as we just add the gauge fixing term into action by hand. To be clear, it can be shown easily that the presence of this extra term breaks gauge symmetry explicitly, thus the gauge fixing procedure has been done. Unfortunately, the result has been not yet completed because FP determinant still remains undetermined.\\
\indent So what we have to do is to calculate the FP determinant. Before doing that, we need to calculate the infinitesimal form of the gauge transformation with infinitesimal parameter $\alpha$. Recalling the gauge transformation \eqref{2.8} then substituting $U(x)$ to be $e^{-i\alpha^a(x)T^a}\xrightarrow{\alpha\ll1}1-i\alpha^a(x)T^a+\mathcal{O}(\alpha^2)\equiv 1-i\alpha(x)+\mathcal{O}(\alpha^2)$ and, of course, an inverse element $U(x)^{-1}=e^{i\alpha(x)}\approx 1+i\alpha(x)$. Finally, we thus have (keeping at only leading order in $\alpha$, i.e. $\mathcal{O}(\alpha^1)$))
\begin{equation}
\label{2.18}
    \begin{split}
        ^\alpha A_\mu &= (1-i\alpha)A_\mu(1+i\alpha)-\frac{i}{g}(\partial_\mu(1-i\alpha))(1+i\alpha) \\
         &= A_\mu+i[A_\mu,\alpha]-\frac{1}{g}\partial_\mu\alpha \\
         &= A_\mu-\frac{1}{g}(\partial_\mu \alpha +ig[\alpha,A_\mu])\\
         &\equiv A_\mu -\frac{1}{g}D_\mu \alpha,
    \end{split}
\end{equation}
where $D_\mu\varphi\equiv \partial_\mu\varphi+ig[\varphi,A_\mu]$, known as a covariant derivative, for any field $\varphi$ transforming under adjoint representation of the gauge group $G$, namely, $\varphi\equiv \varphi^a T^a$ need to be a $\mathfrak{g}$-valued field operator where $\mathfrak{g}$ is a Lie algebra associated with a Lie group $G$. Honestly, we will restrict ourselves into this definition only. Although the definition of the covariant derivative acting on a field which transforms under the fundamental representation is, in fact, different, in this literature, we will focus only on adjoint fields. Note also that, in color indices form, we have $D_\mu^{ab}\varphi^b=(\delta^{ab}\partial_\mu-gf^{abc}A_\mu^c)\varphi^b$.\\
\indent We are ready to compute the FP determinant explicitly by plugging the result \eqref{2.18} combining with the particular form of gauge fixing constraint \eqref{2.16} into the definition \eqref{2.13}. To obtain
\begin{equation}
\label{2.19}
\begin{split}
    \Delta_{FP}&=\left|\det\left( \frac{\delta}{\delta \alpha^b(y)}\left(\partial_\mu A_\mu^a(x)-\frac{1}{g}\partial_\mu D_\mu^{ac} \alpha^c(x)-\omega^a(x)\right)\right)\right|\\
    &=(...)\det(\partial_\mu D_\mu^{ab}\delta^{D}(x-y)),
\end{split}
\end{equation}
up to some constant factor depending on the coupling constant $g$ which can be eliminated out by just simple re-definition of gauge field. Notice here that in the last step of the equation \eqref{2.19}, we have implicitly thrown the absolute operation out by keeping in our mind that we have already require the FP determinant to be a positive value which, when the Gribov ambiguity is taken into account, does not necessarily hold in all possible situation. For future's usage, we can also read off the expression of the FP operator $M^{ab}(x,y)$ as
\begin{equation}
    M^{ab}(x,y)=-\partial_\mu D_\mu^{ab}\delta^{D}(x-y).
\end{equation}
Let us give some remark first that the covariant derivative inside the expression \eqref{2.18} will be reduced into a partial derivative for the case of the Abelian gauge theory due to the simple fact that the Lie bracket or the commutator is trivial, i.e. $[\alpha,A_\mu]=0$. Consequently, in Abelian gauge theory, we end up with
\begin{equation}
    Z(0)=(...)\det(\partial^2)\int \mathcal{D}A \exp{\left[-\left(S_{\mathrm{Maxwell}}+\frac{1}{2\xi}\int d^{D}x\;(\partial_\mu A_\mu)^2\right)\right]}.
\end{equation}
where we have used the fact that the FP determinant in an Abelian gauge model is dependent not to gauge fields, we then pull it out from the integrand. Note that no matter what the value of $\det(\partial^2)$ really is, divergent or not, we absolutely do not care since the result will be just $c$-number-valued quantity or, in human language, it is just a constant dropped out from the path integration. Hence, in Abelian gauge theories, the FP determinant contributes not any further. Unfortunately, in non-Abelian gauge theories, the story is completely different since the determinant is not that trivial. To evaluate such a determinant, we will use the path integration's identity of Grassmann variable as will be derived in an appendix \eqref{A.2}. This yields an interesting result
\begin{equation}
\label{2.22}
    Z(0)=\int\mathcal{D}A\int \mathcal{D}c\int \mathcal{D}\Bar{c}\exp{\left\{-\left[S_{YM}+\int d^{D}x\left(\frac{1}{2\xi}(\partial_\mu A^a_\mu)^2+ \;\Bar{c}^a\partial_\mu D_\mu c^a\right)\right]\right\}},
\end{equation}
up to some (infinite) normalization constant. $c$ and $\Bar{c}$ are fermionic fields since it is Grassmann variable, i.e. it satisfies anti-commutation relation. However, if we look at the equation of motion of these fields more carefully, it takes the same form as the complex scalar fields. As a result, $c$ and $\Bar{c}$ are anti-commutating spin-0 fields which inevitably violate the spin-statistics theorem, one of the most fundamental theorems of QFT, stating that integer-spin fields such as scalar fields and gauge fields must satisfy commutation relation, whereas 
half integer-spin fields such as spinor must satisfy anti-commutation relation (further reading \cite{Duck1998,Streater:1989vi}). Since, generally, spin-statistics theorem plays a significant role for ensuring unitarity, as we will emphasize later in section 2.3, thus both $c$ and $\Bar{c}$ need to be unphysical particles formally known as Faddeev-Popov ghost and anti-ghost, respectively.\\
\indent Let us note further, as we have discussed slightly, even though the FP (anti-)ghosts are very harmful in many senses, it does not imply that they need to be invisible from the theory completely. The word \say{unphysical} really means that they can not appear as an asymptotic state but, honestly, they still can hang around inside the loop diagrams. In particular, FP ghosts are necessary, in loop level calculation, to cancel all gauge dependent parts of gluon 2-point correlation function (only gluon bubble graphs alone can not satisfy Ward-Takahashi identity, the ghost contribution need to be added).
\begin{figure}[h!]
    \centering
 \begin{fmffile}{gluon3}
\begin{fmfchar*}(85,80)
\fmfleft{i}
\fmfright{o}
\fmf{gluon}{i,v1}
\fmf{gluon,right}{v1,v2,v1}
\fmf{gluon}{v2,o}

\end{fmfchar*}
\end{fmffile}
\hspace*{0.5cm}
   \begin{fmffile}{gluon4}
\begin{fmfgraph*}(85,80)
\fmfleft{i}
\fmfright{o}
\fmf{gluon}{i,v}
\fmf{gluon,right}{v,v}
\fmf{gluon}{v,o}

\end{fmfgraph*}
\end{fmffile}
\hspace*{0.5cm}
\begin{fmffile}{ghost}
\begin{fmfgraph*}(105,80)
\fmfleft{i}
\fmfright{o}
\fmf{gluon}{i,v1}
\fmf{ghost,right}{v1,v2,v1}
\fmf{gluon}{v2,o}

\end{fmfgraph*}
\end{fmffile}
    \caption{Gluonic vacuum polarization diagrams: 3-point vertex gluon bubble, 4-point vertex gluon bubble, and ghost bubble contribution. Summation of all these graphs will give rise to the true tensorial structure which satisfies the Ward-Takahashi identity. The absence of any graph will give the wrong result in the end.}
\end{figure}
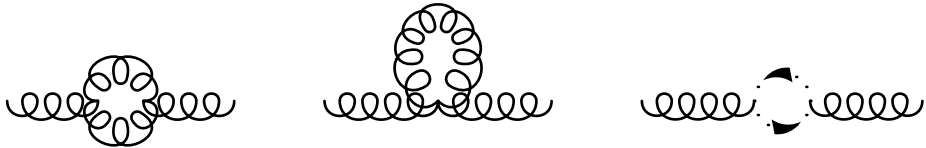\\
\indent To complete this section, we will first write down the effective Lagrangian describing the quantum version of the YM theory. We clearly obtain
\begin{equation}
\label{2.23}
    \mathcal{L}_{QYM}=\frac{1}{4}(F^a_{\mu\nu})^2+\frac{1}{2\xi}(\partial_\mu A^a_\mu)^2+\Bar{c}^a\partial_\mu D_\mu^{ab} c^b.
\end{equation}
This Lagrangian is said to be in linear covariant gauge or $R_\xi$ gauge (R-xi), reducible to other gauge fixing condition. Particularly, for $\xi\rightarrow 0$ it will return to Landau gauge while the choice $\xi=1$ will reduce the theory into a so-called Feynman-'t Hooft gauge condition which is used widely in many standard QFT textbooks since this choice makes the Lorentz covariant part of gluon propagator extremely easy to deal with. Finally, we will read off the relevant Feynman rules from this Lagrangian as will be listed down below (the reader can understand the way to derive and read off the momentum-space Feynman rules in, e.g. \cite{Schwartz:2013pla})
\begin{enumerate}
    \item Gluon propagator. Let us begin by noting that the presence of the gauge fixing term makes the kinetic operator defined in \eqref{2.5} and \eqref{2.6} will change the form into
    \begin{equation}
        \square_{\mu\nu}\rightarrow \Tilde{\square}_{\mu\nu}=\left[\left(1-\frac{1}{\xi}\right)\partial_\mu\partial_\nu-g_{\mu\nu}\partial_\rho\partial_\rho\right].
    \end{equation}
    Thus, the gauge fixing modification version of the equation of motion (linear part) takes the form
    \begin{equation}
      \delta^{ab} \Tilde{\square}_{\mu\nu}A^b_\nu(x) =\delta^{ab}\left[\left(1-\frac{1}{\xi}\right)\partial_\mu\partial_\nu-g_{\mu\nu}\partial_\rho\partial_\rho\right]A^b_\nu(x)=J^a_\mu(x).
    \end{equation}
    Naively speaking, this equation is, in principle, easy to solve by just finding an inverse of the operator $\Tilde{\square}_{\mu\nu}$ which, in fact, is technically difficult. Fortunately, we have very powerful technique to change the operator-valued distribution into the c-number-valued distribution called Fourier transformation. The operator, such as, a partial derivative, in position space will look just like a number inside momentum space. Therefore, we find effectively the transformation $\partial_\mu\rightarrow -ip_\mu$. So we end up with the (number-valued) dynamical equation
    \begin{equation}
        \delta^{ab}\left[g_{\mu\nu}p^2-\left(1-\frac{1}{\xi}\right)p_\mu p_\nu\right]A^b_\nu=J^a_\mu.
    \end{equation}
    The solution is typically just a gluon propagator (Green's function) we are looking for. In any linear covariant gauge choice

 \begin{fmffile}{gluon}
 \begin{equation}
 \label{2.27}
 \begin{gathered}
\begin{fmfgraph*}(60,50)
\fmfleft{i}
\fmflabel{$\nu,b$}{i}
\fmflabel{$\mu,a$}{o}
\fmfright{o}
\fmf{gluon,label=$p$}{i,o}
\end{fmfgraph*}
\end{gathered}
\hspace*{1cm}=\frac{-i\delta^{ab}}{p^2+i\epsilon}\left(g_{\mu\nu}-(1-\xi)\frac{p_\mu p_\nu}{p^2}\right)\equiv\frac{-i\delta^{ab}}{p^2+i\epsilon}\Pi_{\mu\nu}(p,\xi), 
\end{equation}
\end{fmffile}
where the last expression is defined for later advantages. We will re-compute the gluon propagator again in the presence of Gribov problem in the section 3.3 but we will restrict ourselves in the Landau gauge $\xi\rightarrow 0$ in that calculation. Note also that the small complex pole $i\epsilon$ has been inserted reasonably to represent the time ordering in Feynman diagram level which, in the end, can be set to be zero in almost all technical calculations.
\item Ghost propagator. This propagator is much easier to find out since the kinetic part of (anti-)ghost field is the same form as the complex scalar field theory. So the resulting form of propagator is then
\begin{fmffile}{ghostprop}
 \begin{equation}
 \label{2.28}
 \begin{gathered}
\begin{fmfgraph*}(60,50)
\fmfleft{i}
\fmflabel{$b$}{i}
\fmflabel{$a$}{o}
\fmfright{o}
\fmf{ghost,label=$p$}{i,o}
\end{fmfgraph*}
\end{gathered}
\hspace*{1cm}=\frac{i\delta^{ab}}{p^2+i\epsilon} 
\end{equation}
\end{fmffile}
\item The last important (at least in this review article) diagram is an interaction between ghost fields and gluon field. The interaction encoded inside the covariant derivative in the last term of the QYM Lagrangian \eqref{2.23}. Let's focus on this term carefully
\begin{equation}
\label{2.29}
    \mathcal{L}_{int}=-\partial_\mu\Bar{c}^a(-gf^{abc}A^c_\mu)c^b=-gf^{abc}(\partial_\mu \Bar{c}^a)A^b_\mu c^c,
\end{equation}
in the very first step, we perform the integration by part in the exponential of the partition function \eqref{2.22} first. The reason why we can do that is because the boundary term does not affect to the physical situation. After that, the last step, we have used the totally anti-symmetric property of the Lie structure constant $f^{abc}=-f^{acb}$ and re-define color indices slightly to obtain more beautiful result. Then, the Feynman rule reads\\
\begin{fmffile}{ghostver}
 \begin{equation}
 \label{2.30}
 \begin{gathered}
\begin{fmfgraph*}(50,60)
\fmfleft{i}
\fmflabel{$c^c$}{i}
\fmflabel{$\Bar{c}^a$}{o}
\fmfright{o}
\fmftop{t}
\fmflabel{$\mu,b$}{t}
\fmf{gluon}{v,t}
\fmf{ghost,label=$p$}{v,o}
\fmf{ghost}{i,v}
\end{fmfgraph*}
\end{gathered}
\hspace*{1cm}=-gf^{abc}p_\mu,
\end{equation}
\end{fmffile}
where $p_\mu$ actually comes from the Fourier transformation of the partial derivative acting on the anti-ghost field (see \eqref{2.29}). However, in the next chapter, we will work with the Feynman diagram without knowing the specific modified form of the gluon propagator. To remedy this problem, one, instead of treating the gluon as a field, rather treats the gluon field as an external coupling. Thus, we will modify the Feynman rule of this particular vertex as follows\\

\begin{fmffile}{ghostvermod}
 \begin{equation}
 \label{ghostmodified}
 \begin{gathered}
\begin{fmfgraph*}(50,60)
\fmfleft{i}
\fmflabel{$c^c$}{i}
\fmflabel{$\Bar{c}^a$}{o}
\fmfright{o}
\fmftop{t}
\fmflabel{$A^b_\mu$}{t}
\fmf{gluon}{v,t}
\fmf{ghost,label=$p$}{v,o}
\fmf{ghost}{i,v}
\end{fmfgraph*}
\end{gathered}
\hspace*{1cm}=-\frac{g}{\sqrt{V}}f^{abc}p_\mu A^b_\mu.
\end{equation}
\end{fmffile}
Since the gauge field in the momentum space is not a dimensionless quantity, thus the volume $V$ has been introduced to maintain the dimensionality of the Feynman graph. One can think about it as the extra variable obtained when we performing the Fourier transformation from the gauge field in the position space into the momentum space with the periodic boundary condition.

\end{enumerate}
We will use all of these Feynman rules again in section 3.3 to calculate the pole structure of the ghost 2-point function. In the next section, before we start discussing the main problem of the FP traditional method of quantization, we will discuss the first argument of confinement, which is the direct consequence of the FP quantization, due to the works of Kugo and Ojima by analyzing carefully the Fock space of physical states \cite{Kugo1979,Kugo:1979gm}. However, in the next section, we will prepare the basic ingredients before we will cook a tasty dish first in the last section of this chapter.

\section{Ghost and BRST Symmetry}

As we have discussed in the previous section that the gauge symmetry is explicitly broken by the gauge fixing condition during the quantization process is progressing. However, we are studying gauge theories so the gauge transformation is very important. Thus, we expect that the gauge symmetry will be eventually returned after the quantization process has been done. Is that so? the answer is it will be restored with the new attractive form. To warm-up, let's explore first another extra symmetry, easier to explore, besides the ghost generalization of the gauge symmetry.\\
\indent Observe that the form of the ghost sector of the QYM Lagrangian \eqref{2.23} takes almost the same form as the complex scalar field theory (it is not since we need to keep in mind that $c$ and $\Bar{c}$ are not Hermitian conjugation of each other as the ordinary scalar field theory, they are completely different fields with no any relationship). Hence, one might naively guess that there must be the $U(1)$ global symmetry transformating inside this particular sector. Explicitly, we might have
\begin{equation}
    c(x)\rightarrow e^{i\gamma}c(x),\;\;\;\;\;\;\Bar{c}(x)\rightarrow e^{-i\gamma}\Bar{c}(x).
\end{equation}
However, this is not quite true since $U(1)$ transformation will be inconsistent with the particular requirement of the Hermiticity of both ghost and anti-ghost fields. To understand this precise reason, let's consider the Hermitian conjugation of the ghost sector Lagrangian. We thus obtain
\begin{equation}
    \begin{array}{ccc}
         \mathcal{L}_{ghost}^\dagger&=&(c^a)^\dagger\overleftarrow{\partial}_\mu\overleftarrow{D}_\mu(\Bar{c}^a)^\dagger\\
         &=&-(\Bar{c}^a)^\dagger \partial_\mu D_\mu (c^a)^\dagger,
    \end{array}
\end{equation}
where we have used the anti-symmetric property of Grassmann variables in the last step. It is precisely clear that if we demand that $\Bar{c}^\dagger=\Bar{c}$ and $c^\dagger=c$, the ghost sector's Lagrangian will become anti-Hermitian which is unacceptable (once again, try to not misunderstand that $\Bar{c}=c^\dagger$, it is not true at all). Therefore, we assign that anti-ghost is Hermitian while ghost field is anti-Hermitian, i.e. 
\begin{equation}
\label{2.33}
    \Bar{c}^\dagger=\Bar{c},\;\;\;\;\; \;\; c^\dagger =-c.
\end{equation}
This assignment does not allow the $U(1)$ transformation among (anti-) ghosts. Luckily, we still can have the scale transformation by space-time independent real parameter
\begin{equation}
\label{2.34}
    c(x)\rightarrow e^\gamma c(x),\;\;\;\;\;\; \Bar{c}\rightarrow e^{-\gamma}\Bar{c}.
\end{equation}
According to the Noether's theorem, the preservation of the symmetry transformation implies the existence of an associated conserved (Noether) current. The derivation of Noether's current is straightforward which can be seen in many standard field theories textbooks. The conserved current associated with the scaling inside the ghost sector can be further to calculate a so-called associated Noether (conserved) charge as follows
\begin{equation}
\label{2.35}
    Q_c\equiv\int d^{d}x\; J_0^{ghost}.
\end{equation}
Because of the conservation of the charge, this charge can be used to represent the quantum number of the QYM theory. We will call this from now on the ghost number charge used to count the ghost number of any field operator (see the more clean reasoning in the equation \eqref{2.53} below).  \\
\indent After taking an appetizer, let's move to the main dish. Since the presence of the gauge fixing term, which remains even after the quantization has been done, breaks the gauge symmetry among the gauge fields completely. This fact leads us to notice that, the remnant gauge symmetry might complicatedly transform among both gauge fields and ghost fields. Naively thinking, we may take the gauge parameter $\alpha(x)$ to be directly proportional to the ghost field itself (this remark was given in \cite{Schwartz:2013pla}). Explicitly, we guess
\begin{equation}
\label{2.36}
    \alpha(x)=\theta c(x),
\end{equation}
where Grassmann variable $\theta$ has been added to keep the $\alpha$ remain bosonic. We end up with the set of transformation
\begin{equation}
\label{2.37}
    \begin{array}{ccc}
         \delta A^a_\mu &=& \displaystyle{\theta (D_\mu c)^a},\\
         \delta c^a&=&\displaystyle{-\frac{1}{2}\theta gf^{abc}c^bc^c},\\
         \delta \Bar{c}^a&=&-\displaystyle{\theta \frac{1}{\xi}\partial_\mu A^a_\mu}.
    \end{array}
\end{equation}
This set of global symmetry transformation is truly the remnant gauge transformation in the existence of (anti-)ghost modes. This symmetry is known as the BRST symmetry referred to the names of whom we have mentioned in the introduction section. Once again, \say{symmetry implies conservation law}. Let us denote the associated BRST conserved charge as $Q_B$ whose exact form will be left undetermined for a while. One of traditional method to study the symmetry is to investigate the algebraic relations among charges - a charge algebra. To do that, we will perform the BRST transformation upon any field in the theory two times. Consider
\begin{equation}
\label{2.38}
    \begin{split}
       \delta_2\delta_1 c^a  &=\displaystyle{\delta_2 \left(-\frac{1}{2}\theta_1gf^{abc}c^bc^c\right)} \\
         & =-\displaystyle{\frac{1}{2}\theta_1 gf^{abc}(\delta_2c^b)c^c-\frac{1}{2}\theta_1 gf^{abc}c^b(\delta_2c^c)}\\
         &= +\displaystyle{\frac{1}{4}\theta_1\theta_2g^2f^{abc}(f^{bde}c^dc^ec^c+f^{cde}c^bc^dc^e)}\\
         &=\displaystyle{\frac{1}{4}\theta_1\theta_2g^2f^{abc}(f^{bde}c^dc^ec^c-f^{bde}c^dc^ec^c)}\\
         &=0
    \end{split}
\end{equation}
where there are sub-steps changing from the third line to the fourth line in \eqref{2.38}. The first sub-step is to change dummy indices between $c\leftrightarrow b$ in the last term only. After that, we use the totally anti-symmetric property $f^{acb}=-f^{abc}$ so we now get an extra minus sign in the second term. Finally, we just swap the position of ghost fields which are Grassmann, however, the sign remains the same since the position of the ghost fields has been interchange for actual two times. The mathematical implication of the result \eqref{2.38} is very meaningful. It deduces that the algebra among the BRS charges might presumably be of the form (up to trivial constant factor)
\begin{equation}
\label{2.39}
    Q^2_B=0=\{Q_B,Q_B\},
\end{equation}
followed from the nilpotency of a so-called BRST differential $\delta^2=0$. Now let's check our assumption by manipulating the double BRST transformations to other relevant field, i.e. anti-ghost $\Bar{c}$, the calculation is also straightforward to be done
\begin{equation}
\label{2.40}
    \begin{array}{ccc}
        \delta_2\delta_1\Bar{c}^a &=& \displaystyle{\delta_2\left(-\frac{\theta_1}{\xi}\partial_\mu A^a_\mu\right)} \\
         &=&-\displaystyle{\frac{1}{\xi}\theta_1\theta_2 \partial_\mu D_\mu c^a}
    \end{array}
\end{equation}
which is unfortunately non-vanishing in general. Particularly, the result is so familiar somehow. One might immediately notice that the field dependent part of the result is the same form of the equation of motion derived from the functional variation of an anti-ghost field having the form
\begin{equation}
\label{2.41}
    \partial_\mu D_\mu c^a(x)=0.
\end{equation}
Clearly, if we combine the equation of motion \eqref{2.41} with the result in \eqref{2.40}, the contaminated target has been mysteriously eliminated. This fact seems unusual to many people, by the way, it is a very well-known fact to one who is familiar with supersymmetric (SUSY) field theories (reading the section 5.4 of review article \cite{Sohnius:1985qm} will be helpful). Technically speaking, the BRST algebra is said to be closed on-shell meaning that the anti-commutation relation of the BRST charge \eqref{2.39} can not hold by itself unless the equations of motion have already been imposed. One might honestly ask that can we construct the off-shell representation of the BRST algebra. Ideally, it is possible since it is not mathematically forbidden. Formally in SUSY field theories, to construct the closed SUSY algebra, we will introduce the non-dynamical field known as an auxiliary field and let this new field to also transform under such a transformation demanded to be closed. \\
\indent How do we find the actual form of the auxiliary field then? This is a practical difficulty we are worrying about. Fortunately, the auxiliary field in the QYM model is already well-known. One particular method to derive the form of this auxiliary field was given in the useful standard textbook \cite{Weinberg:1996kr} that, in my opinion, is a very natural method. Let us traceback when we try to fix gauge by using the FP method in the previous section. We have selected the class of gauge fixing constraint to be as expressed in \eqref{2.16} with arbitrary (auxiliary) field $\omega^a$. In addition, since the gauge fixing condition inside the delta function is a local expression, we, therefore, average over this field by Gaussian distribution's weight. However, this field $\omega^a$ is not unique in many sense, one can reparametrize it by any constant factor. Nonetheless, we can also perform a Fourier transformation from this field into other field living in its Fourier's space (Fourier transformation, in this context, is not referred to the Fourier transformation from the position space into the momentum space and vice versa but regarded as the Fourier transformation between two sets of auxiliary fields). To be clear about all the statement above, consider the Fourier transformation of the Gaussian weight in \eqref{2.17}
\begin{equation}
\label{2.42}
   \hat{\mathcal{F}}\{e^{-\frac{1}{2\xi}\int d^{D}x(\omega^a)^2}\}= \int \mathcal{D}\omega \; \exp{\left [-\frac{1}{2\xi}\int d^{D}x\;(\omega^a(x))^2-i\int d^{D}x\;\omega^a(x)B^a(x)\right]},
\end{equation}
where $\hat{\mathcal{F}}$ denotes the short hand notation of the Fourier transformation. The expression \eqref{2.42} is again a Gaussian integration \eqref{A.1} yielding other Gaussian distribution with new field variable $B^a(x)$
\begin{equation}
\label{2.43}
  \hat{\mathcal{F}}\{e^{-\frac{1}{2\xi}\int d^{D}x(\omega^a)^2}\}=  e^{-\frac{\xi}{2}\int d^{D}x (B^a)^2}.
\end{equation}
Conversely, this shows that the original Gaussian weight factor can be thought interestingly as the inverse Fourier transformation of the product \eqref{2.43}. Up to normalization constant, once again, dropping out from the partition function, we have
\begin{equation}
\label{2.44}
    e^{-\frac{1}{2\xi}\int d^{D}x(\omega^a)^2}=\int \mathcal{D}B\; \exp{\left[-\frac{\xi}{2}\int  d^{D}x\;(B^a(x))^2+i\int d^{D}x\;\omega^a(x)B^a(x)\right]}
\end{equation}
Since we have worked in the spacetime with Euclidean metric, we expect to have the real valued Lagrangian instead of the complex one. We conventionally redefine the arbitrary field $\omega^a\rightarrow i\omega^a$ while keeping the gauge fixing condition \eqref{2.16} fixed. Let's insert the expression \eqref{2.44} into the partition function \eqref{2.17}, then integrating over auxiliary field $\omega$. Hence,
\begin{equation}
    Z(0)=(...)\int \mathcal{D}A_\mu \int \mathcal{D}B\;\exp{\left[-\left(S_{YM}+\int d^{D}x\; \left[\frac{\xi}{2}(B^a)^2+B^a\partial_\mu A^a_\mu\right]\right)\right]}\Delta_{FP}.
\end{equation}
Following these steps, we just replace the FP determinant by path integrals of the ghost and anti-ghost as we have already done in the previous section. Henceforth, if we are talking about the QYM Lagrangian, we will keep our eyes on the following Lagrangian shown below
\begin{equation}
    \label{2.46}
    \mathcal{L}_{QYM}=\mathcal{L}_{YM}+B^a\partial_\mu A^a_\mu +\frac{\xi}{2}(B^a)^2+\Bar{c}^a\partial_\mu D_\mu c^a,
\end{equation}
where, now, we have already introduced an auxiliary field $B^a(x)$ formally known as a Nakanishi-Lautrup (NL) field. To show that the BRST symmetry within this Lagrangian \eqref{2.46} becomes off-shell representation already, we will start with observing that the set of transformation has already involved the transformation between the anti-ghost mode and the NL auxiliary field as shown below
\begin{equation}
    \label{2.47}
    \begin{array}{ccc}
         \delta A^a_\mu &=& \displaystyle{\theta (D_\mu c)^a},\\
         \delta c^a&=&\displaystyle{-\frac{1}{2}\theta gf^{abc}c^bc^c},\\
         \delta \Bar{c}^a&=&\displaystyle{\theta B^a},\\
         \delta B^a&=&0.
    \end{array}
\end{equation}
The first two transformation rules remains literally the same as in the on-shell version \eqref{2.37}, whereas, the last relation in \eqref{2.37} has already changed. Consequently, this additional transformation rules make the computation of the double BRST transformation of the anti-ghost field closed without imposing the equation of motion at all, i.e.
\begin{equation}
    \delta_2\delta_1\Bar{c}^a=\delta_2(\theta_1 B^a)=0.
\end{equation}
One would ask whether on-shell representation Lagrangian \eqref{2.23} can be restored from the off-shell one in some circumstance? To answer the question, we firstly emphasize the fact that the NL field is non-dynamical, then it can be effectively thought of as slow mode with respect to the rest dynamical degrees of freedom. As a result, the slow mode can be approximately treated as a classical spectrum being able to be integrated out from the quantum path integration through the saddle point approximation. In the action functional's level, saddle point approximation is a domination of the classical equation of motion of $B^a$ field, which is very easy to compute out explicitly. It takes the form
\begin{equation}
    \label{2.49}
    B^a=-\frac{1}{\xi}\partial_\mu A^a_\mu.
\end{equation}
Plugging this equation of motion \eqref{2.49} into the Lagrangian \eqref{2.46} to get precisely the on-shell Lagrangian \eqref{2.23} back. This is what we actually expected since the NL field helps the BRST symmetry to be closed off-shell when we lose it the equation of motion is, therefore, essential again.\\
\indent Since we have already introduced the NL auxiliary field, we will specify the particular forms of both ghost numbers charge $Q_c$ and BRST charge $Q_B$ (also in terms of $B$ field) as the last ingredient to be used in the next section. Beginning with using the formula to derive the general form the Noether current from any quantum field textbook you like. We obtain ghost number current as
\begin{equation}
    \label{2.50}
    J^{ghost}_\mu=-i\left(\Bar{c}\frac{\partial \mathcal{L}}{\partial (\partial^\mu\Bar{c})}+\frac{\partial \mathcal{L}}{\partial(\partial^\mu c)}c\right),
\end{equation}
where $-i$ has been added to future purpose and color indices have been omitted. Observe that we are very careful about the order of the fields inside the expression \eqref{2.50} since both ghosts, anti-ghosts and the derivative of Lagrangian density with respect to them are all Grassmannian. One might be confused with how the derivative of Lagrangian with respect to (anti-)ghost is fermionic, it follows from the fact that the Lagrangian density is bosonic thus its derivative with fermion is fermionic. Next, using the definition \eqref{2.35} to get the ghost number charge of the form 
\begin{equation}
\label{2.51}
    Q_c=-i\int d^{d}x\;\left(\Bar{c}\pi_{\Bar{c}}+\pi_{c}c\right),
\end{equation}
where $\pi_\varphi$ is a canonical momentum conjugate to field $\varphi$. Although we will keep studying this form of charge without specifying the form of canonical momenta to be convenient for calculating the (anti-)commutation relations, the explicit form of the canonical momenta will be sometime clearly useful in some situation. All of the relevant canonical momenta are listed below
\begin{equation}
    \label{2.52}
    \begin{array}{ccc}
\pi_i^a         &=& F^a_{0i}, \\
        \pi^a_{\Bar{c}} &=& (D_0c)^a, \\
        \pi_c^a &=& -\partial_0 \Bar{c}^a,\\
        \pi^a_B &=& -A^a_0.
        \end{array}
\end{equation}
Let's calculate the same-time commutation relation between the ghost number charge and ghost field. It is very straightforward to have (the first term in \eqref{2.51} contributes no ghost field so dropping out)
\begin{equation}
\label{2.53}
   \begin{split}
         [Q_c,c^a(x)]&=-i\displaystyle{\int d^{d}y\;[\pi_c^b(y)c^b(y),c^a(x)]_{x_0=y_0}}  \\
        & =-i\displaystyle{\int d^{d}y\;[\pi_c^b(y)c^b(y)c^a(x)-c^a(x)\pi_c^b(y)c^b(y)]}\\
        &=i\displaystyle{\int d^{d}y\;[\pi_c^b(y)c^a(x)c^b(y)+c^a(x)\pi_c^b(y)c^b(y)]}\\
        &=i\displaystyle{\int d^{d}y\;\{\pi_c^b(y),c^a(x)\}c^b(y)}\\
        &=\displaystyle{i\int d^{d}y(-i\delta^{ab}\delta^{d}(x-y)c^b(y))}\\
        &= c^a(x),
   \end{split}
\end{equation}
where we have used, due to the fact that ghost is a fermion, the canonical anti-commutation relation $\{\pi_c(y)^b,c^a(x)\}=-i\delta^{ab}\delta^{d}(x-y)$.
The meaning of the result in \eqref{2.53} is clearly powerful in the sense that it demonstrates that $Q_c$ represents the operator counting the number of ghosts. It will be more obvious if we assign how this operator acts to the eigenstate $\big|n\big>$ with $n$ ghost numbers. That implies if we demand that the eigenvalue of this state denoting the total numbers of the ghost of that state, everything will be consistent quantum mechanically. We require
\begin{equation}
\label{2.54}
    Q_c\big|n\big>=n\big|n\big>,
\end{equation}
leading to the consequence that the state of form $c\big|n\big>$ will have $n+1$ ghost numbers, i.e.
\begin{equation}
    Q_c(c\big|n\big>)=[Q_c,c]\big|n\big>+cQ_c\big|n\big>=c\big|n\big>+nc\big|n\big>=(n+1)(c\big|n\big>).
\end{equation}
On the other hand, we can also calculate the commutation relation between this charge and anti-ghost field. Without losing any sweat, we get
\begin{equation}
    [Q_c,\Bar{c}]=-\Bar{c},
\end{equation}
deducing that the anti-ghost field carries reasonable $-1$ ghost number.\\
\indent The next relevant Noether charge, the BRST charge, can be also found directly to be
\begin{equation}
\label{2.57}
    Q_B=i\int d^{d}x\;\left[\pi^a_i(D_ic)^a-\frac{g}{2}f^{abc}\pi^a_cc^bc^c+\pi^a_{\Bar{c}} B^a\right],
\end{equation}
where indices $i$ stand for the spatial components of the gauge field $A_i$. Note that the reason why we can ignore the time component is $\pi_0=F_{00}=0$ at the first place. One might also rewrite this form of charge further by using the equation of motion of the gauge field in presence of ghosts as follows
\begin{equation}
    D_\mu^{ab} F_{\mu\nu}^b=-\partial_\nu B^a+gf^{abc}\partial_\nu \Bar{c}^bc^c,
\end{equation}
choosing only the time component $\nu=0$, we have other form of the equation of motion
\begin{equation}
\label{2.59}
    D_i^{ab}F_{i0}^b=-D_i^{ab}\pi_i^b=-\partial_0 B^a+gf^{abc}\pi_c^b c^c,
\end{equation}
where the definition \eqref{2.52} has been used. Following these step, we multiple both sides by $c^a$ into the right hand side and then perform a simple algebra to rewrite the middle expression of \eqref{2.59}, i.e.
\begin{equation}
\label{2.60}
\begin{split}
     -(D_i^{ab}\pi_i^b)c^a&=-(\partial_i\pi^a_i)c^a+gf^{abc}\pi^b_iA^c_ic^a\\
     &=-\partial_i (\pi^a_ic^a)+\pi^a_i\partial_ic^a-\pi^a_igf^{abc}A^c_ic^b\\
     &=-\partial_i(\pi^a_ic^a)+\pi^a_iD_i^{ab}c^b.
\end{split}
\end{equation}
Substitute the result \eqref{2.60}, combining with the equation \eqref{2.59} into the expression of the BRST charge \eqref{2.57} to obtain
\begin{equation}
\label{2.61}
    Q_B=i\int d^{d}x\left[-(\partial_0B)^ac^a+\frac{g}{2}f^{abc}\pi^a_cc^bc^c+\pi^a_{\Bar{c}} B^a\right],
\end{equation}
where the extra total spatial derivative term in \eqref{2.60} has already been integrated out at the boundary of the space.\\
\indent In constrast to the ghost number charge, the BRST charge is fermionic generator, thus it will be very handy later if we introduce a so-called $\mathbb
Z_2-$graded Lie bracket or super Lie bracket, suppose $A$ and $B$ are any fields, defined by
\begin{equation}
    [A,B\}=AB-(-1)^{g(A)g(B)}BA,
\end{equation}
where $g(A)$ denoted the grade of operator $A$ which defined to be $0$ for bosonic operator and $1$ for fermionic one. It satisfies two important properties that are
\begin{enumerate}
    \item Super skew-symmetry
    \begin{equation}
    \label{2.63}
        [A,B\}=-(-1)^{g(A)g(B)}[B,A\}.
    \end{equation}
    \item Super Jacobi identity
    \begin{equation}
        \label{2.64}
        (-1)^{g(A)g(C)}[A,[B,C\}\}+(-1)^{g(B)g(A)}[B,[C,A\}\}+(-1)^{g(C)g(B)}[C,[A,B\}\}=0.
    \end{equation}
\end{enumerate}
This new kind of bracket seems to be quite abstract but it is not that much. We can see that the super Lie bracket will reduce back to the original (anti-)commutator by two sufficient conditions
\begin{equation}
    [A,B\}=
    \begin{cases}
    [A,B]&\mathrm{if\; atleast\;one\;of\;them\;is\;bosonic}\\
    \{A,B\}&\mathrm{if\;both\;of\;them\;are\;fermionic}
    \end{cases}.
\end{equation}
With the new fancy bracket, we can write compactly the $\mathbb{Z}_2-$graded canonical relations among all relevant fields as
\begin{equation}
\label{2.66}
  [\varphi^a_I(x),\varphi^b_J(y)\}=0,\;\;\; \;\;\;\;\;\;\;\; [\pi_{\varphi^a_I}(x),\varphi^b_J(y)\}=-i\delta_{IJ}\delta^{ab}\delta^{d}(x-y),
\end{equation}
where $I,\;J$ are used to label the species of the particles.
Nevertheless, one can find through straightforward calculations that the BRST charge generates the BRST symmetry transformation which follows from the converse argument of the Noether's theorem stating that the Noether charge will generate its associated symmetry transformation. Mathematically, this statement shows by the supercommutation relation
\begin{equation}
\label{2.67}
    [Q_B,\varphi\}=\delta \varphi,
\end{equation}
where $\delta$ is the BRST differential we discussed earlier by redefining the infinitesimal parameter out and $\varphi$ used to represent any field appearing in the quantum YM theory including the NL field also.\\
\indent Let's move to the final supercommutation relations which are the (anti-)commutations among charges themselves. Note that we already have one anti-commutation relation between two BRST charges \eqref{2.39} implying the nilpotency of the BRST differential. By the way, we need to be careful about this argument, it is okay to claim that the nilpotency of the BRST charges implies the nilpotency of the BRST differential but it is not necessary to be true in a converse way. To sketch the proof of our proposition we have claimed, let's consider the super Jacobi identity \eqref{2.64} among $(Q_B,Q_B,\mathcal{O})$ for any operator $\mathcal{O}$. We have
\begin{equation}
    \label{2.68}
    \begin{split}
        (-1)^{g(\mathcal{O})}[Q_B,[Q_B,\mathcal{O}\}\}+(-1)^{g(\mathcal{O})}[\mathcal{O},\{Q_B,Q_B\}\}+(-1)^{1}[Q_B,[\mathcal{O},Q_B\}\}&=0\\
        (-1)^{g(\mathcal{O})}[Q_B,[Q_B,\mathcal{O}\}\}+(-1)^{g(\mathcal{O})}[\mathcal{O},\{Q_B,Q_B\}\}+(-1)^{g(\mathcal{O})}[Q_B,[Q_B,\mathcal{O}\}\}&=0\\
        2[Q_B,[Q_B,\mathcal{O}\}\}+[\mathcal{O},\{Q_B,Q_B\}\}&=0\\
        2\delta^2\mathcal{O}+[\mathcal{O},\{Q_B,Q_B\}\}&=0,
    \end{split}
\end{equation}
where we have noted that $[Q_B[Q_B,\mathcal{O}\}\}=\delta^2\mathcal{O}$ and the super skew-symmetry has been used. We actually see reasonably that if $2Q^2_B=\{Q_B,Q_B\}=0$, then $\delta^2\mathcal{O}=0$ for any operator $\mathcal{O}$. On the other hand, in the converse way, if we demand $\delta^2\mathcal{O}$, it does not essentially imply the nilpotency of the BRST charge $Q_B$, in particular, it can imply that the anti-commutation relation between two $Q_B$ yields any constant number. On the bright side, $\{Q_B,Q_B\}=0$ can be ensured by the direct calculation.
\begin{equation}
    \label{2.69}
    \begin{split}
        \{Q_B,Q_B\}=&\;i\int d^{d}x\;\{Q_B,-(\partial_0B)^ac^a-\frac{g}{2}f^{abc}\partial_0\bar{c}^ac^bc^c+D_0c^a B^a\}\\
        =&\;i\int d^{d}x\;(-[Q_B,\partial_0B^a]c^a-\partial_0 B^a\{Q_B,c^a\}-\frac{g}{2}[Q_B,f^{abc}c^bc^c]\partial_0\bar{c}^a\\
        &-\frac{g}{2}f^{abc}c^bc^c\{Q_B,\partial_0\bar{c}^a\}+[Q_B,B^a]D_0c^a+B^a\{Q_B,D_0c^a\})\\
        =&\; i\int d^{d}x\;\left[-\partial_0B^a\left(-\frac{g}{2}f^{abc}c^bc^c\right)-\frac{g}{2}f^{abc}c^bc^c\partial_0B^a\right]\\
        =&\;0,
    \end{split}
\end{equation}
where we have used \eqref{A.4}, \eqref{2.47}, \eqref{2.52}, \eqref{2.67} and the fact that $\delta(D_\mu c)=0$, which will be verified in the appendix (see \eqref{A.3}). Combining with $\delta(f^{abc}c^bc^c)$ as we have done verifying indirectly in \eqref{2.38}.
\\
\indent The rest of them are also not too hard to calculate by hand. Since $Q_c$ is bosonic charge, the remaining relation will be the commutation relations only. That are
\begin{equation}
\label{2.70}
    \begin{split}
        [Q_c,Q_B] =&-[Q_B,Q_c]  \\
         =&\;i\displaystyle{\int d^{d}x\;[Q_B,\Bar{c}^aD_0 c^a-(\partial_0\Bar{c}^a)c^a}]\\
         =&\;i\displaystyle{\int d^{d}x\;(\{Q_B,\Bar{c}^a\}D_0c^a-\Bar{c}^a\{Q_B,D_0c^a\}-\{Q_B,\partial_0\Bar{c}^a\}c^a}\\
         &+\partial_0\Bar{c}^a\{Q_B,c^a\})\\
         =&\;i\int d^{d}x\;\left[B^aD_0 c^a-\partial_0 B^a c^a+\partial_0\Bar{c}^a\left(-\frac{g}{2}f^{abc}c^bc^c\right)\right]\\
         =&\;Q_B,
    \end{split}
\end{equation}
Finally, the last commutation relation is simpler to be obtained, without deriving, as
\begin{equation}
    \label{2.71}
    [Q_c,Q_c]=0.
\end{equation}
One might not notice how powerful   relations \eqref{2.70} and \eqref{2.71} are. Since we have known from \eqref{2.53} and \eqref{2.54} that the result of the commutator between ghost number charge and any operator will determine the specific ghost number of that operator. Thus, the equation \eqref{2.71} tells us that the ghost number of the ghost number charge must be zero which is, in fact, clear in the sense that the operator that used to count the number of ghosts must contain no any ghost number. To understand our statement, one may think about the number operator in the quantum harmonic oscillation that constructed from the combination between one creation and one annihilation. Strictly speaking, the number operator must create and annihilate nothing.\\
\indent On the other hand, the equation \eqref{2.70} implies the wonderful consequence that the BRST charge carries $+1$ ghost number. However, this is what we expected since, by construction, we set up the BRST transformation parameter to involve the ghost field at the first place (see \eqref{2.36}). The reader who is familiar with differential topology might have noticed that the properties of the BRST charge, that are nilpotency $Q_B^2=0$ and increasing ghost number by one, are quite familiar. Let us remind the reader a little bit about this structure by giving the simplest example of such a structure. Suppose, there are differential $p$-forms, sometime $p$ is called a degree, defined on the manifold $\mathcal{M}$, whose set is denoted by $\Omega^p(\mathcal{M})$ and we can define a so-called exterior derivative $d$ actually map $p$-form to $(p+1)$-form, i.e. $d:\Omega^p(\mathcal{M})\rightarrow \Omega^{p+1}(\mathcal{M})$ and also satisfies the nilpotent property, explicitly, $d^2=0$. Combine all of this ingredient to construct the structure, known as the deRham complex, expressed out as a sequence
\begin{equation}
    ...\xrightarrow{d}\Omega^{p-1}(\mathcal{M})\xrightarrow{d}\Omega^{p}(\mathcal{M})\xrightarrow{d}\Omega^{p+1}(\mathcal{M})\xrightarrow{d}...
\end{equation}
The deRham complex defines the corresponding deRham cohomology as following steps: First, we suppose that there is a so-called closed form defined to be the differential $p$-form that vanishes by the action of exterior derivative or explicitly $d\omega=0,\;\omega\in \Omega^p(\mathcal{M})$. We will denote the set of closed p-forms as $Z^p(\mathcal{M})$ which, in fact, forms a group structure. Due to the nilpotency of the exterior derivative, $d^2=0$, there must be a (normal) subgroup of $Z^p(\mathcal{M})$ denoted as $B^p(\mathcal{M})$ consisting of a so-called exact $p$-forms satisfying $\omega=d\alpha,\;\alpha \in\Omega^{p-1}(\mathcal{M})$. We can therefore define the deRham cohomology group as a quotient space
\begin{equation}
    H^p(\mathcal{M})\equiv Z^p(\mathcal{M})/B^p(\mathcal{M})=\ker d_p/\mathrm{im}\; d_{p-1},
\end{equation}
where $d_r$ is the exterior derivative acting on the differential $r$-forms, $\ker$ and $\mathrm{im}$ stands for kernel and image, respectively.
Note that the dimension of deRham cohomology group is related to the Betti number used to calculate the Euler characteristic class of such manifold. Consequently, the non-trivial cohomology structure implies a non-trivial topological structure of the theory.\\
\indent In QYM theory, $Q_B$ behaves the same way as the exterior derivative but acting on the different structures. In parallel to the construction of the deRham cohomology. Let's suppose that there are operator-valued distribution carrying n-numbers of ghost charge whose set is denoted by $TC^n$ (the symbol might seem not to be related at all, on the bright side, it is truly related. $TC^n$ is generally known as the total complex \cite{Jose}). The corresponding differential, in analogue to exterior derivative, is nothing but the BRST differential $\delta$ defined by $\delta \mathcal{O}\equiv [Q_B,\mathcal{O}\}$ where $\mathcal{O}$ be any operator of $n$ ghost numbers $\in TC^n$ and the (classical) BRST charge $Q_B$ is an element of $TC^1$. These structures, again, define the BRST complex structure as a sequence
\begin{equation}
    ...\xrightarrow{\delta} TC^0\xrightarrow{\delta} TC^1\xrightarrow{\delta}...\xrightarrow{\delta} TC^n\xrightarrow{\delta} ...
\end{equation}
Abstractly, this kind of structure is said to admit a so-called $\mathbb{Z}$-graded Poisson superalgebra structure \cite{Jose} where the state is graded by an integer ($\mathbb{Z}$) which is rigidly an eigenvalue of the ghost charge. Additionally, \say{super} means that the $\mathbb{Z}_2$ gradation still be there in the sense that even(odd) $\mathbb{Z}$ graded state will behave the same way as the degree zero(one) with respect to the $\mathbb{Z}_2$ gradation's perspective.\\
\indent Since we have already shown that the BRST differential is truly nilpotent $\delta^2=0$, hence we can define the BRST cohomology, in analogue to the deRham cohomology, as following way
\begin{equation}
\label{2.75}
    H^n_\delta\equiv \ker \delta_n/\mathrm{im}\;\delta_{n-1}.
\end{equation}
One might loudly shout out immediately that all of this stuff is just an abstract mathematical structure so how important is it physically? This mathematical stuff is generally used to identify the physical space of the state. For the following treatment, we will follow the argument in the reference \cite{Deligne:1999qp}. First of all, let's recall the partition function of the QYM theory in the form that the normalization constant is ignored and the source field is turning off. We thus have
\begin{equation}
    Z(0)=\int \mathcal{D}A \int \mathcal{D}B\int \mathcal{D}c\int \mathcal{D}\bar{c}\; e^{-S_{QYM}},
\end{equation}
and the correlation function of an operator $\mathcal{O}$ defined to be 
\begin{equation}
    \big<\mathcal{O}\big>\equiv\frac{1}{Z(0)}\int \mathcal{D}A \int \mathcal{D}B\int \mathcal{D}c\int \mathcal{D}\bar{c}\; e^{-S_{QYM}}\mathcal{O}.
\end{equation}
By the way, if we require $\mathcal{O}$ to be physical, we need to expect that $\mathcal{O}$ is also gauge invariant (BRST invariant in the quantum regime), explicitly, it need to sufficiently satisfy the condition $\delta \mathcal{O}=0$. Mathematically, it said to be that operator is an element of the kernel of the BRST differential map, $\mathcal{O}\in \ker \delta$. The ensemble average of such an operator can be seen easily that it is also invariant under the BRST transformation, i.e.
\begin{equation}
\label{2.78}
\begin{split}
    \delta\big<\mathcal{O}\big>&=\delta\left(\frac{1}{Z(0)}\int \mathcal{D}A \int \mathcal{D}B\int \mathcal{D}c\int \mathcal{D}\bar{c}\; e^{-S_{QYM}}\mathcal{O}\right)\\
    &=\frac{1}{Z(0)}\int \mathcal{D}A \int \mathcal{D}B\int \mathcal{D}c\int \mathcal{D}\bar{c}\; e^{-S_{QYM}}\delta\mathcal{O}=0.
\end{split}
\end{equation}
One might notice suddenly that we have required two sub-conditions to make \eqref{2.78} hold consistently. The first requirement is the quantum YM action is invariant under the BRST transformation which is extremely obvious since if it were not, we would have no idea what meaningless things we have been dealing with so far. The second one is quite not too obvious but we need to pretend that it really holds in the present case. The second condition requires all of path integration measures to be Haar measures with respect to the BRST transformation. Strictly speaking, the measure needs to be invariant under the BRST transformation, in other words, it is lack of the BRST anomalies. The discussion of the situation that BRST symmetry is anomalous can be seen in, for example, \cite{L'Yi2007}. \\
\indent Suppose that these two essential conditions are satisfied, let's consider the BRST exact operator of n degrees $\mathcal{\Tilde{O}}\in \mathrm{im}\;\delta$, i.e. $\mathcal{\Tilde{O}}=\delta \mathcal{O}'$ for any operator $\mathcal{O}'$ carrying $n-1$ ghost numbers. We end up with the fact that an expectation value of $\mathcal{O}$ stays zero no matter what (can be observed directly from \eqref{2.78}). Frankly speaking, it implies further that the probability (expectation value) to find a non-vanishing operator $\mathcal{O}$ in any situation is precisely zero. Thus, this shows that all gauge-invariant physical states belong to the BRST cohomology group \eqref{2.75} since all BRST exact states are generally unphysical by its nature. To be more specific, the physical states only belong to the 0th-order BRST cohomology or, in this sense, it means that the physical state must carry only zero number of ghost charges. Actually, this necessary condition will be clarified in the next section below.

\section{Kugo-Ojima Quartet Mechanism}
As we have already emphasized once in section 2.1 that ghosts and anti-ghosts have to be unphysical degrees of freedom since they do violate the spin-statistics theorem inevitably, we will try to convince the reader to accept that this fact is definitely unacceptable for unitary quantum field theories. First of all, we will start with performing the plane wave modes expansion of these two quantum fields out as follows
\begin{equation}
    \label{2.79}
    \begin{split}
        c^a(x)&=\int \frac{d^{d}k}{(2\pi)^{d}}\frac{1}{\sqrt{2\omega_k}}\;(-c^a_k e^{-i\mathbf{k}\cdot\mathbf{x}}+(c^a_k)^\dagger e^{i\mathbf{k}\cdot\mathbf{x}}),\\
         \bar{c}^a(x)&=\int \frac{d^{d}k}{(2\pi)^{d}}\frac{1}{\sqrt{2\omega_k}}\;(\bar{c}^a_k e^{-i\mathbf{k}\cdot\mathbf{x}}+(\bar{c}^a_k)^\dagger e^{i\mathbf{k}\cdot\mathbf{x}}),
    \end{split}
\end{equation}
where $\omega_k$, $\varphi_k$ and $\varphi_k^\dagger$ are an associated energy, annihilation operator and creation operator, respectively. Note that the signs in the expressions \eqref{2.79} are chosen to be consistent with the Hermiticity assignment of ghosts \eqref{2.33}. Plugging these expressions into the super-commutation relations \eqref{2.66} to obtain the relationship among the creation-annihilation modes as
\begin{equation}
    \label{2.80}
    \{(\bar{c}^a_k)^\dagger,c^b_q\}=\{(c^a_k)^\dagger,\bar{c}_q^b\}=-\delta^{ab}\delta^{d}(\mathbf{k}-\mathbf{q}),
\end{equation}
while the remaining all vanish. So far we are focusing on the operator formalism, here let's move to the state consideration to explore the structure of the vector space associated to this quantum theory. Let's firstly define the simplest state called a vacuum state $\big|0\big>$ by requiring this state to be annihilated by an annihilation operator of any value of momentum, i.e. $\varphi^a_k\big|0\big>=0$ and satisfying the normalization condition $\big<0\big|0\big>=1$. After we define this state, all particle states of momentum $k$ and color $a$, $\big|\varphi^a_k\big>$, can be constructed directly by acting the creation operator into the vacuum one, namely, $\big|\varphi^a_k\big>=(\varphi^a_k)^\dagger\big|0\big>$. Unfortunately, the story about the theory involving ghosts is too far from the word \say{similar} to the ordinary theory in QFT 101. To understand our argument, let's consider the inner product of two states, e.g.
\begin{equation}
    \label{2.81}
    \begin{split}
        \big<\bar{c}^a_k\big|c^b_q\big>&=\big<0\big|\bar{c}^a_k (c^b_q)^\dagger\big|0\big>\\
        &=\big<0\big|\{\bar{c}^a_k, (c^b_q)^\dagger\}\big|0\big>-\big<0\big| (c^b_q)^\dagger \bar{c}^a_k\big|0\big>\\
        &=-\delta^{ab}\delta^{d}(\mathbf{k}-\mathbf{q})\big<0\big|0\big>=-\delta^{ab}\delta^{d}(\mathbf{k}-\mathbf{q}),
    \end{split}
\end{equation}
and we can always find easily that $\big<c^a_k\big|\bar{c}^b_q\big>$ also give the same result. The simple computation \eqref{2.81} implies really strong consequence that is the vector space of the quantum YM theory at the beginning has an indefinite metric. Hence, such a vector space is absolutely not a Hilbert space (so Fock space). So? What kind is the problem anyway? Physically speaking, if the metric's definiteness of the vector space associated to the quantum field theory is not guaranteed, there will be some situation in the real world that have negative probability! This obviously leads to the violation of the unitarity without question. \\
\indent At this point, let us convince the reader even more that this situation truly follows from the violation of the spin-statistics theorem. Suppose that (anti-)ghosts does not violate the spin-statistics theorem, this implies that (anti-)ghost, which is the spin-0 particle, must satisfy the commutation relations 
\begin{equation}
    \label{2.82}
    [(\bar{c}^a_k)^\dagger,c^b_q]=[(c^a_k)^\dagger,\bar{c}_q^b]=-\delta^{ab}\delta^{d}(\mathbf{k}-\mathbf{q}).
\end{equation}
We see precisely that the computation \eqref{2.81} by using the relations \eqref{2.82} yields the positive-definite result
\begin{equation}
    \label{2.83}
    \begin{split}
        \big<\bar{c}^a_k\big|c^b_q\big>&=\big<0\big|\bar{c}^a_k (c^b_q)^\dagger\big|0\big>\\
        &=\big<0\big|[\bar{c}^a_k, (c^b_q)^\dagger]\big|0\big>+\big<0\big| (c^b_q)^\dagger \bar{c}^a_k\big|0\big>\\
        &=\delta^{ab}\delta^{d}(\mathbf{k}-\mathbf{q})\\&\geq 0,
    \end{split}
\end{equation}
This shows that the unitarity is safe in the light of the satisfaction of the spin-statistics theorem. Thus we have finished the sketch of proof of our statement.\\
\indent To summarize, due to the violation of the spin-statistics theorem, ghost and anti-ghost states generate the indefinite metric that makes Hilbert(Fock) space ill-defined in both mathematical and physical senses. Once upon a time, people have been figured the actual way to solve this difficulty properly, we will just follow their path of understanding from now on. The way they do is to suppose that the true physical state, $\big|\psi\big>$ said, spans only the subspace $\mathcal{V}_{phys}$ of the total space of all states in the QYM theory denoted by $\mathcal{V}$. By the help of a so-called subsidiary condition chosen to be of the form
\begin{equation}
\label{2.84}
    Q_B\big|\psi\big>=0,
\end{equation}
as firstly suggested by Curci and Ferrari \cite{Curci1976}, we will be able to project $\mathcal{V}$ into its physical subspace $\mathcal{V}_{phys}$. One might ask curiously how could people figure out such a condition? In fact, this condition means that the physical states must be gauge invariant, translating into the requirement of BRST invariance in the quantum level, which is quite making sense intuitively.\\
\indent Note further that the non-Abelian subsidiary condition \eqref{2.84} can be used to restore the traditional subsidiary condition in QED in some circumstances. In particular, in the Abelian limit that $f^{abc}=0$, the BRST charge \eqref{2.61} will reduce to
\begin{equation}
    Q^{Abelian}_B=\int d^{d}x\;:B(x)\partial_0 c(x)-(\partial_0B(x))c(x):
\end{equation}
where :: is a normal ordering operation, defined such that all annihilation operators in this operation will be relocated into the right of creation operators, inserted into the definition of the Noether charge to get rid of the ambiguity about the order of the operators after quantized. Once again, the NL field can be also expanded in terms of its creation-annihilation modes. The calculation of the time derivative of any operator can be done easily in the Heisenberg picture (in the end, we can work in any picture since all representations are related to one another through unitary transformation according to the Stone-von Neumann theorem \cite{Stone1932,vNeumann1932}) of such a QFT. By assigning the NL field to be a Hermitian operator-valued field, as it should be, we end up with 
\begin{equation}
    \label{2.86}
    \begin{split}
        \partial_0c(x)&=i\int \frac{d^{d}k}{(2\pi)^{d}}\sqrt{\frac{\omega_k}{2}}\;(c_ke^{-i\mathbf{k}\cdot\mathbf{x}}+c^\dagger_ke^{i\mathbf{k}\cdot\mathbf{x}}),\\
        \partial_0B(x)&=i\int \frac{d^{d}k}{(2\pi)^{d}}\sqrt{\frac{\omega_k}{2}}\;(-B_ke^{-i\mathbf{k}\cdot\mathbf{x}}+B^\dagger_ke^{i\mathbf{k}\cdot\mathbf{x}}).
    \end{split}
\end{equation}
The explicit computation of the BRST charge is quite tricky but straightforward. Let's try evaluate term by term, e.g. the first term of the product $\int d^{d-1}x:B\partial_0c:$
\begin{equation}
    \begin{split}
        &=\frac{i}{2}\int d^{d}x\int\frac{d^{d}k}{(2\pi)^{d}}\int\frac{d^{d}q}{(2\pi)^{d}}\sqrt{\frac{\omega_q}{\omega_k}}:B_kc_q:e^{-i\mathbf{(k+q)\cdot x}}\\
        &=\frac{i}{2}\int\frac{d^{d}k}{(2\pi)^{d}}\int\frac{d^{d}q}{(2\pi)^{d}}\sqrt{\frac{\omega_q}{\omega_k}}:B_kc_q:(2\pi)^d\delta^d(\mathbf{k+q})\\
        &=\frac{i}{2}\int\frac{d^{d}(-k)}{(2\pi)^{d}}\sqrt{\frac{\omega_{-k}}{\omega_k}}:B_kc_{-k}:\\
        &=\frac{i}{2}\int \frac{d^dk}{(2\pi)^d}:B_kc_{-k}:,
    \end{split}
\end{equation}
where we have used the integral representation of a Dirac delta function \eqref{A.20}, the fact that $\omega_{-k}=\omega_{k}$ (can be seen explicitly by expressed the energy out using the dispersion relation) and the measure $d^dk$ is invariant under the reflection symmetry $k\rightarrow-k$ since this integral measure has already integrate over the whole space. Through the long calculations, we end up with the simple result
\begin{equation}
\label{2.88}
    Q_B^{Abelian}=i\int\frac{d^dk}{(2\pi)^d}\;(c^\dagger_k B_k-B^\dagger_k c_k).
\end{equation}
Next substituting the result \eqref{2.88} into \eqref{2.84} combining also with the fact that the ghost fields in an Abelian gauge theory are constant modes contributing nothing to the physical state. Intuitively, the ghost's annihilation operator will instantly annihilate the physical states while the creation operator of such a field will shift the physical states by a meaningless constant. Thus, we deduce the subsidiary condition for an Abelian gauge theory to be of the form
\begin{equation}
\label{2.89}
    i\int\frac{d^dk}{(2\pi)^d}\;B_k\big|\psi\big>\equiv B^{(+)}\big|\psi\big>=0.
\end{equation}
This condition is sometimes called the Nakanishi-Lautrup quantization condition which is the subsidiary condition used for canonically quantizing an Abelian gauge theory in any linear covariant gauge fixing condition \cite{Lautrup:1967zz,Nakanishi1966,Nakanishi1967}. This condition is still not familiar for many people since, in QFT 101, we formally work with the canonical quantization of gauge theory in the Feynman-'t Hooft gauge. In particular, we can reduce our consideration from the linear covariant gauge into the specific Feynman-'t Hooft gauge by setting the Gaussian width to be unity, i.e. $\xi=1$. The subsidiary condition reduces into the most familiar form called the Gupta-Bleuler condition \cite{Bleuler1950,Gupta1950}
\begin{equation}
\label{2.90}
    \partial_\mu A_\mu^{(+)}\big|\psi\big>=0,
\end{equation}
where we have substituted the equation of motion of the NL field (2.49) into \eqref{2.89} to obtain \eqref{2.90}. \\
\indent The time has come, it is time to study the asymptotic behaviors of all these fields we focusing on. To study this, we first derive the following relations, by considering
\begin{equation}
\label{2.91}
\begin{split}
    [Q_B,\hat{T}(\mathcal{O}_1(x_1)...\mathcal{O}_n(x_n))\}&=\delta(\hat{T}(\mathcal{O}_1(x_1)...\mathcal{O}_n(x_n)))\\
    &=\sum^n_{i=1}\hat{T}(\mathcal{O}_1(x_1)...\delta\mathcal{O}_i(x_i)...\mathcal{O}_n(x_n)),
\end{split}
\end{equation}
where the operator $\hat{T}$ is, known as the time ordering operator, used to re-align the operators inside the expectation value into the right chronological order. It is very useful to know the explicit definition of the time ordering operator which is defined as the following form
\begin{equation}
    \label{2.92}
    \hat{T}(\mathcal{O}(x)\mathcal{O}(y))=\theta(x^0-y^0)\mathcal{O}(x)\mathcal{O}(y)\pm \theta(y^0-x^0)\mathcal{O}(y)\mathcal{O}(x),
\end{equation}
the particular sign of \say{$\pm$} will be chosen to be minus sign if both operators are fermionic, otherwise, it will choose to be plus sign and the operator $\theta(x)$ is a Heaviside step-function defined as
\begin{equation}
    \label{2.93}
    \theta(x)=\begin{cases}
    0,&x<0\\
    1,&x\geq 0.
    \end{cases}
\end{equation}
The next step is sandwiching the expression \eqref{2.91} by the vacuum states $\big|0\big>$, we thus obtain
\begin{equation}
    \label{2.94}
        \sum^n_{i=1}\big<\mathcal{O}_1...\delta\mathcal{O}_i...\mathcal{O}_n\big>=\big<0\big|[Q_B,\hat{T}(\mathcal{O}_1...\mathcal{O}_n)\} \big|0\big>=0,
\end{equation}
where $\big<...\big>\equiv \big<0\big|\hat{T}(...)\big|0\big>$ and we have used the assumption that the vacuum state is a physical state, as it expected to be, so $Q_B\big|0\big>=0$. This relation is sometime called a Ward-Takahashi (WT) identity for a Green function. We will consider the two special cases of this WT identity. That are
\begin{equation}
    \label{2.95}
        \big<0\big|\{Q_B,\hat{T}(B^a(x)\bar{c}^b(y))\}\big|0\big>=\big<B^a(x)B^b(y)\big>=0,
\end{equation}
\begin{equation}
    \label{2.96}
        \big<0\big|\{Q_B,\hat{T}(A^a_\mu(x)\bar{c}^b(y))\}\big|0\big>=\big<D_\mu c^a(x)\bar{c}^b(y)\big>+\big<A^a_\mu(x)B^b(y)\big>
        =0.
\end{equation}
Consider second derivative with respect to the spacetime $x^\mu$ of the first term of the middle expression in \eqref{2.96}. Let's carefully evaluate step-by-step, for first derivative, we have
\begin{equation}
\label{2.97}
\begin{split}
    \frac{\partial}{\partial x^\mu}\big<D_\mu c^a(x)\bar{c}^b(y)\big>=&\;\big<0\big| \frac{\partial}{\partial x_\mu}\theta(x^0-y^0)D_\mu c^a(x)\bar{c}^b(y)\big|0\big>\\
    & -\big<0\big| \frac{\partial}{\partial x_\mu}\theta(y^0-x^0)\bar{c}^b(y)D_\mu c^a(x)\big|0\big>\\
    =&\;\delta(x^0-y^0)\big<\{D_\mu c^a(x),\bar{c}^b(y)\}\big>\\
    =&-i\delta^{ab}\delta^d(x-y).
\end{split}
\end{equation}
Note that the derivative with respect to $x^\mu$ of $D_\mu c^a(x)$ definitely vanishes with the help of the equation of motion \eqref{2.41} so we therefore did not care about it in the calculation above \eqref{2.97}. Moreover, the identity $\frac{\partial}{\partial x}\theta(x)=\delta(x)$, \eqref{2.52} and \eqref{2.66} have been used in the sub-sequential steps. For the second derivative, it is easy to get
\begin{equation}
    \label{2.98}
    \square_x \big<D_\mu c^a(x)\bar{c}^b(y)\big>=-i\delta^{ab}\frac{\partial}{\partial x^\mu}\delta^d(x-y),
\end{equation}
after this step, we will perform the Fourier transformation in both sides of the relation \eqref{2.98}. Effectively, it is the same as we translate from $\partial_\mu\rightarrow -ip_\mu$. Hence,
\begin{equation}
\label{2.99}
    \hat{\mathcal{F}}\big<D_\mu c^a(x)\bar{c}^b(y)\big>=\delta^{ab}\frac{p_\mu}{p^2}.
\end{equation}
We can also compute other relation by, once again, taking second derivative with respect to $x^\mu$. This time, we will use the equation of motion of the NL field (2.49) with the WT identity \eqref{2.95}. In the end of all processes, it yields
\begin{equation}
    \label{2.100}
    \hat{\mathcal{F}}\big<A^a_\mu(x)B^b(y)\big>=-\delta^{ab}\frac{p_\mu}{p^2}.
\end{equation}
One might ask, we sure reader will, that so far what is the purpose of doing all of these stuffs? The physical meaning of the momentum space expectation value between two fields, a two-point function, is nothing but a momentum space propagator. Thus, we can read off the pole structures from both relations \eqref{2.99} and \eqref{2.100}. As a result, this shows that the fields in the quantum YM theory admit the massless pole structure. Asymptotically, we can then write all fields in terms of other massless fields
\begin{equation}
    \label{2.101}
    \begin{split}
        A^a_\mu(x)&\rightarrow \partial_\mu \chi^a(x)+...\\
    B^a(x)&\rightarrow \beta^a(x)+...\\
   (D_\mu c)^a(x)&\rightarrow \partial_\mu \gamma^a(x)+...\\
    \bar{c}^a(x)&\rightarrow \bar{\gamma}^a(x)+...
    \end{split}
\end{equation}
\say{...} denotes the term which has different pole structure. An asymptotic version of the super commutation \eqref{2.67} reads
\begin{equation}
\label{2.102}
    \begin{split}
        [Q_B,\chi^a]=\gamma^a,&\;\;\;\;\;\{Q_B,\bar{\gamma}^a\}=\beta^a,\\
        \{Q_B,\gamma^a\}=0,&\;\;\;\;\;[Q_B,\beta^a]=0.
    \end{split}
\end{equation}
Since we are living in the operator level, we can instead work in state level by performing the mode expansion as before. We have obviously the (anti-)commutations among annihilation operators
\begin{equation}
\label{2.103}
     \begin{split}
        [Q_B,\chi^a_k]=\gamma^a_k,&\;\;\;\;\;\{Q_B,\bar{\gamma}^a_k\}=\beta^a_k,\\
        \{Q_B,\gamma^a_k\}=0,&\;\;\;\;\;[Q_B,\beta^a_k]=0.
        \end{split}
\end{equation}
On the other hand, the (anti-)commutation relations among creation operator can be found from taking the Hermitian conjugation into each relation in \eqref{2.103}, then using the Hermiticity properties of (anti-)commutators, i.e. $[A,B]^\dagger=-[A^\dagger,B^\dagger]$ and $\{A,B\}^\dagger=\{A^\dagger,B^\dagger\}$ combining with the Hermiticity of the BRST charge $Q^\dagger_B=Q_B$. We will obtain the right (anti-)commutation relations we have requested. Explicitly,
\begin{equation}
    \label{2.104}
    \begin{split}
        [Q_B,(\chi^a_k)^\dagger]=-(\gamma^a_k)^\dagger,&\;\;\;\;\;\{Q_B,(\bar{\gamma}^a_k)^\dagger\}=(\beta^a_k)^\dagger,\\
        \{Q_B,(\gamma^a_k)^\dagger\}=0,&\;\;\;\;\;[Q_B,(\beta^a_k)^\dagger]=0.
        \end{split}
\end{equation}
We already have creation operators of each asymptotic field, we can construct the state consistently by acting the creation operator to the vacuum state. We define
\begin{equation}
    \label{2.105}
    \begin{split}
        \big|k,n\big>&\equiv \chi^\dagger_k\big|0\big>,\\
        -\big|k,-n\big>&\equiv \beta^\dagger_k\big|0\big>,\\
        \big|k,n+1\big>&\equiv \gamma^\dagger_k\big|0\big>,\\
        -\big|k,-(n+1)\big>&\equiv \bar\gamma^\dagger_k\big|0\big>,
    \end{split}
\end{equation}
where $k$ and $n$ label dynamical quantum number (such as momentum) and the ghost number, respectively.
The particular reasons we define the state this way can be understood with the following logic. First of all, we begin with assign the longitudinal mode $\chi_k$ of the gauge potential to have a reference value of the ghost number $n$ and reference momentum $k$. We see precisely from the first commutation relation of \eqref{2.103} that the BRST generates the transformation from the longitudinal gluon into the asymptotic ghost mode of the same momentum $\gamma_k$. Following the fact that the BRST charge carries one ghost number, thus, there is no wonder that the resulting ghost field must carry \say{$n+1$} ghost numbers. Besides, due to the conservation of the ghost number following from the fact that the ghost charges $Q_c$ is a Noether charge, there must also exist their pairs that carry the opposite values of ghost numbers, i.e. $-n$ and $-n-1$ which are the NL field and anti-ghost mode, respectively. Note that this formalism is not used only in the quantum YM theory, however, it becomes the general formalism to construct the theory with the same structure (quantum theory with complicated constraints), e.g. in quantum gravity \cite{Mielke2017} and topological field theory \cite{hep-th/9206061,Witten1988}. This procedure is formally known as the Batalin-Vikovisky or field-anti-field (BV-BRST) formalism \cite{Batalin1984,Batalin1981}.\\
\indent Once again, this all states form the vector space that also admits the BRST complex structure. As we have said before that the BRST complex further admits the $\mathbb{Z}$-graded Poisson superalgebra, it implies that we can decompose the vector space of states in the quantum YM theory as the direct sum among the vector space of each degree. Mathematically speaking, we can write $\mathcal{V}_{phys}=\bigoplus_n\mathcal{V}_n$ (remember that we have already projected from the total vector space into the physical vector subspace). To do so practically, we will first define the $n$-ghost numbers projection operator $P^{(n)}$ which can be constructed recursively. We guess the form of it as
\begin{equation}
    \label{2.106}
    P^{(n)}\equiv \frac{1}{n}(\beta^\dagger_kP^{(n-1)}\mathcal{O}_1+\chi^\dagger_kP^{(n-1)}\mathcal{O}_2+\gamma^\dagger_kP^{(n-1)}\mathcal{O}_3+\bar\gamma^\dagger_kP^{(n-1)}\mathcal{O}_4),
\end{equation}
where $\mathcal{O}_i$ are any operator which can be found explicitly by imposing the properties of this projector. The question is what kind of the projection operator we expected to have. First property will be the trivial property that every projection operator of any kind must essentially have, that is 
\begin{equation}
    \label{2.107}
    P^{(n)}P^{(m)}=P^{(m)}P^{(n)}=\delta_{mn} P^{(n)}, \;\;\;\mathrm{implying}\;\;\;(P^{(n)})^2=P^{(n)}.
\end{equation}
This property is obvious since once we act with the $n$-projector, that state will, one hundred percent, carry $n$-ghost numbers already if we act once by the same projector, it will give the same result for sure. In fact, this property implies a very deep consequence that is the projection operator of any degree has to be a bosonic generator unless the second time or more action of such an operator will give non-sense zero. Consequently, we can deduce instantly that $\mathcal{O}_1$ and $\mathcal{O}_2$ are bosonic while $\mathcal{O}_3$ and $\mathcal{O}_4$ are fermionic.\\
\indent The second property follows from the fact that we have expected the right decomposition which is consistent with the graded algebra embedded inside the considering theory. We expect so much that the projection operator must satisfy the completeness relation
\begin{equation}
    \label{2.108}
    1=\sum^\infty_{n=0}P^{(n)}=P^{(0)}+\sum^\infty_{n=1}P^{(n)}.
\end{equation}
This allows us to decompose any physical state $\big|\psi\big>\in\mathcal{V}_{phys}$ into the sum of that state projected by projection operator of each degree. Clearly, we have
\begin{equation}
    \label{2.109}
    \big|\psi\big>=P^{(0)} \big|\psi\big>+\sum^\infty_{n=1}P^{(n)} \big|\psi\big>.
\end{equation}
However, the state we are focusing on is a physical state which must be subjected to the subsidiary condition \eqref{2.84}. Nevertheless, since the vector is now decomposed into the direct sum of each degree, we also expect that the projected states are linearly independent to their friends. All of these arguments imply that the projection operator must inevitably commute with the BRST charge, namely,
\begin{equation}
    \label{2.110}
    [Q_B,P^{(n)}]=0.
\end{equation}
Plug the expression \eqref{2.106} into the constraint \eqref{2.109} to determine the relations among $\mathcal{O}_i$ with the help of \eqref{A.14}, \eqref{A.15} and \eqref{2.104}, it yields
\begin{equation}
    \label{2.111}
    \begin{split}
        0=&\;\beta^\dagger_kP^{(n-1)}\delta \mathcal{O}_1-\gamma_k^\dagger P^{(n-1)}\mathcal{O}_2+\chi^\dagger_kP^{(n-1)}\delta\mathcal{O}_2-\gamma^\dagger_kP^{(n-1)}\delta\mathcal{O}_3\\
        &\;+\beta^\dagger_kP^{(n-1)}\mathcal{O}_4-\bar{\gamma}^\dagger_kP^{(n-1)}\delta\mathcal{O}_4.
    \end{split}
\end{equation}

If we demand the equation \eqref{2.111} to hold, we require consequently as follows
\begin{equation}
    \label{2.112}
        \delta\mathcal{O}_1=-\mathcal{O}_4,\;\delta\mathcal{O}_4=0,\;
        \delta\mathcal{O}_3=-\mathcal{O}_2,\;\delta\mathcal{O}_2=0.
\end{equation}
The possible choice we can choose to satisfy the requirement \eqref{2.112} is
\begin{equation}
    \label{2.113}
    \mathcal{O}_1=-\chi_k,\;\mathcal{O}_2=-\beta_k,\;\mathcal{O}_3=\bar\gamma_k,\;\mathcal{O}_4=\gamma_k.
\end{equation}
Let us note that the choice, i.e. \eqref{2.113} is not unique in general. In the original paper \cite{Kugo1979,Kugo:1979gm}, they have chosen the slightly different choice by choosing instead $\mathcal{O}_1=-\chi_k-\omega_{kl}\beta_l$, where $\omega_{kl}$ is a coefficient of the matrix defined inside that paper, which also satisfies the requirement \eqref{2.112} without hesitation.\\
\indent It was suggested firstly by Fujikawa \cite{Kugo:1979gm} that such a projection operator of $n\geq 1$ degree can be rewritten into the anti-commutator between the BRST charge and something else. Specifically, we can write
\begin{equation}
    \label{2.114}
    P^{(n)}=\{Q_B,R^{(n)}\},\;\;\;\;\;\mathrm{where}\;\;R^{(n)}=-\frac{1}{n}(\bar\gamma^\dagger_kP^{(n-1)}\chi_k+\chi^\dagger P^{(n-1)}\bar\gamma_k).
\end{equation}
This form leads to very significant consequence. To understand how important it is, let's define the set of state, called zero-norm state, as $\mathcal{V}_0$. The element of this kind of set can be defined through the action of the $n\geq 1$ projection operator $P^{(n)}\big|\psi\big>\equiv\big|\psi_0\big>\in \mathcal{V}_0$. We can easily show that the zero-norm state will be orthogonal to the physical state
\begin{equation}
    \label{2.115}
    \big<\psi\big|\psi_0\big>=\big<\psi\big|P^{(n)}\big|\psi\big>=\big<\psi\big|\{Q_B,R^{(n)}\}\big|\psi\big>=0,
\end{equation}
where we have used the expression \eqref{2.114} with the subsidiary condition \eqref{2.84}. What does the result show us? It means physically that two different states $\big|\psi\big>$ and $\big|\psi+\psi_0\big>$ represent the same physics. In other words, $\big|\psi_0\big>$ is an unphysical state because of this reasoning. After we have already traced out all zero-norm states out from the physical world, the remaining states are equipped by a positive-definite metric. Thus, we can now define the physical Hilbert space for the QYM theory, guaranteeing the unitarity to be safe since all harmful ghosts are gone, to be 
\begin{equation}
\label{2.116}
    P^{(0)}\mathcal{V}\equiv\mathcal{H}_{phys}=\mathcal{V}_{phys}/\mathcal{V}_0.
\end{equation}
In summary, the unitarity of the (perturbative) quantum YM theory will be ensured after we have already integrated out all four elementary unphysical particles which are longitudinal polarization gluon, NL field, ghost and anti-ghost as defined in \eqref{2.105}. These unphysical particles form a family so-called a quartet (quartet means four). Strictly speaking, Kugo and Ojima used all the statements above to prove the unitarity of the quantum YM theory by claiming that all particles belonging to the quartet are confined. This mechanism, celebrating the name Kugo-Ojima (KO) quartet mechanism, completed the first-ever proof of the ghost confinement. Unfortunately, the transverse polarization modes of gluon still be free from the cage and still not yet proving to be confined. The discussion about the possibility that transverse gluon can be also formed a quartet found in, for example, \cite{1102.3119,1301.5292}.\\
\indent Before ending this section, we will give two important remarks about this mechanism. The first one is that the Hilbert space of physical states in the QYM theory defined in \eqref{2.116} can be shown to be identical to the zeroth BRST cohomology defined in \eqref{2.75}. We think it is not too hard to convince the reader that the unphysical state above is clearly a $Q_B$-closed and also an exact one. Reasonably speaking, the closed property can be understood from the imposition of the subsidiary condition \eqref{2.84} while the exact property, in the light of the form \eqref{2.114}, can be shown directly as
\begin{equation}
\label{2.117}
    \big|\psi_0\big>=P^{(n)}\big|\psi\big>=(Q_BR^{(n)}+R^{(n)}Q_B)\big|\psi\big>=Q_B(R^{(n)}\big|\psi\big>),
\end{equation}
do not forget that we have already imposed the subsidiary condition so the last term in the middle expression of \eqref{2.117} typically vanishes. Finally, it implies amazingly that the physical state, which carries zero number of ghost charge unless it will become zero-norm mode, belongs to the zeroth-order BRST cohomology $H^0_{Q_B}$. One might notice immediately that the considering cohomology is not the same cohomology we considered in the last subsection that is instead the cohomology with respect to the BRST differential $\delta$. However, we have already done sketching a proof that the nilpotency of the BRST charge truly implies the nilpotency of the BRST differential. To conclude, the BRST cohomology $H^0_{Q_B}$ is homomorphic to $H^0_\delta$ without wondering.\\
\indent The second remark is a much more serious one since the correctness of the KO quartet mechanism depends strongly on this remark. If we ask the reader that what is the most essential ingredient in the recipe to construct this mechanism. What is in the reader's mind? For us, we will answer the subsidiary condition since, as we have discussed earlier, this is an essential condition to project the indefinite-metric vector space to the semi-definite vector space before we can do the last step to finish the mechanism. The key point is a subsidiary condition supposing that the physical state is a BRST singlet or BRST invariant. Thus, we need to be sure that the BRST charge is well-defined in the sense that it truly represents the right BRST generator. To be more clear, $Q_B$ must suffer not to any symmetry breaking which is not always the case. As already mentioned once in the previous subsection, the BRST symmetry can be anomalous due to the non-invariance of the measure under the BRST transformation. Anomalous correction can generate the explicit symmetry breaking term, making the BRST charge ill-defined. Nonetheless, at the beginning of the section, we have done discussing that the FP method is incomplete. The modification of the FP quantization in the presence of the Gribov ambiguity requires the restriction to the Gribov horizon (see in detail in the next section). The restriction to the finite region makes the BRST transformation can not be defined globally. This obstruction leads to the breaking, interpreted to be spontaneously broken type by Maggiore and Schaden \cite{hep-th/9310111}, of the BRST symmetry \cite{0806.4348}. It can be shown (read section 3.4 of \cite{1202.1491}) that the symmetry breaking term is the operator of mass dimension 2. According to the renormalization group language, in 4-dimensional spacetime, it said to be super-renormalizable which will produce no any further an infinite loop correction. That means the symmetry breaking term modifies only the physics in low-energy (IR) regime but does not modify things living a high-energy (UV) scale. Thus, it is called soft (this terminology is usually used in the context of the supersymmetry breaking model). To summarize, in the deconfinement phase or UV region where the perturbative calculation is fine (follows from the asymptotic freedom), the KO mechanism is doing very well. Whereas, in the confinement phase or IR region, this mechanism fails since the BRST charge is not well-defined at first glance. Hence, the unitarity of non-perturbative quantum YM theory can not be proven with the same argument and still not be proven yet. On the bright side, the impossibility of proving the unitarity of the quantum YM theory in the confinement phase shines a new signal to physics' society. It suggests us to propose that the BRST violation can somehow lead to the confinement phenomenon. (This kind of argument can be found in, e.g. \cite{leticia}).

\newpage
\chapter{Gribov Ambiguity}
Even though the FP quantization with the resulting BRST symmetry seems to be very successful in many senses, as we have discussed once in the introduction, it is still not complete yet. What is the origin of all messes? Frankly speaking, it follows from the fact that we demand the gauge fixing condition to be very ideal. We require that the gauge orbit essentially crosses the gauge fixing constraint surface once and only once. In a realistic situation, how can we be so sure about that? Unfortunately, the actual answer is no, we cannot. In this chapter, we will start the main dish by studying the ambiguity of the covariant gauge fixing and the most successful (but yet not complete also) possible resolution of the problem.
\section{Gribov Ambiguity}
Recall first that the condition that there is only one solution of the gauge fixing equation has been implicitly used to force the identity \eqref{2.12} becoming the reliable one. Otherwise, we rather need to write the form of unity as
\begin{equation}
    \label{3.1}
    1=\frac{1}{\sum\frac{1}{\Delta_{FP}}}\int \mathcal{D}U\;\delta(G(^UA)),
\end{equation}
where the summation symbol is understood to be a sum over the all possible solutions of the constraint equation $G(A)=0$. In particular, we can rewrite the relation \eqref{3.1} further in functional form by the help of the new variable $N(G[A])$ denoted the numbers of the solution of the constraint equation, i.e. $N(G[A])\equiv |\{U\in G|G(^UA)=0\}|$. Thus, we have
\begin{equation}
    \label{3.2}
    1=\frac{1}{N(G[A])}\int \mathcal{D}U\;\delta(G(^UA))\Delta_{FP},
\end{equation}
where $\Delta_{FP}$ is a usual FP determinant which is now thought as a functional-valued quantity. Before we go forward, we will do a terminology for a bit. Along an orbit, if there are two $A_\mu$ and $A'_\mu\not=A_\mu$ which both satisfy the same gauge constraint $G[A]=0$, we will henceforth say $A_\mu$ and $A_\mu'$ are Gribov copy to each other. Now let's denote the numbers of all Gribov copies inside the one particular orbit as $N_{GC}(G[A])$. It can be seen simply that 
\begin{equation}
    \label{3.3}
    N(G[A])=|Z(A)|\cdot N_{GC}(G[A]),
\end{equation}
where $Z(A)$ is called a stabilizer subgroup of $G$ defined by $Z(A)\equiv \{U\in G|^UA=A\}$. By applying the so-called orbit-stabilizer theorem \cite{1712.01710}, we end up with the condition
\begin{equation}
    N(G[A])=\frac{|G|}{|[A]|}\cdot N_{GC}(G[A]).
\end{equation}
This shows that the functional $N(G[A])$ is truly orbit-dependent making things harder to evaluate in the path integration. Most importantly, it precisely shows that if Gribov copies really exist, the gauge degrees of freedom still not yet be fixed completely leading to over-counting the degree of freedom. Hence, the FP procedure fails eventually.
\\
\indent Nevertheless, it can be more harmful than we can ever imagine. To see that, suppose that $A_\mu$ and $A'_\mu$ both satisfy the Landau gauge fixing condition $\partial_\mu A^a_\mu=0$. Since $A'_\mu$ is a gauge field, there must always exist the element of gauge group, $U\in G$ said, such that $A'_\mu=^UA=UA_\mu U^{-1}-\frac{i}{g}(\partial_\mu U)U^{-1}\approx A_\mu-\frac{1}{g}D_\mu \alpha$. Thus, the gauge condition leads to
\begin{equation}
    \label{3.5}
    0=\partial_\mu A'_\mu=\partial_\mu A_\mu -\frac{1}{g}\partial_\mu D_\mu \alpha=-\frac{1}{g}\partial_\mu D_\mu \alpha,
\end{equation}
implying that the FP operator $-\partial_\mu D_\mu$ contains zero modes which is not sensitive to the gauge constraint. What we can imply more about this? First let's observe that the FP operator is actually Hermitian in the light of the Landau gauge. Explicitly,
\begin{equation}
\label{3.6}
\begin{split}
    \partial_\mu D_\mu\alpha&=\partial_\mu (\partial_\mu\alpha+ig[A_\mu,\alpha])\\
    &=\square\alpha+ig[\partial_\mu A_\mu,\alpha]+ig[A_\mu,\partial_\mu \alpha]\\
    &=\partial_\mu(\partial_\mu \alpha)+ig[A_\mu,\partial_\mu\alpha]\\
    &=D_\mu\partial_\mu \alpha,
\end{split}
\end{equation}
where we have used the linearity properties of both Lie bracket and partial derivative to perform the Leibniz's rule in the second line and the Landau gauge condition has been imposed in the third sub-step. The Hermiticity of the FP operator implies that it has only real eigenvalues. However, the perturbation around the zero modes can generate the negative eigenvalue of the FP operator. Hence, the positive definite condition of the FP determinant cannot be used and the introduction of ghost fields in \eqref{2.22} is meaningless (since we have used the positive definite property of the FP determinant to carelessly change from $|\det M^{ab}|$ into $\det M^{ab}$ in that step).\\
\indent However, the Gribov ambiguity does not affect all situations of the YM theory. In high enough energy where the coupling constant is sufficiently small, effectively we can think that the contribution of the gauge field is typically small $A_\mu\rightarrow 0$. The zero modes of the FP operator as expressed in \eqref{3.5} will reduce to the zero modes of the d'Alembertian operator $\square$ which is nothing but the well-known relativistic wave equation. The solution is, of course, a trivial plane wave. To have a local QFT, we require that the gauge field must be normalizable, hence vanishing at the spatial infinity. The plane wave solution is impossible to fulfill this condition, thus such an $\alpha$ cannot exist in a high energy limit and then the Gribov problem is gone. Unfortunately, in the confinement phase, the Gribov ambiguity is still haunted. \\
\indent Note further that the argument above helps us to conclude that an Abelian gauge theory is free from the Gribov ambiguity since, in Abelian limit, $f^{abc}=0$, the FP operator also reduces into the relativistic wave operator as also happened in the case of a weakly coupled YM theory. This makes sure that the traditional QED still works well.
\\
\indent In fact, the presence of the Gribov ambiguity does not harmfully affect only on analytical calculations but also affects numerical computations especially in lattice gauge theories. To understand this problem, we will first note that the unity we have inserted into the partition function \eqref{2.12} can be thought in an alternative way. Instead of thinking that we add the unity of the non-trivial form, we will rather think that we insert the additional partition function whose value turns out to be one by the help of an appropriate normalization factor. Namely, the unity of the form \eqref{2.12} can be rewritten to be
\begin{equation}
    \label{3.7}
    \begin{split}
        Z_{gf}&=\int \mathcal{D}U\;\delta(G(^UA))\det M_{ab}\\
        &=\int \mathcal{D}B\int\mathcal{D}U\int\mathcal{D}c\int\mathcal{D}\bar{c}\;e^{-S_{gf}},
    \end{split}
\end{equation}
where $S_{gf}$ denotes the action functional of both gauge fixing term and ghost term. Note that here we have pretended to think that the traditional way of quantization is still fine by naively neglecting the absolute symbol of the FP determinant.\\
\indent In particular, the partition function can be also rewritten further by using the functional generalization of the Dirac's delta function identity \eqref{2.9} to obtain
\begin{equation}
    \label{3.8}
    \begin{split}
        Z_{gf}&=\int \mathcal{D}U\;\sum_i\frac{\delta(U-U_i)}{\Delta_{FP}(^{U_i}A)}\det M_{ab}(^U A)\\
        &=\sum_i\frac{\det M_{ab}(^{U_i}A)}{|\det M_{ab}(^{U_i}A)|}.
    \end{split}
\end{equation}
However, we have known from the analysis before that the existence of the Gribov copies can be translated into the existence of the zero modes of the FP operator leading to the vanishing determinant of the FP operator. As a result, we can deduce that the naive FP method yields a non-sense $0/0$ result as appeared in computational lattice calculations. This organizes to other well-known problem in this field which is known to be the Neuberger $0/0$ problem \cite{Neuberger:1986xz}.\\
\indent To understand where did the problem actually comes from. We need to change the mindset about how we think about the partition function first. We need not treat the partition function in the same way we have treated in the usual path integration in QFT 101 and statistical physics. Conversely, if we are trying to compute the partition function of the form \eqref{3.7}, it means that we are amount to compute the topological invariant of the specific group manifold, interpreted the same way as the Witten index in SUSY non-linear sigma model \cite{Witten:1982df}, which is found to excitingly coincide with the Euler-Poincar\'e characteristic $\chi(G)^{\mathrm{numbers\;of\;lattice\;sites}}$. Remember that the gauge group $G$ is a Lie group, thus it is also a manifold without wondering. So the problem now lies down into the studying of the gauge group's structure itself.\\
\indent Recall that in the standard model, especially in the QCD sector, we formally use the $SU(N)$ as a gauge group (for IR-QCD's extensions, we use at most $SO(N)$ and exceptional $G_2$). Thus, we will try to compute the Euler characteristic of the $SU(N)$ group, however, it is quite hard to be done since the triangulation, which is the beginning step to evaluate the homology group and associated Betti number, of such a group manifold, is quite difficult to do. Fortunately, we can further reduce our consideration on the gauge group into a much simpler manifold. Note that a sphere of $2N-1$ dimension can be constructed as the quotient space between two unitary group $U(N)$ of different dimensions, i.e.
\begin{equation}
    \label{3.9}
    S^{2N-1}=U(N)/U(N-1).
\end{equation}
Then, consider the Cartesian product of the spheres
\begin{equation}
    \label{3.10}
    \begin{split}
        S^{2N-1}\times S^{2N-3}\times ...\times S^3=&\;U(N)/U(N-1)\times U(N-1)/U(N-2)\times ...\\&
        \; \times U(2)/U(1)\\
        \cong&\; U(N)/U(1)\\
        \cong&\; SU(N).
    \end{split}
\end{equation}
This implies that $SU(N)$ group can be written into the Cartesian product of spheres, which are easier to compute the associated Euler characteristic, of appropriate dimensions. Due to the consequence of the K\"unneth formula \eqref{b.19}, we will obtain the Euler characteristic of the $SU(N)$ group as
\begin{equation}
    \label{3.11}
    \chi(SU(N))=\chi(S^{3})\chi(S^{5})...\chi(S^{2N-1}).
\end{equation}
By direct computation of homology classes and associated Betti numbers of spheres, one will eventually end up with the result $\chi(S^n)=1+(-1)^n$ (see \eqref{B.22}). In other words, the Euler characteristic of $n$-sphere will be zero for spheres of odd dimensions and two for even-dimensional spheres. Combining this fact with the relation \eqref{3.11} to get the reasonable consequence that the Euler characteristic of $SU(N)$ group for which $N\geq 2$ case is precisely zero.\\
\indent Flashback to the origin of why we need to study the Euler characteristic, we deduce from this consequence simply that, instead of adding the unity of the form \eqref{3.7}, we have just inserted the partition function whose value is generally zero and then carelessly divided out by normalization factor which is also zero. Loosely speaking, in the case that Gribov ambiguity is not free, the partition function of the form \eqref{3.7} is misunderstood to be the unity as we usually thought in \eqref{2.12}. Its true identity is $0/0$ as we found in the Neuberger situation.
\\
\indent There are several possible resolutions to remedy the Neuberger problem. We will discuss two of them in a very qualitative sense. The first one is, to not have the indefinite $0/0$ form at the first glance, we can introduce the mass dimension regulator, called the Curci-Ferrari (CF) mass $m_{CF}$ said, into the partition function \cite{hep-lat/0501016}. As a result, in this case, the expectation value of any operators in a so-called CF model is well-behaved in the sense that it has been already free from the Neuberger problem. To restore the original YM result, we just take a massless limit $m_{CF}\rightarrow 0$ and then using the L'Hospital rule to calculate such a thing. It was suggested by Kondo \cite{1202.4162} that, due to the specific form of non-linear transformation from the original gauge field, we can eventually write the massive YM theory with correct degrees of freedom, i.e. two transverse and one longitudinal polarization modes. The new kind of gauge field, which will be reduced into the usual massive Proca field in the Abelian limit, precisely gauges invariant. However, what we refer to the word gauge invariance is the invariance under the modified version of BRST symmetry (the CF model is no longer invariant under original BRST symmetry but modified one). Note also that the modified BRST symmetry has already lost its nilpotency which shows a sign of the failure of the KO's unitarity proof. Thus, even though the CF model is shown to be renormalizable \cite{hep-th/9510167} but it is not perturbatively unitary. The particular reason why unitarity is violated in the CF model was also argued in \cite{1209.3994} that the CF model is not valid in high energy since it can be only treated as a low energy effective field theory of QCD existing below some energy scale cutoff. One way to make it valid in high energy is to take the limit $\xi\rightarrow \infty$, which is equivalent to the unitary gauge in the well-known YM-Higgs model, where all unphysical fields are decoupled out from the theory. To summarize, these all attractive features of the CF model leads people to conclude that CF model does play a role to be an alternative massive YM model without introducing any Higgs field.\\
\indent Another way to overcome the Neuberger problem is suggested from the analysis above showing that the vanishing of the Witten index is the real problem here. The alternative method is to perform the partial gauge fixing known as the Abelian projection. As shown in \cite{0801.1274} (and in more detail in \cite{1409.1599}), the element $g$ of the group $G$ can be uniquely decomposed into $h$ multiplied by $\xi$ which belong to the maximal stabilizer subgroup $\tilde{H}$ and its quotient space $G/\tilde{H}$, respectively. By the help of the formula \eqref{b.19}, we can reduce the problem of computing the Euler characteristic of the full non-Abelian group $\chi(G)$ into merely computing $\chi(G/\tilde{H})\chi(\tilde{H})$. Let's study the simplest possible model that is not trivial, the $SU(2)$ QCD, we have that the maximal stabilizer group uniquely coincides with the maximal torus group $U(1)$. Now we see that the Euler characteristic reduces into just a product of the Euler characteristics that are much easier to evaluate, i.e. $\chi(SU(2))=\chi(SU(2)/U(1))\chi(U(1))$. The former one is nothing but the Euler characteristic of the 2-sphere which is well-known already (see also \eqref{B.22}), that is, $\chi(SU(2)/U(1))=\chi(S^2)=2$. The problem is that the remaining contribution of the Euler characteristic class still vanishes. On the bright side, we have seen precisely where does the problem actually come from. If we fix the problem of the vanishing Witten index of the compact $U(1)$ group, we will completely solve the Neuberger's problem for $SU(2)$ case. It was suggested, by von Smekal \cite{vonSmekal:2008ws}, that such a problem can be resolved by performing the stereographic projection from $U(1)\cong S^1\rightarrow \mathbb{R}$ which transforms the topological space of genus one into zero. Accordingly, the Euler characteristic now becomes non-zero and the Neuberger's problem for the case of $SU(2)$ group has gone. For the more general case, for $SU(N)$ case, one can choose the maximal option of the Abelian projection to project the full non-Abelian group $SU(N)$ into the maximal torus group $U(1)^{N-1}$ and the so-called flag manifold $F_{N-1}$ \cite{Perelomov:1987va} (In more abstract language, such a manifold is the special case of the so-called generalized flag variety). We can check the consistency by comparing with the $SU(2)$ case, the maximal torus group is trivially consistent and also $F_1$ turns out to be isomorphic to the 1-dimensional complex projective space, widely known as the Riemann sphere, $\mathbb{C}P^1\cong S^2$ agreeing with the previous result. Once again, the Euler characteristic of the coset space is not the real face of the problem but the compact torus subgroup $U(1)^{N-1}$. However, we can also repeat the same method by stereographic projecting the $N-1$-torus into the $\mathbb{R}^{N-1}$ and we are done.
\\
\indent We wish that all of these kinds of stuff are enough to convince readers to understand that the Gribov ambiguity is a very dangerous thing that can destroy everything we have believed so far without mercy. However, by far, we have just only assumed that the Gribov ambiguity does exist. How many percent sure we can believe about its existence. According to the Singer's proof \cite{Singer:1978dk}, it is one-hundred percent inevitable for any covariant gauge fixing condition. Honestly speaking, the proof given by Singer is very mathematically difficult to be understood easily. We will, therefore, follow the simplified version appeared in \cite{Nair:2005iw}.\\
\indent Note that the Gribov ambiguity appeared as a non-trivial structure of the gauge orbit. Thus, we need to reach a deeper understanding of the gauge orbit. We begin with recalling that the gauge orbit $[A]$ is defined to be a quotient space of the set of overall connection 1-form $\{A\}$ modulo out by gauge transformation $\mathcal{G}$ (mathematically known as the automorphism of the corresponding principal fibre bundle). In differential topology language, the set of overall gauge 1-form $\{A\}$ can be realized alternatively as a principal fibre bundle over a base space $[A]$ with associated fibre $\mathcal{G}$. Obviously, this principal fibre bundle admits a structure group $G$ to be a considering gauge group since the transition function defined on this principal bundle is actually an element of such a group (gauge one-forms related to one another via adjoint gauge transformation with group element $U\in G$). In summary, this defines the fibration of the kind
\begin{equation}
\label{3.13}
    \mathcal{G}\hookrightarrow\{A\}\rightarrow [A]
\end{equation} (this sequence map demonstrates the inclusion from the fibre to the corresponding fibre bundle, then fibrating down to its base space).\\
\indent Now, to see the origin of the Gribov copies, we need to impose the gauge fixing constraint into the bundle structure. Mathematically, the gauge fixing is just choosing a specific representative out from the equivalence class $[A]$. Alternatively, by using the language of fibre bundle theory, picking a representative is equivalent with defining a so-called section which does assign a point on $\{A\}$ associated to each point on its base space $[A]$. If the section is globally defined, it means that there exists the unique representative, hence, lack of the Gribov copies. Strictly speaking, showing the existence of the Gribov copies is about to prove that the fibre bundle $\{A\}$ is a trivial bundle $\cong [A]\times \mathcal{G}$ (there is a theorem stating that a bundle will be a trivial bundle if and only if it admits a global section, see, e.g. Theorem 6 in the appendix B.3 \cite{Nakahara:2003nw}).\\
\indent To prove the non-triviality of the fibre bundle, one first need to extract the important information from the fact that $\{A\}$, by definition, is constructed to be a convex set, i.e. any element in $\{A\}$ can be written as a so-called affine combination of other elements $A_{\mu 1}$ and $A_{\mu 2}$. Namely, we can always write
\begin{equation}
    A_\mu(\tau,x)=\tau A_{\mu 1}(x)+(1-\tau)A_{\mu 2}(x).
\end{equation}
Convex set implies the contractibility of any loop inside that set. The deformation properties of loops define an equivalence class, which actually forms a group, known as homotopy class $\pi_n(\{A\})$. In this case, it implies that homotopy group of the set $\{A\}$ is trivial, i.e. $\pi_n(\{A\})\cong\{e\}$ where $\{e\}$ denoted trivial set for any value of $n$.\\
\indent Next, let's study the contractibility of fibre $\mathcal{G}$. Naturally, we can introduce a sequence of gauge transformation (gauge transformation trivially forms a group structure) parameterized by a real variable $s$. The usual element of the gauge group undergoes the triviality condition at infinity to maintain the finiteness of the gauge theory. Therefore, we demand $U(s,x)$ to enjoy the condition $U(s,x)\rightarrow 1$ as $s\rightarrow \pm\infty$. Pictorially, this can be also looked like a closed curve in $\mathcal{G}$ whose base point is located at an identity transformation $1$. Consequently, the homotopy class of the type $\pi_1(\mathcal{G})$ has been introduced accordingly (loop is topologically equivalent to a circle $S^1$). In the usual sense, the gauge element can be also realized as a transition function that performs a map from space(time) manifold which, in this case, there is an extra dimension $s$. Thus, $U(s,x)$ represents a map from $\mathbb{R}^{D+1}$ to a structure group $G$. Once again, the finiteness of the YM energy implies that the gauge transformation at far infinity must be trivial, this can be effectively thought as a 1-point compactification since all points at infinity of space(time) manifold are mapped into the same point. This leads to the compactification from $\mathbb{R}^{D+1}\rightarrow S^{D+1}$. Finally, the gauge transformation of the finite gauge theory is a map $S^{D+1}\rightarrow G$ defining the homotopy class $\pi_{D+1}(G)$. In the real world that has only three spatial dimensions, we end up with the wonderful result
\begin{equation}
    \pi_1(\mathcal{G})\cong \pi_5(G).
\end{equation}
For $SU(3)$ QCD, we have $\pi_5(SU(3))\cong \mathbb{Z}$ (See also Bott periodicity theorem for unitary group \eqref{Bott unitary}). So? All of these kinds of stuff tell us anything? Absolutely! The short exact sequence of fibre structure, which we have known in the name fibration, \eqref{3.13} can be used to imply the long exact sequence of corresponding homotopy class \cite{Hatcher:478079} of the form
\begin{equation}
\label{3.16}
    ...\rightarrow \pi_2(\{A\})\rightarrow \pi_2([A])\rightarrow \pi_1(\mathcal{G})\rightarrow \pi_1(\{A\})\rightarrow ...
\end{equation}
Since we have shown earlier that the homotopy group of the overall gauge potential $\{A\}$ are all trivial. Thus, this exact sequence \eqref{3.16} melts down into
\begin{equation}
    \pi_2([A])\cong\pi_1(\mathcal{G})\cong\pi_5(G),
\end{equation}
which is isomorphic to $\mathbb{Z}$ in the case of $SU(3)$ QCD. From the theorem 5 in the appendix B.3, we have known that if the base space is contractible to a point, then the corresponding fibre bundle is trivial. However, this shows that the fibre bundle $\{A\}$ is precisely non-trivial meaning that the section, which is described as the gauge fixing condition, cannot be defined globally. Hence, Gribov ambiguity!\\
\indent How about $SU(2)$ sub-dynamics of QCD? It is quite obvious that it must also suffer from the Gribov ambiguity since we have seen in the case of lattice gauge theories that $SU(2)$ QCD undergoes the Neuberger problem. To see that again in the language of Singer, we will pick one specific choice of gauge transformation $U_0(x)\equiv U(s_0,x)$ out of the sequence. In this particular case, the transition function is topologically equivalent to a homotopy class mapping from some fixed point $s=s_0$ on $\mathcal{G}$ into itself. This leads to the introduction of the zero degree homotopy class $\pi_0(\mathcal{G})$ which is currently isomorphic to $\pi_4(G)$. For $G=SU(2)$, we have $\pi_4(SU(2))\cong \mathbb{Z}_2$ (this result can be obtained by performing the following steps. First is to use the isomorphism $SU(2)\cong S^3$ combining with the consequence of \eqref{B.26} to have the isomorphism $\pi_4(S^3)\cong\pi_1(SO(3))$. In the end, just using the Bott periodicity condition for an orthogonal group \eqref{Bott orthogonal} to get the result we requested). Finally, we just have to repeat our analysis of the long exact sequence again and obtain the same consequence as same as the $SU(3)$ case that there exists an obstruction that forbids the fibre bundle $\{A\}$ to be trivial for such a gauge group. Eventually, we have thus finished the Singer's proposal of proving the existence of the Gribov copies for such a gauge group.\\
\indent Note that there is also a stronger mathematical argument to ensure the existence of the Gribov ambiguity due to the independent work of Narasimhan and Ramadas \cite{Narasimhan1979}. They calculated the holonomy group (the set of the closed loop parallel transportation induced by the connection on the given bundle) on the horizontal space of set of all possible gauge fields $\{A\}$ (i.e. the set of tangent vectors $A^T$ that satisfy the equation $D_\mu A^T_\mu=0$). They found that such a holonomy group is typically non-trivial. What is the consequence to that then? Following figure 3.1, it can be seen that the transition amplitude between two points on $[A]$ is deeply related to that holonomy group. The non-triviality of the holonomy group leads to the non-uniqueness of the parallel transport. Hence, there is the Gribov ambiguity. Moreover, in that  paper, the way how to prove the twisted property of the principal bundle is to the relevant matrix model for QCD gluon. In this particular case, $SU(2)$ QCD, the gluon field is constructed to be the real $3\times 3$ (3 follows from the total numbers of the adjoint degrees of freedom of the associated group) matrix, sometimes, called the matrix corresponds to the three-body bundle. In particular, the practical construction and further generalization to $SU(3)$ case ($8\times 8$ real matrix) of the QCD matrix model has been done properly in, for example, \cite{balach2014matrix}.
\begin{figure}[h!]
    \centering
    \includegraphics[width=6cm]{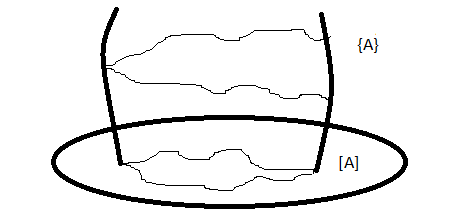}
    \caption{The closed loop parallel transportation comparing between on $[A]$ and on $\{A\}$.}
\end{figure}\\
\indent For readers who did not understand the mathematical proofs above, it is okay! Since from now on, we will discuss the alternative argument, given by Gribov himself \cite{Gribov1978}, to guarantee the existence of the Gribov ambiguity. Gribov studied the Gribov ambiguity appearing in the $SU(2)$ YM theory (QCD subdynamics) with the Coulomb gauge $\partial_i A_i=0$ which is effectively equivalent to the Landau gauge by assigning an additional temporal gauge condition $A_0=0$, i.e. he focused on the static field configuration. In a vacuum, where the field strength tensor naturally vanishes, the gauge field is in the pure gauge form, namely, $A_\mu=U^{-1}\partial_\mu U$ where $U\in G$.\\
\indent Gribov proposed the static, spherically symmetric ansatz to demonstrate his argument. Explicitly, we choose the ansatz of the form
\begin{equation}
    \label{3.18}
    U(\mathbf{x})=\exp \left(i\frac{f(r)}{2}\frac{\mathbf{x}\cdot \sigma}{r}\right)\equiv \exp\left(i\frac{f(r)}{2} \hat{n}\cdot \sigma\right),
\end{equation}
where $\sigma$ is the generator of $SU(2)$ gauge group which is nothing more but the Pauli matrices, and $r$ is spatial radius $r\equiv\sum_i^3 x^2_i$. In particular, we can perform the Taylor expansion of such an ansatz to rewrite the ansatz \eqref{3.18} further such that
\begin{equation}
    \label{3.19}
    \begin{split}
        U&=1+i\frac{f(r)}{2}\frac{\mathbf{x}\cdot \sigma}{r}-\frac{f^2}{8}\frac{(\mathbf{x}\cdot \sigma)^2}{r^2}+...\\
        &=1+i\frac{f(r)}{2}\frac{\mathbf{x}\cdot \sigma}{r}-\frac{f^2}{8}\frac{x_ix_j\sigma_i\sigma_j}{r^2}+...\\
        &=1+i\frac{f(r)}{2}\frac{\mathbf{x}\cdot \sigma}{r}-\frac{f^2}{16}\frac{x_ix_j\{\sigma_i,\sigma_j\}}{r^2}+...\\
        &=1+i\frac{f(r)}{2}\frac{\mathbf{x}\cdot \sigma}{r}-\frac{f^2}{8}\frac{x_ix_j\delta_{ij}}{r^2}+...\\
        &=(1-\frac{1}{2}\frac{f^2}{4}+...)+i(\frac{f(r)}{2}+...)\hat{n}\cdot \sigma\\
        &=\cos \frac{f(r)}{2}+i\sin \frac{f(r)}{2} \hat{n}\cdot \sigma,
    \end{split}
\end{equation}
where the anticommutation relation of the Pauli matrices has been used $\{\sigma_i,\sigma_j\}=2\delta_{ij}$. (Note that, in this situation, we will use the spatial indices $i$ and the color indices $a$ simultaneously, since both indices run from one to three.) By the help of identities $\partial_i r=n_i$ and $\partial_i n_j=\frac{1}{r}(\delta_{ij}-n_in_j)$, we can find the full expression of the gauge field in this ansatz as
\begin{equation}
    \label{3.20}
    \begin{split}
         A_i&=U^{-1}\partial_i U\\
         &=\frac{\hat{n}_i(\hat{n}\cdot \sigma)}{2}f'+\frac{(\delta_{ij}-n_in_j)\sigma_j}{2r}\sin f+\frac{1-\cos f}{2r}\epsilon_{ijk}n_j\sigma_k,
    \end{split}
\end{equation}
$'$ labels the radial derivative $\frac{\partial}{\partial r}$.
Thus, the Coulomb gauge condition $\partial_i A_i=0$ in this ansatz can be expressed as follows
\begin{equation}
    \label{3.21}
    r^2f''+2rf'-2\sin f=0.
\end{equation}
This equation is quite simple and it can be more simplified by introducing $\ln r=t$, we end up with very familiar equation
\begin{equation}
\label{Gribov pendulum}
\ddot{f}+\dot{f}-2\sin f=0.
\end{equation}
The equation \eqref{Gribov pendulum} is actually an equation describing damped (forced by external periodic force) harmonic oscillation known sometime to be the Gribov pendulum equation. With the appropriate boundary condition, we can solve this differential equation systematically. There exists three possible solutions: $f=0$, $f=f_{1/2}(r)$ and $f=-f_{1/2}(r)$, respectively. The non-trivial solution $f_{1/2}(r)$ behaves such that $f_{1/2}(0)=0$ (to avoid singularity) and $f_{1/2}(\infty)=-\pi/2$. Remarkably, each solution has its own topological charge, to see this, we substitute the ansatz \eqref{3.18} into the definition of Chern-Simon number \eqref{CS number pure}, we obtain the relation
\begin{equation}
    \nu=-\frac{1}{\pi}\left[f-\frac{1}{2}\sin(2f)\right]_{r=0}^{r=\infty}.
\end{equation}
Thus, each solution carries $\nu=0,\frac{1}{2}$ and $-\frac{1}{2}$, respectively. Intuitively, the first solution $f=0$ is a trivial solution interpreted to be our usual vacuum carrying no topological charge while the rest solutions $f=f_{1/2}(r)$ and $f=-f_{1/2}(r)$ are topologically heavy vacuum solutions, which will be known from now on as the Gribov vacua, expected to relate with the usual vacuum through somewhat mechanism. As observed from both the Singer and Gribov arguments, we expect that such a mechanism must relate to the topological structure of the YM theory itself.\\
\indent Finding the (vacuum) solutions of the $SU(2)$ YM theory is equivalent to finding the gauge field configuration which satisfies the classical equation of motion. In other words, the classical solution of the YM theory is extremized configuration of the YM action \eqref{2.3}. To extremize the YM action, it is more convenient to slightly rewrite the YM action, one has
\begin{equation}
\label{3.24}
    S_{YM}=\frac{1}{8}\int d^{D}x\;(F^a_{\mu\nu}\pm \Tilde{F}^a_{\mu\nu})^2\mp\frac{1}{4}\int d^{D}x\;F^a_{\mu\nu}\Tilde{F}^a_{\mu\nu},
\end{equation}
where $\Tilde{F}_{\mu\nu}$ is a component of a so-called dual field strength tensor defined in \eqref{B.12}. One can check very easily that this form of the YM action \eqref{3.24} is truly equivalent to the original form of the YM action \eqref{2.3}. We see that the cross term of the first term actually cancels with the last term of the expression \eqref{3.24} but the monster whose true form need to be revealed is the following term
\begin{equation}
    \begin{split}
        \Tilde{F}^a_{\mu\nu}\Tilde{F}^a_{\mu\nu}&=\frac{1}{4}\epsilon_{\mu\nu\alpha\beta}F^a_{\alpha\beta}\epsilon_{\mu\nu\gamma\delta}F_{\gamma\delta}\\
        &=\frac{1}{2}(\delta_{\alpha\gamma}\delta_{\beta\delta}-\delta_{\alpha\delta}\delta_{\beta\gamma})F^a_{\alpha\beta}F_{\gamma\delta}\\
        &=\frac{1}{2}(F^a_{\alpha\beta}F^a_{\alpha\beta}-F^a_{\alpha\beta}F^a_{\beta\alpha})\\
        &=F^a_{\mu\nu}F^a_{\mu\nu}.
    \end{split}
\end{equation}
So we have everything we supposed to have. Now let's observe that the first term of the new YM action is clearly positive definite. Thus we end up with the fact that the YM theory must satisfy the bound known as the Bogomolnyi-Prased-Sommerfield (or BPS for short) bound \cite{osti_7309001,Prasad1975}. The saturation of such a bound will yield the classical YM solution. To be specific, the YM action will be extremized when the YM field strength is either self-dual or anti-self-dual, i.e.
\begin{equation}
    F_{\mu\nu}=\Tilde{F}_{\mu\nu}.
\end{equation}
Mysteriously, any self-dual anti-symmetric tensor, $S_{\mu\nu}$ can be written as 
\begin{equation}
    S_{\mu\nu}=\eta^a_{\mu\nu}c^a,
\end{equation}
where $c^a$ is any Lie-algebra-valued function and $\eta^a_{\mu\nu}$ is known formally as the 't Hooft tensor \cite{tHooft1976} whose components are
\begin{equation}
    \label{'t Hooft tensor}
    \eta^a_{\mu\nu}=\epsilon_{a\mu\nu}+\delta^a_{\mu}\delta_{0\nu}-\delta^a_{\nu}\delta_{0\mu}.
\end{equation}
In fact, this symbol did not come from the heaven but did come from the brilliant mind of 't Hooft. To understand the origin of this symbol, let's recall first that a self-dual antisymmetric tensor belongs to $(1,0)$-irreducible representation of Lorentz group (to be more specific, the projective representation of the universal covering group $SL(2,\mathbb{C})$), whereas a 4-vector lies within the $(\frac{1}{2},\frac{1}{2})$ representation of the same group. Therefore, the 't Hooft tensor is nothing but a Clebsch-Gordon coefficients used to construct $(1,0)$-representation out from tensor product between two four-vector $(\frac{1}{2},\frac{1}{2})\otimes(\frac{1}{2},\frac{1}{2})$. Similarly, we can also write any anti-self-dual antisymmetric tensor $A_{\mu\nu}$ as $A_{\mu\nu}=\bar{\eta}^a_{\mu\nu}$ where the anti-'t Hooft tensor $\bar{\eta}^a_{\mu\nu}$ is almost the same as $\eta^a_{\mu\nu}$ beside the component $\bar{\eta}^a_{0i}=\eta^a_{i0}$. Before we continue our discussion, let us write down the useful properties of 't Hooft tensor \cite{Smilga:2001ck} first
\begin{enumerate}
    \item \begin{equation}
    \label{3.29}
    \epsilon^{abc}\eta^b_{\mu\rho}\eta^c_{\nu\sigma}=g_{\mu\nu}\eta^a_{\rho\sigma}-g_{\mu\sigma}\eta^a_{\rho\nu}+g_{\rho\sigma}\eta^a_{\mu\nu}-\delta_{\rho\nu}\eta^a_{\mu\sigma}.
\end{equation}
This identity is very meaningful, it shows that, by the help of the 't Hooft tensor, we can represent the generator associated to the $SU(2)$ group as the Lorentz algebra.
\item \begin{equation}
    \eta^a_{\mu\nu}\eta^a_{\mu\nu}=12.
\end{equation}
\end{enumerate}
To summarize what we have done so far, by introducing the 't Hooft tensor, we can simply write the YM field strength, which is the solution of the classical YM theory, as the 't Hooft tensor times any function. Because of this, Corrigan, Fairlie \cite{Corrigan1977}, 't Hooft \cite{tHooft1976} and Wilczek \cite{Wilczek1977} proposed the choice of ansatz to be 
\begin{equation}
    \label{phi4}
    A^a_\mu=-\eta^a_{\mu\nu}\partial_\nu \ln{\phi(x)}.
\end{equation}
If we substitute this ansatz, named after them, into the YM equation of motion
\begin{equation}
    \label{YM eom}
    D_\mu F^a_{\mu\nu}=0.
\end{equation}
Although it is very tricky but straightforward to be done, we will end up with
\begin{equation}
    \phi^3\partial_\mu(\phi^{-3}\square \phi)=0,
\end{equation}
implying
\begin{equation}
    \label{3.34}
    \phi^{-3}\square \phi=\mathrm{constant}\equiv -\lambda\;\;\mathrm{or}\;\;\square\phi+\lambda \phi^3=0,
\end{equation}
which is merely an equation of motion of the massless Klein-Gordon equation with self-interaction of the form $-\frac{\lambda}{4}\phi^4$ (usually known from QFT101 as $\phi^4$ (phi to the four) theory). Strictly speaking, this ansatz reduces the complicated problem on non-linear YM equation into a very well-understood equation of the real scalar field. Therefore, this ansatz is sometimes called the $\phi^4$-ansatz. The solutions of the massless Klein-Gordon equation are not too difficult to be found. We have two possible solutions \cite{Fushchich1993} that are
\begin{equation}
    \label{meron}
    \phi(x)=\frac{1}{\sqrt{\lambda x^2}},
\end{equation}
and 
\begin{equation}
    \label{instanton}
    \phi(x)=\sqrt{\frac{8}{\lambda}}\frac{\rho}{x^2+\rho^2}.
\end{equation}
($\rho$ is actually an arbitrary constant) Each solution leads to different non-trivial YM connection. The former solution leads to the YM solution called deAlfaro-Fubini-Furlan (AFF) (one) meron solution \cite{DeAlfaro1976}
\begin{equation}
    \label{AFF meron}
    A^a_\mu(x)=\frac{\eta^a_{\mu\nu}x_\nu}{x^2},
\end{equation}
while the latter gives the well-known Belavin-Polyakov-Schwartz-Tyupkin (BPST) instanton \cite{Belavin1975}
\begin{equation}
    \label{BRST instanton}
    A^a_\mu(x)=\frac{2\eta^a_{\mu\nu}x_\nu}{x^2+\rho^2}.
\end{equation}
Henceforth, we will study how these two YM vacuum solutions generate the Gribov copies one-by-one. Historically, it was pointed out by Sciuto \cite{Sciuto1977}, independently to Abbott and Eguchi \cite{Abbott1977} that the BPST instanton solution demonstrates the possible relationship between two Gribov vacua. After that, Chiu, Kaul and Takasugi \cite{Chiu1978} used the same argument to study the role of AFF meron solution to this story. However, in this review article, we will study in the converse timeline since the AFF one meron is, of course, simpler than the BPST instanton.\\
\indent Let's begin by noting that the AFF meron solution can be rewritten as follows
\begin{equation}
    \label{meron pure}
    A_\mu=\frac{1}{2}U^{-1}\partial_\mu U,
\end{equation}
where we just throw out the factor of $i$ by, e.g. redefining the gauge field or generator of the gauge group and
\begin{equation}
\label{3.40}
    U(x)=\frac{x_0-i\sigma\cdot \mathbf{x}}{\sqrt{x^2}}\equiv \exp(i\beta(x)\hat{n}\cdot \sigma),
\end{equation}
with $\beta(x)\equiv -\tan^{-1}\left(\frac{r}{x_0}\right)$ (be careful about the convention that $r$, we used here, is as same as the $r$ appeared before. Namely, $r$ does not involve time component $x_0$ in it). Observe that the gauge configuration \eqref{meron pure} is of the pure gauge form up to some constant factor but still not in the Coulomb gauge as expected from the Gribov example. However, since the gauge field is not unique at the first glance, we have a freedom to choose a different gauge field that has almost the same properties as the AFF meron has but, instead, lies within the Coulomb gauge fixing condition. The so-called transverse meron solution can be related with the AFF meron through the suitable gauge transformation. Namely,
\begin{equation}
    \label{transverse meron}
    A_\mu^T=V^{-1}A_\mu V+\frac{1}{2}V^{-1}\partial_\mu V,
\end{equation}
where, of course, $\partial_i A_i^T=0$. By introducing the new variable
\begin{equation}
    G(x)=U^{\frac{1}{2}}(x)V(x)\equiv \exp\left[i\left(\alpha(x)-\frac{\pi}{2}\right)\frac{\mathbf{x}\cdot\sigma}{r}\right],
\end{equation}
we can express out the transverse gauge field out explicitly as
\begin{equation}
\label{transverse pure}
    A^T_\mu=\frac{1}{2}[(U^{1/2}G)^{-1}\partial_\mu(U^{1/2}G)+(U^{-1/2}G)^{-1}\partial_\mu(U^{-1/2}G)].
\end{equation}
Plug the gauge configuration \eqref{transverse pure} into the transverse Coulomb condition to obtain the differential equation with very familiar form \cite{Chiu1978}
\begin{equation}
\label{3.44}
    \alpha''+\frac{2}{r}\alpha'-\frac{x_0}{r^2\sqrt{x^2}}\sin 2\alpha=0.
\end{equation}
In asymptotic limit such that $|t|\ll r$, the differential equation \eqref{3.44} will reduce into the Gribov pendulum equation 
\eqref{3.21}. Thus, the gauge field of the form \eqref{transverse pure} is clearly what we are looking for. The analysis given in the original paper \cite{Chiu1978} tells us that there will be four types of solution to this equation that are
\begin{equation}
\label{3.45}
    \begin{split}
        1.\;&\alpha(x)=\begin{cases}
        \frac{\pi}{2}&t<0\\
        \pi-\gamma(x)&t>0
        \end{cases}\\
        2.\;&\alpha(x)=\begin{cases}
        \frac{\pi}{2}+\gamma(x)&t<0\\
        \pi&t>0
        \end{cases}\\
        3.\;&\alpha(x)=\begin{cases}
        \frac{\pi}{2}&t<0\\
        \gamma(x)&t>0
        \end{cases}\\
        4.\;&\alpha(x)=\begin{cases}
        \frac{\pi}{2}-\gamma(x)&t<0\\
        0&t>0
        \end{cases}
    \end{split}
\end{equation}
$\gamma$ has been introduced in such a way that $\alpha=\pm \gamma+n\pi$ to help us understanding the behavior of solutions. The behavior of $\gamma$ itself is as follows:
\begin{enumerate}
    \item $\gamma(r=0,t)=0$.
    \item $\gamma(r,t\rightarrow \infty)\rightarrow 0$.
    \item $\gamma(r\rightarrow \infty,t)\rightarrow \frac{\pi}{2}$.
\end{enumerate}
The meaning will be clear when we consider the topological charge of such a solution. Let's use the Chern-Simon number \eqref{CS number} to represent the appropriate topological charge in this theory. Since the Chern-Simon charge must be evaluated at the boundary of the underlying manifold. Let's choose the boundary condition such that the manifold is effectively compactified into the cylinder of the type $S^d\times \mathbb{R}$ where $\mathbb{R}$ represents the time axis while the cross-section of the cylinder is the spatial part $S^d$ (once again, the spatial coordinates are compactified into the sphere to maintain the finiteness of the YM energy). Thus, there will be three boundary surfaces in this case. Two of them are the base at $t=\pm\infty$ of the cylinder while another one is the side at $r=\infty$ of the cylinder. The associated Chern-Simon charges evaluated at time infinities take the form
\begin{equation}
    \nu_\pm=-\frac{1}{\pi}\lim_{T\rightarrow\pm\infty}\left[\alpha-\frac{1}{2}\frac{x_0}{\sqrt{x^2}}\sin 2\alpha\right]^{r=\infty}_{r=0},
\end{equation}
while
\begin{equation}
    \eta=\frac{1}{\pi}\lim_{T\rightarrow\infty}\left[\alpha-\frac{1}{2}\frac{x_0}{\sqrt{x^2}}\sin 2\alpha\right]^{T}_{-T}
\end{equation}
represents the topological charge at the spatial infinity's boundary. The total topological charge is therefore expressed as
\begin{equation}
    \nu=\nu_+-\nu_-+\eta.
\end{equation}
This, in differential topology language, is a simple example of the well-known mathematical theorem known as the Atiyah-Patodi-Singer index theorem \cite{Atiyah:1975jf} which is the generalization of the Atiyah-Singer theorem \cite{Atiyah1963} to the case that contribution from the spatial boundary of the manifold is concerned, i.e. $\eta$-invariant term. We can check that for every possible situation in \eqref{3.45} $\eta$ will always vanish. Intuitively, this is not coincident since one knows that, for example \cite{Nakahara:2003nw}, that the $\eta$ term represents the gauge anomalies on the spatial boundary. Concisely speaking, $\eta=0$ implies that no gauge anomalies are living on the spatial boundary surface of this theory. The non-trivial cases such that the $\eta$-invariant is not vanishing, but will be perfectly cancelled, can be found in several situations, e.g. M-theory compactification \cite{deAlwis1997}, lattice chiral QCD and topological insulators \cite{1712.03679}.\\
\indent Now let's compute the topological charges of the fourth case (one of my favourite) of the solution \eqref{3.45} to study the physical picture of the model. One find through direct calculation that $\nu_+=0, \;\nu_-=\frac{1}{2}$ and $\eta=0$, this means that such a solution describes the quantum tunnelling from the Gribov vacuum carrying $1/2$ charge into the traditional topologically charged-less vacuum. To understand more intuitive picture of this, we consider the temporal gauge fixing $A_0=0$ condition of the meron solution. We will see that \cite{Actor1979} the single meron solution demonstrates physically the unstable Wu-Yang monopole. Strictly speaking, the fourth solution \eqref{3.45} can be visualized as follows: at the beginning of time, there is a Wu-Yang monopole located inside the vacuum, later asymptotically, such a Wu-Yang monopole decays completely into another vacuum satisfying the same gauge fixing condition as the same one at the far past. Hence, this ensures the Gribov ambiguity.\\
\indent For the BPST instanton \eqref{BRST instanton} solution, we can repeat all analyses in long-distance limit. Why the long-distance limit is needed? Observe that, by using the same configuration \eqref{3.40}, we can rewrite the BPST instanton in more attractive way
\begin{equation}
    A^a_\mu=\frac{x^2}{x^2+\rho^2}U^{-1}\partial_\mu U.
\end{equation}
This form of the BPST instanton is almost pure gauge but it is still not. The particular reason is the BPST has a core of finite size, i.e. finite radius $\rho$ which differs from the case of the meron solution. However, in the long-distance limit $x\gg \rho$, the size of the instanton's core will be considerably negligible, hence, it becomes pure gauge in this limit. By the same analysis, we can show that the BPST instanton also describes the tunneling solution interpolates between two different Gribov vacua (in this case, the total topological charge is one, not one-half, then the transverse BPST instanton jumps from $\nu_-=-1/2$ to $\nu_+=1/2$ and vice versa). Remarkably, due to these behaviors, one believes that the BPST instanton composes of two merons (meron, in fact, means \say{a part of}) and we also know that meron is merely a Wu-Yang monopole. These all pieces of stuff reveal a deep connection between instanton and monopole as conjectured in several papers (see, e.g. \cite{1501.06517,hep-lat/9710049}).
\\
\indent Here I will remark about the Gribov's argument to show the existence of the Gribov copies. Actually, the extension to the higher $SU(N)$ group is not that necessary since the argument is almost general at the first glance. To understand the remark, let's recall that the unitary group $SU(N)$ with $N\geq 3$ can be written as the product of the spheres (see \eqref{3.10}), thus we can consistently compute the 3rd homotopy group of the $SU(N)$ as follows
\begin{equation}
    \pi_3(SU(N))\cong \pi_3(S^{(2N-1)})\times...\times\pi_3(S^3)\cong \pi_3(SU(2))\cong \mathbb{Z},
\end{equation}
where we have used the fact that the homotopy group of the Catesian product space can be split into the homotopy group of each one and $\pi_k(S^n)$ is trivial group if $n>k$ (intuitively, the triviality can be thought as following sense: the way we can embed the sphere onto the sphere of higher dimension is always trivial, hence the corresponding homotopy class is also trivial). In fact, we can end up with the same result by using alternatively the Bott periodicity theorem for the unitary groups \eqref{Bott unitary}. The meaning of the relation is that, along the appropriate direction on the Lie group manifold, the instanton and meron solutions for higher $SU(N)$ group are precisely equivalent to the solutions for the well-studied $SU(2)$ case. Thus, the generalization is not that essentially as claimed.

\section{Gribov Horizon and Fundamental Modular Region}
In the previous section, we study how the Gribov ambiguity arises in the non-Abelian gauge theories and how harmful its consequence is. In this section, we will now study the old but gold method suggested first by Gribov in his original paper. Gribov proposed that, at the beginning, the gauge orbit flows across from the place of which FP operator has positive eigenvalues, when it is about to cross into the dangerous zone of negative modes, the gauge orbit is immediately cut right there. In the other words, the space of gauge fields, in this sense, is restricted to the everywhere bounded (as will be showed soon enough) region known as the Gribov region, $\Omega$ said. Mathematically, the Gribov region is defined as follows
\begin{equation}
    \Omega\equiv \{A^a_\mu|\;\partial_\mu A^a_\mu=0, \; M^{ab}(A)\equiv -\partial_\mu D^{ab}_\mu >0\}.
\end{equation}
After restricting the path-integral measure within this region, the FP operator turns out to be positive definite so the problem of introducing the ghost fields as we have done in the FP procedure is resolved. Obviously, we can see it precisely that the boundary of the Gribov region is a collection of the zero modes of the FP operator, as will be called the (first) Gribov horizon and denoted by $\partial \Omega$.\\
\indent So far, the idea is quite clear logically but it turns to give us another big question, that is, how can we determine the shape of this region or how can we locate the location of the Gribov horizon? This, in fact, translates into the problem of differential topology which aims to study the topological property of some topological space. The well-known method we will use henceforth is called a Morse theory that studies the properties of the topological space by investigating the critical behavior of the so-called Morse function(al). To be specific, we will choose the Morse functional to have attractive minima. Generically, we choose that functional to be a chiral-like action expressed down below
\begin{equation}
\label{3.51}
    S[U]\equiv ||^UA||^2\equiv \mathrm{Tr}\int d^{D}x\;\left(\frac{i}{g}U^{-1}\partial_\mu U+U^{-1}A_\mu U\right)^2,
\end{equation}
where the gauge field here is treated as a fixed variable. Due to the gauge's freedom, we can always choose the configuration such that the local minimum point of such a functional is at $U=1$ which is the simplest choice possible. In the end, we reduce the problem into finding the critical behavior of the functional called the Hilbert-Schmidt norm defined on the gauge space. Explicitly, we are about to investigate the functional of the form
\begin{equation}
\label{3.52}
    ||A||^2\equiv \int d^{D}x\;\mathrm{Tr}(A_\mu A_\mu)=\frac{1}{2}\int d^{D}x\; A^a_\mu(x) A^a_\mu(x). 
\end{equation}
Let's explore what will happen if this functional has a relative extremum. Consider an infinitesimal variation with respect to the gauge transformation (we are trying to explore the infinitesimal variation in the space of gauge fields)
\begin{equation}
    \begin{split}
        \delta||A||^2&=\delta\left(\frac{1}{2}\int d^{D}x\;A^a_\mu A^a_\mu\right)\\
        &=\int d^{D}x\;\delta(A^a_\mu) A^a_\mu\\
        &=-\int d^{D}x\; D^{ab}_\mu \alpha^b A^a_\mu\\
        &=-\int d^{D}x\; (\delta^{ab}\partial_\mu -gf^{abc}A^c_\mu) \alpha^b A^a_\mu\\
        &=-\int d^{D}x\; \partial_\mu \alpha^a A^a_\mu\\
        &=\int d^{D}x\;\alpha^a(\partial_\mu A^a_\mu).
    \end{split}
\end{equation}
If this configuration is truly the local extremum of this function, that implies $||A||^2$ must vanish for any value of infinitesimal parameter $\alpha$. This implies further that the gauge field which extremizes the Hilbert-Schmidt functional must be the gauge configuration that satisfies the Landau gauge fixing condition $\partial_\mu A^a_\mu=0$.\\
\indent Next, if we demand more that the considering configuration does not only extremize the functional but also minimize it, we will end up with
\begin{equation}
    \begin{split}
        \delta^2||A||^2&=-\int d^{D}x\; \partial_\mu \alpha^a \delta A^a_\mu\\
        &=\int d^{D}x\;\partial_\mu\alpha^aD^{ab}_\mu \alpha^b\\
        &=-\int d^{D}x\;\alpha^a(\partial_\mu D^{ab}_\mu)\alpha^b\\
        &=\int d^Dx\; \alpha^aM^{ab}\alpha^b>0.
    \end{split}
\end{equation}
That is, Requiring that the gauge field do minimize the functional is equivalent to require that the FP operator $M^{ab}=-\partial_\mu D^{ab}_\mu$ is positive definite. In summary, the specific gauge configuration minimizing the functional \eqref{3.52} will be the gauge field inside the Gribov region satisfying the Landau gauge constraint.\\
\indent Before we go through this, let us note the interesting analogue has been also introduced in the context of the SUSY field theory by Witten \cite{witten1982}. In the SUSY field theory, the relevant Morse function coincides with the so-called superpotential. In general, the presence of the SUSY leads to the degeneracy of the ground state since the numbers of the fermion and boson are equal in the same supermultiplet. However, in the real world, the superpartner has been not yet found in the currently reachable energy scale, thus the SUSY must be spontaneously broken. Particularly, the superpotential takes the responsibility for this purpose. In our situation, the Morse function(al) which is chosen to be the Hilbert-Schmidt norm of the gauge field has been also introduced to eliminate the degeneracy arisen from the cancellation between the negative and positive modes of the FP operator. In other words, one can think the negative-positive modes cancellation in the Gribov case as the same way one think about the cancellation between the fermionic and bosonic modes in the supersymmetric field theory case.\\
\indent So we can already locate the Gribov horizon, let's move to the discussion of the geometrical properties of such a region
\begin{enumerate}
    \item It can be proved rigorously \cite{dell'antonio1991}, by using the mathematical branch called analysis, that every gauge orbits will pass through the gauge fixing hyperplane at least once. Strictly speaking, if we restrict the path-integration into this region, it ensures that the gauge condition can single out the redundant degrees of freedom.
    \item Configurations $A_\mu\rightarrow0$ live within the Gribov region, this proposition can be proved very easily by using the fact that, in this regime, $D_\mu\rightarrow \partial_\mu$ makes the FP operator positive definite, i.e. $-\partial^2\delta^{ab}=p^2\delta^{ab}>0$. This property ensures that the perturbative calculation must work well as it always be.
    \item Since the space of gauge fields is expected to be a convex set meaning that, for every $A_{\mu1},A_{\mu2}$ belonging to the gauge space, the affine combination $A_\mu=\tau A_{\mu1}+(1-\tau)A_{\mu2}$ where $\tau\in[0,1]$ still belongs to the gauge space. Thus, it is very natural to think that the same thing must occur in the Gribov's case. Let's see
    \begin{equation}
        \begin{split}
            M^{ab}(\tau A_{\mu1}+(1-\tau)A_{\mu2})=&-\partial_\mu(\delta^{ab}\partial_\mu-gf^{abc}(\tau A^c_{\mu1}+(1-\tau)A^c_{\mu2}))\\
            =&-\partial_\mu(\delta^{ab}\tau\partial_\mu -gf^{abc}\tau A^c_{\mu1}))\\
            &\:-\partial_\mu(\delta^{ab}(1-\tau)\partial_\mu-gf^{abc}(1-\tau)A^c_{\mu2}))\\
            =&\;\tau M^{ab}(A_{\mu1})+(1-\tau)M^{ab}(A_{\mu2}).
        \end{split}
    \end{equation}
    This shows that if both $A_{\mu1}$ and $A_{\mu2}$ are living inside the Gribov region, in the other words, $M^{ab}(A_{\mu1})$ and $M^{ab}(A_{\mu2})$ are both positive definite, so $\tau A_{\mu1}+(1-\tau)A_{\mu2}$. Therefore, the Gribov region is convex.
    \item The Gribov region is bounded from every direction. To show that, recall that the second part of the FP operator, to be specific $g\partial_\mu(f^{abc}A^c_\mu)$, is traceless since the color ($SU(N)$) generator is traceless by its definition. Thus, the sum of the eigenvalues of this part of the FP operator yields zero meaning that there exists the gauge configuration $A_\mu$ representing the negative mode. Explicitly, we write
    \begin{equation}
        \int d^{D}x\;\alpha^a[g(f^{abc}A^c_\mu)]\alpha^b\equiv \kappa<0.
    \end{equation}
    Now consider the following gauge configuration
    \begin{equation}
    \label{3.57}
    \begin{split}
        \int d^{D}x\alpha^aM^{ab}(\lambda A_\mu)\alpha^b&=\int d^{D}x\;\alpha^a[-\partial_\mu(\delta^{ab}\partial_\mu-gf^{abc}\lambda A^c_\mu)]\alpha^b\\
        &=\int d^{D}x\;\alpha^a(-\partial_\mu^2)\alpha^a+\lambda \kappa.
    \end{split}
    \end{equation}
    Since the first term of the right-hand side of the \eqref{3.57} is positive definite, thus, with sufficiently large $\lambda$, nothing guarantees that the FP operator of the gauge configuration above will be positive-definite. Consequently, if it is not positive definite, then it locates outside the Gribov region. In the other words, if we begin with some gauge field $A_\mu$ that belongs to the Gribov region, there exists some large constant number $\lambda$ such that the gauge field of the form $\lambda A_\mu$ flows far enough until it literally lives outside the Gribov safe-zone already.
\end{enumerate}
After we have reviewed the properties of the Gribov region, we hope that it is enough to convince readers to feel safe to restrict our gauge orbit inside this region. However, everything has both pros and cons. The most disadvantage of this region is it is not stringent enough to expel all of the Gribov copies. Let's focus on the gauge field at the first Gribov horizon, we have known that things live on the boundary of the Gribov region is the collection of zero modes of the FP operator, hence, we can deduce easily that such a gauge configuration differs from the one belongs to the bulk of the Gribov region in the sense that they are not essentially relative minima of the Hilbert-Schmidt functional \eqref{3.52} which implies that, suppose the gluon field on the Gribov horizon denoted by $A_\mu$, there will always exist the gauge transformation such that the transformed field is typically less than the non-transformed one. Strictly speaking, there exists $U$ which makes the following inequality hold
\begin{equation}
    ||A||^2>||^UA||^2.
\end{equation}
Notice that this degeneracy is not that harmful since this gauge field located at the boundary of the Gribov region meaning that, under the integration over the gauge field restricted to the Gribov region, this configuration is then regarded as the endpoint of the integrand contributing nothing to the result of the integration. By the way, observe that there is another gauge field of the form $(1-\varepsilon)A_\mu$ for an arbitrary small parameter $\varepsilon$. Due to the fact that the Gribov region is convex, the new gauge potential $(1-\varepsilon)A_\mu=(1-\varepsilon)A_\mu+\varepsilon \times0$ also belongs to the Gribov region but, in this situation, away from the Gribov horizon (remind ourselves that the perturbative gauge field $A_\mu=0$ always lives within the Gribov region.) Thus, the $(1-\varepsilon)A_\mu$ must be a local minimum of the functional by its construction. However, by continuity condition, we have
\begin{equation}
    ||(1-\varepsilon)A||^2>||(1-\varepsilon)^UA||^2.
\end{equation}
Thus, $(1-\varepsilon)A_\mu$ is not an absolute minimum. Hence, Gribov copies arise in the Gribov region $\Omega$.\\
\indent To remedy this, of course, we must restrict to a more stringent region containing only absolute minima of the functional \eqref{3.52} contrasting with the Gribov region that can have the relative minima inside it. We call this region the fundamental modular region (FMR) denoted by $\Lambda$. In literature, after restricting the gauge configuration inside this region, it is said to be working in the minimal Landau gauge. Let's begin with the explicit definition of the FMR as shown below \cite{SemenovTyanShanskii1986}
\begin{equation}
    \Lambda\equiv\{A^a_\mu|\;||^UA||^2\geq||A||^2,\forall U\in G\},
\end{equation}
where $||\cdot||$ is the Hilbert-Schmidt norm as defined in \eqref{3.51} and \eqref{3.52}. Since only absolute minima are allowed to belong in the FMR, thus, it is kind of obvious that the FMR is smaller than the Gribov region, i.e. $\Lambda\subset \Omega$.\\
\indent Once again, we will end this section by discussing the properties of the FMR. Note first that here, in some properties, we will not go through it in detail since it is the same properties as also appeared in the case of the Gribov region.
\begin{enumerate}
    \item Since all gauge orbits intersect with the Gribov horizon, so the boundary of the FMR $\partial \Lambda$.
    \item Perturbative gauge fields live inside the FMR.
    \item $\Lambda$ is convex. We are attempt to prove this since this particular property is not that trivial to see as another one. Let's recall our chiral action \eqref{3.51}
    \begin{equation}
        \begin{split}
            ||^UA||^2&=\mathrm{Tr}\int d^{D}x\;\left(U^{-1}A_\mu U+\frac{i}{g} U^{-1}\partial_\mu U\right)\left(U^{-1}A_\mu U+\frac{i}{g} U^{-1}\partial_\mu U\right)\\
            &=||A||^2+\frac{2i}{g}\mathrm{Tr}\int d^{D}x\;\left(U^{-1}A_\mu \partial_\mu U+\frac{1}{g^2}\partial_\mu U^{-1}\partial_\mu U\right),
        \end{split}
    \end{equation}
    where the cyclic property of trace operation $\mathrm{Tr}(ABC)=\mathrm{Tr}(BCA)$ has been implicitly used in the sub-sequential steps. Henceforth,
    it will be shown easily that \begin{equation}
     \frac{2i}{g}\mathrm{Tr}\int d^{D}xU^{-1}[\tau A_{\mu1}+(1-\tau)A_{\mu2}] \partial_\mu U+\frac{1}{g^2}\partial_\mu U^{-1}\partial_\mu U\geq 0
    \end{equation}
    or $||^U(\tau A_{\mu1}+(1-\tau)A_{\mu2})||^2\geq||(\tau A_{\mu1}+(1-\tau)A_{\mu2})||^2$ if $||^UA_{\mu1}||^2\geq||A_{\mu1}||^2$ and $||^UA_{\mu2}||^2\geq||A_{\mu2}||^2$ are independently hold. In summary, any affine combination of two gauge fields belonging to the FMR also belongs to the FMR. In the other words, we have completed the proof that $\Lambda$ is convex.
    \item Again, since the FMR is typically smaller than the Gribov region, thus, it is bounded from every direction.
\end{enumerate}
The final remark about this region is it still contains the Gribov copies. Fortunately, at this point, the Gribov copies are allowed to live only on the boundary of the FMR not inside it. Therefore, the restriction into the FMR is then solved the Gribov ambiguity completely in the way that the Gribov copies contribute no more to the path-integral and the FP operator is now positive definite. The most unsatisfactory thing is that currently there is no analytical method to do so. As good as we can get, in the next section, we will study the Gribov's way to restrict the path integration inside the Gribov region \cite{Gribov1978}, i.e. Even through the Gribov copies are still there, at least, we can introduce the FP ghost in the usual way.
\section{Gluon Propagator's Enhancement}
The time has come, we will start this section to compute the new form of the gluon propagator due to the restriction on the Gribov region. Let's see what are things we can do. Recalling the partition function \eqref{2.22}
\begin{equation}
    Z(0)=\int_\Omega\mathcal{D}A\int \mathcal{D}c\int \mathcal{D}\Bar{c}\exp{\left\{-\left[S_{YM}+\int d^{D}x\left(\frac{1}{2\xi}(\partial_\mu A^a_\mu)^2+ \;\Bar{c}^a\partial_\mu D_\mu c^a\right)\right]\right\}}.
\end{equation}
However, at this moment, we deform the range of the integration from the whole gauge space into the Gribov region $\displaystyle{\int\mathcal{D}A}\rightarrow \displaystyle{\int_\Omega\mathcal{D}A}$. The ultimate objective of us now is to explore what is the most suitable condition to evaluate the deformed integral. Observe that the FP operator is nothing but the kinetic differential operator of the ghost fields which has nothing to be curious about since the ghost fields were introduced to evaluate the FP determinant at the first place. What is this observation tells us? Instead of thinking that the FP operator is the kinetic operator of the ghost field, we can rather think that it can be treated as the inverse operator of the ghost propagator. Thus, equivalent to the condition that the FP operator is positive definite inside the Gribov region, we can alternatively look for the condition that the ghost propagator has no non-trivial pole.\\
\indent The observation suggests us strongly to extract the pole structure of the ghost propagator. The easiest way to single out the pole structure from any propagator is to calculate its two-points one-particle irreducible (1PI) Feynman diagram. For one who is not familiar with the stated argument, this kind of situation can be found also in the QFT101, e.g. The electron self-energy contributing to the extra pole electron mass which leads further to the radiative correction of the Lamb shift effect, the shift between $2s$ and $2p$ orbitals of the Hydrogen atom (See, for example, \cite{Schwartz:2013pla}.)\\
\indent In our case, we are about to evaluate the 2-points 1PI ghost graph, especially, at 1-loop level which is, in particular, enough to obtain what we want. Generally, the ghost propagator can be written in term of the geometric series as follows
 \begin{equation}
   \begin{gathered}
  iG^{ab}_{ghost}=\hspace*{1cm}
  \begin{fmffile}{ghostprop2}
\begin{fmfgraph*}(35,25)
\fmfleft{i}
\fmflabel{$b$}{i}
\fmflabel{$a$}{o}
\fmfright{o}
\fmf{ghost}{i,o}
\fmfforce{(-0.2w,0.0h)}{i}
\fmfforce{(1.1w,0.0h)}{o}
\end{fmfgraph*}
\end{fmffile}
\hspace*{1cm}+\hspace*{1cm}
\begin{fmffile}{ghostprop1loop}
\begin{fmfgraph*}(70,25)
\unitlength = 2.5mm
  \fmfleft{i} \fmfright{o}
  \fmf{ghost}{i,v1} \fmf{ghost}{v2,o}
  \fmf{ghost}{v1,v2}
  \fmf{gluon,right}{v2,v1}
  \fmfforce{(-0.2w,0.0h)}{i}
  \fmfforce{(0.25w,0.0h)}{v1}
  \fmfforce{(0.65w,0.0h)}{v2}
  \fmfforce{(1.1w,0.0h)}{o}
  \fmflabel{$b$}{i}
\fmflabel{$a$}{o}
\end{fmfgraph*}
\end{fmffile}
\hspace*{1cm}+\;\cdots,
\end{gathered}
\end{equation}
where $\cdots$ denotes higher-loops diagrams which can be obtained simply after we have a 1-loop diagram. Hence, we will focus only on the 1-loop graph, for simplicity, we will firstly truncate the external leg and will add it back later,
\begin{figure}[h!]
    \centering
    \begin{fmffile}{ghostprop1loopexpressed}
\begin{fmfgraph*}(190,60)
\unitlength = 2.5mm
  \fmfleft{i} \fmfright{o}
  \fmf{ghost,label=$k$}{i,v1} \fmf{ghost,label=$k$}{v2,o}
  \fmf{ghost,label=$k-p$}{v1,v2}
  \fmf{gluon,label=$p$,right}{v2,v1}
  \fmfforce{(-0.2w,0.1h)}{i}
  \fmfforce{(0.25w,0.1h)}{v1}
  \fmfforce{(0.65w,0.1h)}{v2}
  \fmfforce{(1.1w,0.1h)}{o}
  \fmflabel{$b$}{i}
\fmflabel{$a$}{o}
\fmflabel{$A^c_\mu,e$}{v1}
\fmflabel{$A^d_\nu,f$}{v2}
\end{fmfgraph*}
\end{fmffile}
\end{figure}\\
along with the help of the Feynman rules \eqref{2.28} and \eqref{ghostmodified}, we can read off the Feynman diagram above as
\begin{equation}
\label{3.65}
    \begin{split}
        iG^{ab}_{ghost,1-loop}(k^2)&=\frac{g^2}{V}f^{adf}f^{ecb}\int\frac{d^{D}p}{(2\pi)^{D}}\;\frac{i\delta^{ef}}{(k-p)^2}k_\nu (k-p)_\mu A^d_\nu(-p)  A^c_\mu(p)\\
        &=\frac{ig^2}{V}f^{ade}f^{ecb}\int\frac{d^{D}p}{(2\pi)^{D}}\; A^d_\nu(-p)  A^c_\mu(p)\frac{k_\nu (k-p)_\mu}{(k-p)^2},
    \end{split}
\end{equation}
Now we will perform very wonderful trick, in particular, we can guess the true form of the product $A^d_\nu(-p)A^c_\mu(p)$ by using the following argument. Since the propagator is legitimately Lorentz invariant and the product above is dependent only on $p$, thus, the product is expected to be written in the Lorentz covariant tensorial structure taking the unique form (up to the Lorentz scalar function $C(A^2)$)
\begin{equation}
\label{3.66}
    A^d_\nu(-p)A^c_\mu(p)=C(A^2)\left(g_{\mu\nu}-\frac{p_\mu p_\nu}{p^2}\right)\equiv C(A^2)\Pi_{\mu\nu}(p),
\end{equation}
where $\Pi_{\mu\nu}(p)$ is understood as $\Pi_{\mu\nu}(p,\xi=0)$ as defined once in the expression \eqref{2.27}. Obviously $\Pi_{\mu\nu}$ is the right Lorentz covariant tensorial structure in the sense that $\Pi_{\mu\nu}\rightarrow \Lambda_\mu\!^\rho \Lambda_\nu\!^\sigma \Pi_{\rho\sigma}$ under the Lorentz transformation (this will be cancelled precisely with the Lorentz transformation matrix from $k^\mu\rightarrow \Lambda^\mu\!_\nu k^\nu$). Moreover, the Lorentz scalar $C(A^2)$ is easy to determine by contacting both sides of the equation \eqref{3.66} by $g^{\rho\mu}$ then summing over indices $\nu$ and $\rho$. So we get
\begin{equation}
\label{3.67}
    C(A^2)=\frac{1}{d}A^d_\mu(-p)A^c_\mu(p).
\end{equation}
Plug the result of analysis \eqref{3.66} and \eqref{3.67} into \eqref{3.65} to obtain
\begin{equation}
    iG^{ab}_{ghost,1-loop}(k^2)=\frac{ig^2}{Vd}f^{ade}f^{ecb}\int\frac{d^{D}p}{(2\pi)^{D}}\; A^d_\rho A^c_\rho\frac{k_\nu (k-p)_\mu}{(k-p)^2}\Pi_{\mu\nu}.
\end{equation}
The next step is to observe that the integration here is the integration over the whole momentum space, thus, $p_\mu$ in $(k-p)_\mu$ can be ignored since if we split the integrand into two separate terms, the term depending on the $p_\mu$ will be the odd function on the momentum integral over the space of $p$. One more thing that can use the help of the oddity of the integrand to simplify is the product $p_\mu p_\nu$ inside $\Pi_{\mu\nu}$, it can be understood clearly that the integrand will be odd function unless $\mu=\nu$. Thus,
\begin{equation}
   \int d^{D}p\;\Pi_{\mu\nu}(...)= \int d^{D}p\;\left(g_{\mu\nu}-\frac{p_\mu p_\nu}{p^2}\right)(...)=g_{\mu\nu}\left(1-\frac{1}{D}\right)\int d^{D}p\;(...).
\end{equation}
The last thing we need to evaluate is the color indices which are not a big deal at all since the true expression of ghost propagator must be averaged over the color space. Consequently, we need to sum over all possible color of the final state then dividing out by prefactor representing the total numbers of the color, i.e. $N^2-1$ for $SU(N)$. Finally, we have
\begin{equation}
    \begin{split}
        iG_{ghost}(k)&=\frac{\delta^{ab}}{N^2-1}iG^{ab}_{ghost}(k)\\
        &=\frac{\delta^{ab}}{N^2-1}\left(\frac{i\delta^{ab}}{k^2}+\frac{i}{k^2}iG^{ab}_{ghost,1-loop}(k)\frac{i}{k^2}+\mathcal{O}(g^4)\right)\\
        &=\frac{i}{k^2}-\frac{ig^2}{VD}\frac{f^{ade}f^{eca}}{N^2-1}\frac{1}{k^2}\int\frac{d^{D}p}{(2\pi)^{D}}\; A^d_\mu A^c_\mu\frac{1}{(k-p)^2}+\mathcal{O}(g^4)\\
        &=\frac{i}{k^2}+\frac{ig^2}{VD}\frac{N\delta^{cd}}{N^2-1}\frac{1}{k^2}\int\frac{d^{D}p}{(2\pi)^{D}}\; A^d_\mu A^c_\mu\frac{1}{(k-p)^2}+\mathcal{O}(g^4)\\
        &=\frac{i}{k^2}\left(1+\frac{g^2}{VD}\frac{N}{N^2-1}\int\frac{d^{D}p}{(2\pi)^{D}}\; \frac{(A^a_\mu)^2}{(k-p)^2}+\mathcal{O}(g^4)\right)\\
        &\equiv\frac{i}{k^2}(1+\sigma(k^2,A)+\mathcal{O}(g^4)),
    \end{split}
\end{equation}
where the identity $f^{dae}f^{cae}=N\delta^{cd}$ has been used. As we have mentioned once, that the ghost propagator, in fact, can be expressed as the geometric series of the 1PI graph. Thus, we expect that the $n$th-ordered graph must be of the form $\frac{i}{k^2}\sigma^n$. Using the formula of the geometric series $1+\sigma+\sigma^2+...=\frac{1}{1-\sigma}$. We therefore can write
\begin{equation}
    G_{ghost}(k^2)=\frac{1}{k^2}\frac{1}{1-\sigma(k^2,A)}.
\end{equation}
There it is, the pole structure has been revealed itself already. After we are done with the tricky calculation, let's grab some tea and discuss the implication of the result. Here, we see that there are two poles of the ghost propagator: one is $k^2=0$ and another one is $1-\sigma=0$. The former is very meaningful in the sense that such a pole is a signature that all massless field's propagators must have. In other words, this pole tells us that the ghost field is massless. Additionally, this pole is good since it is, in general, positive definite following from the fact that the ghost field is not tachyonic. On the other hand, the latter is not a well-behaved pole since nothing can guarantee that it will be positive definite. In the end, it helps Gribov to find the appropriate condition to restrict the path integration into his Gribov region, i.e. We require that
\begin{equation}
    1-\sigma(k^2,A)>0.
\end{equation}
This condition, in literature, is known as the Gribov no-pole condition. In particular, we can simplify this condition furthermore by looking carefully at the expression of $\sigma$. Let's see
\begin{equation}
    \sigma(k^2,A)=\frac{g^2N}{VD(N^2-1)}\int\frac{d^{D}p}{(2\pi)^{D}}\frac{(A^a_\mu)^2}{(k-p)^2},
\end{equation}
it can be seen that by slight increasing of the $k$, the integrand becomes smaller and smaller. Thus, we deduce reasonably that $\sigma(k^2,A)$ is a decrease function of $k^2$. One might not feel clear about this, so let's prove it in the simplest case, in 2-dimensional case ($D=2$). We have
\begin{equation}
\label{3.74}
    \begin{split}
        \sigma(k^2,A)&=\frac{g^2N}{2V(N^2-1)}\int\frac{d^2p}{(2\pi)^2}\frac{(A^a_\mu)^2}{(k-p)^2}\\
        &=\frac{g^2N}{8\pi^2V(N^2-1)}\int^\infty_0 dp\;p(A^a_\mu(p))^2\int^{2\pi}_0d\theta\;\frac{1}{k^2+p^2-2pk\cos\theta}\\
        &=\frac{g^2N}{8\pi^2V(N^2-1)}\int^\infty_0 dp\;p(A^a_\mu(p))^2\left[\frac{2}{k^2-p^2}\tan\left(\frac{k+p}{k-p}\tan^{-1}\left(\frac{\theta}{2}\right)\right)\right]^{2\pi}_{0}\\
        &=\frac{g^2N}{4\pi V(N^2-1)}\int^\infty_0 dp\;\frac{p}{k^2-p^2}(A^a_\mu(p))^2.
    \end{split}
\end{equation}
The derivative respect to $k^2$ will show whether \eqref{3.74} is increasing function or decreasing function. The derivative is very straightforward to be done
\begin{equation}
    \frac{\partial \sigma}{\partial k^2}=-\frac{g^2N}{4\pi V(N^2-1)}\int^\infty_0 dp\;\frac{p}{(k^2-p^2)^2}(A^a_\mu(p))^2<0.
\end{equation}
Indeed, the higher-dimensional generalization, which is harder to perform the integration, is needed to be sure but we will presume that the result will hold in any dimension. By the usage of this argument, it is good enough to reduce the Gribov no-pole condition into
\begin{equation}
    1-\sigma(0,A)>0.
\end{equation}
To insert this constraint into the path integral, we rewrite it into a more systematical form, i.e. By using the Heaviside step-function. To be explicit, we can rewrite the partition function of the QYM to be
\begin{equation}
    \begin{split}
         Z(0)=&\int_\Omega\mathcal{D}A\int \mathcal{D}c\int \mathcal{D}\Bar{c}\;e^{-\left[S_{YM}+\int d^{D}x\left(\frac{1}{2\xi}(\partial_\mu A^a_\mu)^2+ \;\Bar{c}^a\partial_\mu D_\mu c^a\right)\right]}\\
         =&\int\mathcal{D}A\int \mathcal{D}c\int \mathcal{D}\Bar{c}\;\theta(1-\sigma(0,A))e^{-\left[S_{YM}+\int d^{D}x\left(\frac{1}{2\xi}(\partial_\mu A^a_\mu)^2+ \;\Bar{c}^a\partial_\mu D_\mu c^a\right)\right]}\\
         =&\int\mathcal{D}A\int \mathcal{D}c\int \mathcal{D}\Bar{c}\int^{\infty+i\varepsilon}_{-\infty+i\varepsilon}\frac{d\beta}{2\pi i}\;\frac{e^{\beta(1-\sigma(0,A))}}{\beta}e^{-\left[S_{YM}+\int d^{D}x\left(\frac{1}{2\xi}(\partial_\mu A^a_\mu)^2+ \;\Bar{c}^a\partial_\mu D_\mu c^a\right)\right]},
    \end{split}
\end{equation}
where, in the final step, instead of writing the Gribov no-pole condition in term of the Heaviside step-function, we are decide to write it in term of its integral representation \eqref{A.23} since it is sometime easier to be evaluated.\\
\indent So far so good, it is time to compute the gluon propagator corresponding to the new partition function. Since, in this case, the ghost sector and gluon self-interaction terms are not useful, we will implicitly ignore them for now. Moreover, the performance in the momentum space is more convenient. We therefore have the momentum-space partition function of the form (see also in appendix A.3)
\begin{equation}
\label{3.78}
    Z^{quad}_{gluon}(J)=\int \frac{d\beta}{2\pi i}\frac{e^\beta}{\beta}\int \mathcal{D}A\exp\left\{-\left[\int\frac{d^{D}p}{(2\pi)^{D}}\left(\frac{1}{2}A^a_\mu\Delta^{ab}_{\mu\nu}A^b_\nu-A^a_\mu J^a_\mu\right)\right]\right\},
\end{equation}
where we have redefined a bit from $A\rightarrow -A$ without changing anything but the last term in the expression and the new differential operator $\Delta^{ab}_{\mu\nu}(p)\equiv\delta^{ab}( p^2\Pi_{\mu\nu}(p,\xi)+\frac{2\beta g^2N}{VD(N^2-1)}\frac{g_{\mu\nu}}{p^2})$. The first part of the new operator comes from the usual propagator while the extra contribution comes from the Gribov no-pole condition. The integral is, once again, a Gaussian integral which can be computed through the formula \eqref{A.1}. The result is then
\begin{equation}
    \begin{split}
        Z^{quad}_{gluon}(J)&=\int \frac{d\beta}{2\pi i}\frac{e^\beta}{\beta}\frac{1}{\sqrt{\det\Delta^{ab}_{\mu\nu}}}\exp\left[\int\frac{d^{D}p}{(2\pi)^D}J^a_\mu(-p)(\Delta^{ab}_{\mu\nu})^{-1}J^b_\nu(-p)\right]\\
        &=\int \frac{d\beta}{2\pi i}\frac{e^\beta}{\beta}\frac{1}{\sqrt{\det\Delta^{ab}_{\mu\nu}}}\left(1+\int\frac{d^{D}p}{(2\pi)^{D}}J^a_\mu(-p)(\Delta^{ab}_{\mu\nu})^{-1}J^b_\nu(-p)+\mathcal{O}(J^4)\right).
    \end{split}
\end{equation}
After obtaining the true expression of the partition function, we can use this to compute the gluon propagator straightforwardly as follows
\begin{equation}
\label{3.80}
   \begin{split}
        \big<A^a_\mu(p)A^b_\nu(0)\big>&=\frac{1}{Z^{quad}_{gluon}(0)}\frac{\partial^2 Z^{quad}_{gluon}(J)}{\partial J^a_\mu(-p)\partial J^b_\mu(0)}\bigg|_{J=0}\\
        &=\frac{1}{Z^{quad}_{gluon}(0)}\delta^{D}(p)\int\frac{d\beta}{2\pi i}\frac{e^\beta}{\beta}\frac{1}{\sqrt{\det \Delta^{ab}_{\mu\nu}}}(\Delta^{ab}_{\mu\nu})^{-1}\\
        &=\frac{\displaystyle{\delta^{D}(p)\int\frac{d\beta}{2\pi i}e^{f(\beta)}(\Delta^{ab}_{\mu\nu})^{-1}}}{\displaystyle{\int \frac{d\beta}{2\pi i}e^{f(\beta)}}},
   \end{split}
\end{equation}
where $f(\beta)\equiv \beta-\ln\beta-\frac{1}{2}\mathrm{Tr}\ln(\Delta^{ab}_{\mu\nu})$.
We come into the problem, the last big one, how we can evaluate the integral over the parameter $\beta$. This is when the thermodynamics' argument is taken into account. Recalling that the partition function $Z(0)\sim e^{F}$ where $F$ is a free energy of the system. We have also known that the free energy is an extensive quantity meaning it is directly proportional to the size of the system $V$. Thus, no matter the result of integration will look like, in the limit $V\rightarrow \infty$, the configuration that extremizes the function $f(\beta)$ will dominate the integral. This method is known as the saddle point approximation widely used in, for example, the mean-field theory of the Ising model. Suppose the value of $\beta=\beta_0$ is the value that extremizes the function $f(\beta)$, we can then approximate (up to negligible higher-order fluctuation)
\begin{equation}
    \int\frac{d\beta}{2\pi i}e^{f(\beta)}=e^{f(\beta_0)}\int\frac{d\beta}{2\pi i}(1+...)\equiv e^{f(\beta_0)}\mathcal{N}.
\end{equation}
Thus, we end up with (assume that the $\Delta^{-1}$ does not oscillate too strong)
\begin{equation}
    \big<A^a_\mu(p)A^b_\nu(0)\big>=\delta^{d+1}(p)(\Delta^{ab}_\mu)^{-1}\bigg|_{\beta=\beta_0}.
\end{equation}
So the climax of the story is to calculate the inverse of the operator
\begin{equation}
\label{3.83}
    \Delta^{ab}_{\mu\nu}(p)\bigg|_{\beta=\beta_0}=\delta^{ab}\left( p^2\Pi_{\mu\nu}(p,\xi)+\frac{2\beta_0 g^2N}{VD(N^2-1)}\frac{g_{\mu\nu}}{p^2}\right)\equiv\delta^{ab}\left( p^2\Pi_{\mu\nu}(p,\xi)+\gamma^4\frac{g_{\mu\nu}}{p^2}\right).
\end{equation}
One can check that the inverse of the operator above takes the form
\begin{equation}
    (\Delta^{ab}_{\mu\nu})^{-1}(p)\bigg|_{\beta=\beta_0}= \delta^{ab}\left(\frac{p^2}{p^4+\gamma^4}\Pi_{\mu\nu}(p)+\frac{\xi p^2}{\xi\gamma^4+p^4}\frac{p_\mu p_\nu}{p^2}\right).
\end{equation}
In the Landau gauge condition, i.e. $\xi=0$, we eventually obtain the gluon propagator of the form
\begin{equation}
\label{Gribov propagator}
     \big<A^a_\mu(p)A^b_\nu(0)\big>=\delta^{D}(p)\delta^{ab}\frac{p^2}{p^4+\gamma^4}\Pi_{\mu\nu}(p).
\end{equation}
To end this chapter, let us give remarks about the new form of the gluon propagator, from now on, called the gluon propagator of the Gribov type. First of all, let's start with the most important observation that the gluon propagator now changes its form. Strictly speaking, it effectively becomes a massive propagator with the mass $\gamma$, called the Gribov mass from now on. Where does it come from actually? Recall that, at the beginning, the gluon field is classically constructed to be a massless field. The evidence is the classical YM theory, by construction, is generically scaleless or conformal invariant. Amazingly, the effect of the quantization makes very interesting consequence. In the quantum regime, the massless gluon gains its mass miraculously. In particular, the generation of the mass scale is not that mysterious at all. Tracing back into the resolution of Gribov to resolve the Gribov ambiguity, Gribov has introduced the Gribov region whose horizon is bounded from every direction (see the final property of the Gribov region). At that moment, we have already introduced the specific length scale into the gauge space which can be translated into the mass-scale inside its Fourier world. Strictly speaking, by introducing the Gribov region, we have also introduced the mass scale into the theory without hesitation. If the result from Gribov is true, the QYM theory thus have the mass gap which will be shown further in the next chapter that the presence of this mass scale do naturally generate the confinement of the gluon.\\
\indent In fact, the result is not totally quantum. Let's remind readers that the result is clearly obtained when the size of the system is sufficiently large or $V\rightarrow \infty$. In this particular limit, the finite size effect from the periodic boundary condition will be negligible. Roughly speaking, the box quantization tells us that the momentum mode will be quantized into the discrete value which is inversely proportional to the size of the system. Thus, in the large volume limit, the discrete distribution of the momentum mode will become continuous and approximately classical. Hence, the limit is therefore called the semiclassical limit. However, the semiclassical approximated used by Gribov will make sense if and only if the $f(\beta)\sim V$. Let's see how it works actually. As we have discussed in the above paragraph that the $\gamma$ represents the mass scale of the QYM theory, thus, we believe so much that the scale must be finite and not equal to zero. The reason can be deduced intuitively if the mass scale is infinite, it is equivalent to the fact that the radius of the Gribov region is typically zero. This can not be solved anything since the zero radius Gribov region will forbid any gauge orbit to exist in its own space. On the other hand, for the case that the Gribov mass is zero, it implies that the first Gribov horizon locates at the far infinity in the space of gauge fields. This situation will not solve any ambiguity also since if the Gribov region has infinite radius, the Gribov region can not stop any gauge orbit from passing into the negative eigenmodes' zone. In fact, it can be understood with simpler logic, if the Gribov mass is typically zero, it means that the mass gap is not generated at all, thus, the result will merely coincide with the classical theory. To see the behavior of the Gribov mass, let's recall the full expression of the Gribov mass
\begin{equation}
\label{3.86}
    \gamma^4=\frac{2\beta_0g^2N}{VD(N^2-1)}.
\end{equation}
Thus, to have non-zero and finite Gribov mass, the value $\beta_0$ must run as $\beta_0\sim V$. As a result, $f(\beta_0)\sim V$ as expected.\\
\indent In the end, one might ask that so how we can determine the specific value of the Gribov mass scale $\gamma$. Observe from the expression \eqref{3.86} that what we need to do is to determine the value of $\beta_0$, to be specific, the finite ratio between $\frac{\beta_0}{V}$. After that, we can use the result to estimate the value of the Gribov mass (the rest of them are just the parameters depending on the details of the theory). To compute it out, we need to think first that what really is the $\beta_0$. Right! $\beta_0$ is the configuration that makes the function $f(\beta)$ become extremum. Thus, we can then simply find out the value of $\beta_0$ through the calculus 101 condition $\frac{df}{d\beta}\big|_{\beta=\beta_0}=0$ which is widely known in literature as the gap equation. Finally, we are done.

\newpage

\chapter{Gluon Confinement}
In this final chapter, we will use the result in the previous chapter to show that the gluon should be confined due to the violation of the axioms of the QFT. By the way, this topic is kind of difficult since it requires a lot of basics knowledge in both mathematics and physics. Thus, we will take a serious discussion on the basics in the first section and then use it in the last section.
\section{Axioms of Quantum Field Theories}
Axiomatic QFT (AQFT for short) is an effort to assign rigorous mathematical descriptions to QFT. The subject begins with the first question, simple but not too trivial, at all: which object can be used to define a \say{quantum field}? To answer this simple question, let's start our discussion with an important physical observation. Historically, what we have known as \say{field} was introduced to solve the problem called the action at a distance appearing in the naive notion of the classical force, e.g. Newtonian gravity and Coulomb's electrostatics. Instead of mentioning the quantity like force, we will rather focus on the quantity like a field that is distributed everywhere in the finite local region in spacetime. This motivates us to introduce the notion of field as the distribution instead of the naive function. Hence, we will first study the distribution theory. Straightforwardly, the (tempered) distribution is a continuous linear functional on a so-called space of Schwartz test functions \begin{equation}
\label{Schwartz test space}
    \mathcal{S}\equiv\{f(x)\in C^\infty(\mathbb{R}^D)|\sup_{x\in \mathbb{R}^D}|x^\alpha \partial^\beta f(x)|<\infty\}
\end{equation}
where $x^\alpha\equiv x^{\alpha_0}_0...x^{\alpha_d}_d$ and $\partial^\beta\equiv \frac{\partial^{\beta_0+...\beta_d}}{(\partial x^0)^{\beta_0}...(\partial x^d)^{\beta_d}}$ \cite{Dybalski2018}. In other word, the distribution $\varphi$ is a linear functional $\varphi:\mathcal{S}\rightarrow \mathbb{C}$ and we will denote the collection of the distribution as $\mathcal{S}^*$. In fact, any function $\varphi$ that grows no faster than any polynomial (called polynomially bounded) defines a (regular) tempered distribution through the formula
\begin{equation}
    \varphi(f)\equiv \int d^Dx\;\varphi(x) f(x).
\end{equation}
The famous example of the singular type is a Dirac delta distribution $\delta(f)\equiv f(0)$. Let us remark that even though the value at the specific point in the spacetime of any distribution, defined as the functional, cannot be found, we can still talk about the local value of it. Specifically, there is a very important terminology need to be mentioned here. We will say that the distribution vanishes, $\varphi=0$, in the neighborhood of the point $x_0$ if $\varphi(f)=0$ for all test functions $f(x)$ which vanish outside this neighborhood. The complement of the set of points at which a distribution $\varphi$ vanishes with respect to the entire spacetime will be called the support of $\varphi$ \cite{Streater1976}.\\
\indent Now we just obtain the definition of the classical field. Obviously, the next thing we have to do is to promote the classical field into the quantum field. Recall that Heisenberg's uncertainty principle suggests that the $c$-number-valued distribution essentially extends to the operator-valued distribution. Explicitly, the field has been promoted into the map of the kind
\begin{equation}
    \varphi: \mathcal{S}\rightarrow \mathrm{Lin}(\mathcal{D},\mathcal{H}),
\end{equation}
where $\mathrm{Lin}(\mathcal{D},\mathcal{H})$ denotes the set of all linear ($\mathrm{Lin}$ stands for linear) maps, formal definition of operators, from a dense domain $\mathcal{D}$ to the Hilbert space $\mathcal{H}$ with the inner product $\big<\big|\big>$ and norm $||\cdot||\equiv \sqrt{\big<\cdot\big|\cdot\big>}$. In particular, it is natural to expect that the common dense domain $\mathcal{D}$ is contained inside the domain of the field distribution $D(\varphi(f))$. In quantum theory, we mainly deal with the Hermitian, or more strictly, self-adjoint operator since the spectral values, the generalization of the eigenvalues, of such operators are all real. The dense domain of adjoint operator $O$ is defined as $D(O^\dagger)\equiv \{\psi\in\mathcal{H}|\exists\tilde\psi\in \mathcal{H}:\forall\psi'\in D(O),\big<\tilde\psi\big|\psi'\big>=\big<\psi\big|O\psi'\big>\}$ where $\mathbb{P}(\mathcal{H})$ is the projective Hilbert space which we shall clarify it further below. Consequently, inside this dense domain, we can say that the operator $O$ is a Hermitian operator when $D(O)\subset D(O^\dagger)$ and $O^\dagger\psi=O\psi$ for all possible $\psi\in D(O)$ and it will be called the self-adjoint operator if $D(O)=D(O^\dagger)$. For such an operator, writing as $\big<\psi\big|O\psi'\big>=\big<\psi\big|O\big|\psi'\big>$ is sometimes acceptable.\\
\indent Only operators in Hilbert space $\mathcal{H}$ is not enough to define QFT, not even QM. In experiment, the measurement apparatus, i.e. initial state and final state, is represented by an element in the projective Hilbert space $\mathbb{P}(\mathcal{H})$ defined to be an equivalence class $[\psi]=\{\psi'\in \mathcal{H}|\psi'\sim\psi\}$ where $\psi'$ is said to be equivalent with $\psi$, denoted as $\psi'\sim \psi$, if there exists some non-zero constant $\lambda\in \mathbb{C}$ such that $\psi'=\lambda\psi$. The requirement of the equivalence relation is introduced to make all elements in the equivalence class represent the same physical situation (all elements in the same equivalence class will give the same probability amplitude).\\
\indent Yet, this is still not the complete description of QFT since, as mentioned at the beginning of the section that field was initially introduced to solve the action at a distance problem, thus we necessarily further demand that our field must obey the relativistic properties. In other words, the quantum field must transform covariantly under the Poincar\'e transformation forming the group of transformation known as, of course, the Poincar\'e group $\mathcal{P}$ whose element is of the form $(\Lambda,a)\in \mathcal{L}\rtimes T$ where $\mathcal{L}\equiv SO^\uparrow_+(1,d)$ and $T\sim \mathbb{R}^D$ are the proper orthochronous Lorentz group and translational group, respectively. Note that the reason why the product between the Lorentz group and translational group must be the semi-direct product is the fact that the double action of the Poincar\'e group can be expressed as follows
\begin{equation}
    \label{doublePoincare}
    (\Lambda,a)(\Lambda',a')=(\Lambda\Lambda',\Lambda a'+a)
\end{equation}
Following the above analysis, the theory of the relativity is strongly based on the assumption that the physical system is invariant under the frame-changing or Poincar\'e symmetry. In the level of state, the argument can be written shortly as 
\begin{equation}
    \big|\big<\psi\big|\psi'\big>\big|=\big|\big<u(\Lambda,a)\psi\big|u(\Lambda,a)\psi'\big>\big|,
\end{equation}
where $u$ is a representation of Poincar\'e group, i.e. $u$ denotes a map of the form $u:\mathcal{P}\rightarrow GL(D,\mathbb{R})$ and the absolute symbol $||$ is essentially inserted to eliminate the ambiguity of phase, i.e. there is still an extra $U(1)$ degree of freedom that makes the projective Hilbert space non-unique depending on the choice of this phase. 
\\
\indent In more general situation, suppose $g$ denotes an element of the group $G$ and $u(g)$ is its representation. If $G$ is truly the symmetry of the quantum system, we therefore have
\begin{equation}
\label{4.4}
   \big| \big<\psi\big|\psi'\big>\big|=\big|\big<u(g)\psi\big|u(g)\psi'\big>\big|.
\end{equation}
According to the equation \eqref{4.4}, we expect that the representation of the element of that group must satisfy the unitary condition $u^\dagger u=\mathbb{1}$ which means the appropriate representation must be a unitary representation. We will denote the set of all unitary operators defined in the Hilbert space $\mathcal{H}$ as $\mathcal{U}(\mathcal{H})$. However, this condition is too weak, in general, it will be stronger if we define the transition amplitude between the elements in the projective Hilbert space $\mathbb{P}(\mathcal{H})$ as $p([\psi],[\psi'])=\frac{|<\psi|\psi'>|}{<\psi|\psi><\psi'|\psi'>}$ then requiring that the transition probability satisfies $p([\psi],[\psi'])=p(u[\psi],u[\psi'])$ is much more stronger.\\
\indent Let us note that there is an important theorem showing that the action of $\mathcal{P}$ on the projective Hilbert space $\mathbb{P}(\mathcal{H})$ induces the unitary (anti-unitary) representation of that group element. The theorem states that \cite{Robert2016}
\begin{theorem}[Wigner's Theorem] Suppose that there are $\psi,\;\psi'\in\mathbb{P}(\mathcal{H})$ and the bijective map $u:\mathbb{P}(\mathcal{H})\rightarrow \mathbb{P}(\mathcal{H})$ which preserves the transition amplitude. Then there exists a linear (anti-linear) map $U:\mathcal{H}\rightarrow \mathcal{H}$ which is unitary (anti-unitary) such that $u([\psi])=[U(\psi)]$.
\end{theorem}
We will call $U$ as a lift of $u$ to $\mathcal{H}$, denoted by $U=\hat{\gamma}(u)$.\\
\indent Now the main problem turns out to be to determine the appropriate lift of the group. The problem will be resolved by the theorem due to Bargmann stating as follows \cite{Schottenloher2008}
\begin{theorem}[Bargmann's Theorem]
Suppose $G$ be a simply-connected and finite-\\dimensional Lie group with vanishing $2^{nd}$ Lie algebra cohomology. Then every continuous projective representation $u:G\rightarrow\mathcal{U}(\mathbb{P}(\mathcal{H}))$ has a lift as a unitary representation $U:G\rightarrow \mathcal{U}(\mathcal{H})$.
\end{theorem}
Unfortunately, the Poincar\'e group $\mathcal{P}=\mathcal{L}\rtimes T$ is not the case. Even though $T$ is simply-connected, $\mathcal{L}$ is not. Thus, we need to enlarge the Lorentz group into the bigger connected group that can cover the whole Lorentz group, i.e. its universal covering group, in this case, the universal covering group of $\mathcal{L}$ is $Spin(1,d)$.\\
\indent The analysis of the relativistic properties gives us more information. Since we have known already that the covariance property of the quantum field encodes inside the unitary representation of the Poincar\'e group. With the help of the following theorem, we can find out explicitly the specific form of that unitary representation.
\begin{theorem}[Stone's theorem]
If U is a strongly continuous unitary representation of group $\mathbb{R}$ in the unitary group of Hilbert $\mathcal{H}$, then there exists a self-adjoint operator $O$ in dense domain $D$ such that, for all $t\in \mathbb{R}$, we have $U(t)\big|_D=e^{itO}$. 
\end{theorem}
We call $O$ as the infinitesimal generator of $U(t)$. As a result, we obtain
\begin{equation}
    U(\omega,a)=\exp(i\omega_{\mu\nu}J^{\mu\nu}+ia^\mu P_\mu)=U(\mathbb{1},a)U(\omega,\mathbb{1})
\end{equation}
where $\omega_{\mu\nu}$ is the infinitesimal form of the finite angle of the Lorentz group $\Lambda_{\mu\nu}$, i.e. $\Lambda_{\mu\nu}\approx g_{\mu\nu}+\omega_{\mu\nu}$. In particular, $P_\mu$ and $J_{\mu\nu}$ denotes the infinitesimal generators of translational group $T$ and Lorentz group $L$, respectively. By imposing the physical interpretation, $P_\mu$ is treated to be the 4-momentum operator. Since all physical systems tend to evolve to the lowest possible energy configuration, thus, to maintain the stability of the system, we expect that the zeroth component of $P_\mu$ or energy is, therefore, essentially bounded from below. Mathematically, we say that it respects the so-called spectral condition.
\begin{definition}
\begin{enumerate}
    \item Joint spectrum $Sp(O_1,...,O_n)$ is the set of $(\lambda_1,...,\lambda_n)$ such that, for all $(t_1,...,t_n)$, $\sum_{i=1}^n\lambda_i t_i$ is in the spectrum of $\sum_{i=1}^n t_iO_i$ (obviously, all $O_i$ need to be compatible to each other since non-compatible operators cannot have simultaneous spectral value).
    \item The spectral condition is expressed as follows
    \begin{equation}
        Sp(P_0,P_1,...P_d)\equiv \{(p_0,p_1,...,p_d)\subset V^+\},
    \end{equation}
    where $V^+$ is known as the closed future lightcone, all points in it are timelike separated and evolve from present to future, defined mathematically as $\{(x_0,\mathbf{x})\in \mathbb{R}^{1,d}|x_0\geq 0,x^2_0-\mathbf{x}^2\geq 0\}$.
\end{enumerate}
\end{definition}
Notice that there is still the problem here. Observe that the Lorentz transformation affects inevitably the 4-momentum operator since the Lorentz transformation affects directly to the translational group (see \eqref{doublePoincare}). First of all, the causal structure is preserved under such a transformation meaning that there is no need to worry about the consistency of the spectral condition. However, the classification of the particle's spectrum will be difficult to do. Thus, this fact leads Wigner to use rather the subgroup of the Lorentz group of which all elements leave the 4-momentum operator unchanged. Such a subgroup is known as the little group or stabilizer subgroup. Due to the use of the little group, we can construct the induced representation of the Poincar\'e group that is well-behaved in the sense that the energy-momentum's spectra are unaffected. The resulting induced representation leads to the Wigner's classification of the physical particles. Wigner shows that only mass $m$ and spin $J$ are enough to characterize the particle (this fact, intuitively, follows from the presence of two Casimir operators, that are mass operator $p^2$ and Pauli-Lubanski spin operator $W^2$, of the Lorentz group) which is the representation of the Lorentz group (while the additional translational group will constrain the spectrum condition). Hence, the types of physical spectra can be listed below
\begin{enumerate}
    \item Massive particle is a state with positive mass $m>0$, spin $J$ and positive energy $E>0$.
    \item Massless particle is a state with zero mass $m=0$ but non-zero energy $E>0$ and helicity $h$ (we cannot practically measure the spin of the massless particle since the massless particle always travels at the speed of light that means it is impossible to catch the massless particle then measuring its spin).
    \item Vacuum is a state with vanishing mass $m=0$, vanishing spin $J=0$ and, of course, vanishing energy $E=0$.
\end{enumerate}
Following from the spectrum condition, the last type of the physical state called vacuum state. However, in the Wigner's theory, this is just the description used in the relativistic QM. By the way, this motivates us to define the physical subspace of the Hilbert space of the QFT, i.e. the vacuum state, denoted by $\Omega$, which is interpreted to be the ground state of the underlying QFT. In this sense, the vacuum state is believed to be the state that contains no particle at all. Thus, it is also expected to be singlet or completely invariant under the Poincar\'e transformation. Namely, we have
\begin{equation}
    U(\Lambda,a)\Omega=\Omega,\;\;\;\forall(\Lambda,a)\in \mathcal{P}.
\end{equation}
So far so good, we currently have all useful ingredient to construct the QFT.
\begin{definition}
The Wightman QFT \cite{Wightman} is a quadruple $(\mathcal{H},\mathcal{U},\varphi,\mathcal{D})$, that is composed of
\begin{enumerate}
    \item Separable Hilbert space $\mathcal{H}$ and its projective space $\mathbb{P}(\mathcal{H})$.
    \item Strongly continuous unitary representation of the universal covering of the Poincar\'e group $\mathcal{U}(\mathcal{H})$.
    \item The Hermitian operator-valued distribution $\varphi$ defined on the dense domain $\mathcal{D}$ in the Hilbert space $\mathcal{H}$.
    \item 1-dimensional vacuum $\Omega$ subspace of the projective Hilbert space $\mathbb{P}(\mathcal{H})$ which is unique with respect to the Poincar\'e symmetry.,
\end{enumerate}
which satisfies all following Wightman axioms
\begin{itemize}
    \item \textbf{W1} (\textbf{Covariance}) Since \begin{equation}
        \begin{split}
            \big<\psi\big|\varphi(f)\big|\psi'\big>&=\big<^{(\Lambda,a)}\psi\big|\varphi(^{(\Lambda,a)}f)\big|^{(\Lambda,a)}\psi'\big>\\
            &=\big<\psi\big|U^\dagger(\Lambda,a)\varphi(^{(\Lambda,a)}f)U(\Lambda,a)\big|\psi'\big>,
        \end{split}
    \end{equation}
    where $^{(\Lambda,a)}f\equiv f(\Lambda^{-1}(x-a))$. Therefore,
    \begin{equation}
        \varphi(^{(\Lambda,a)}f)=U(\Lambda,a)\varphi(f)U^\dagger(\Lambda,a).
    \end{equation}
    This axiom requires that the quantum field must be relativistic.
    \item \textbf{W2} (\textbf{Locality}) \begin{equation}
        [\varphi(f_1),\varphi(f_2)]=0
    \end{equation}
    if its support on the test functions $f_1$ and $f_2$ are spacelike separated, that is, if $f_1=f_1(x_1)$ and $f_2=f_2(x_2)$, then $(x_1-x_2)^2<0$ implying that the information cannot be sent between two fields at different events separated by spacelike interval. In other words, no information can travel faster than light.
    \item \textbf{W3} (\textbf{Spectral condition}) The joint spectrum of the 4-momentum operator is contained in the future lightcone. This condition is to maintain the bounded from below property of the energy.
    \item \textbf{W4} (\textbf{Uniqueness of vacuum}) All elements in $\mathbb{P}(\mathcal{H})$ which is invariant under translation are the vacuum state multiplied by some scalar. This implies the uniqueness of vacuum, as its name, since the scalar multiplies with vacuum $\Omega$ looks no different to the pure $\Omega$ in the rays on Hilbert space.
\end{itemize}
\end{definition}
Note that, in this case, we consider only the scalar field. For the more general case, we need to include the contribution of transforming the tensor and spinor indices into the W1 axiom also. Note further that the scalar field is a bosonic field, however, the generalization to the fermionic field can be done by merely changing the commutator into the super-commutator in W2. \\
\indent
This is very understandable why Wightman defines QFT in this way. However, the next discussion is the unexpected one. Recall that $f\mapsto\varphi(f)$ is an operator-valued distribution defined in the Hilbert space $\mathcal{H}$. Thus, the inner product on the Hilbert space is therefore defined appropriately. Namely, we have, for every $\psi'\in D(\varphi(f))$ 
\begin{equation}
    \big<\psi\big|\varphi(f)\psi'\big>\in \mathbb{C}.
\end{equation}
Whereas, the number $|\big<\psi\big|\varphi(f)\psi'\big>|^2\in \mathbb{R}$ represents the probability of finding the value of operator $\varphi(f)$ measured in the final state $\psi$ out from the prepared initial state $\psi'$. Let assume further that the observable is self-adjoint operator (as an observable should be), i.e. We can write $\varphi(f)^\dagger=\varphi(f)$, accordingly, we can introduce a very important quantity called the Wightman function(al) or $n$-points correlation function as a separatedly continuous, multi-linear map
\begin{equation}
    \label{Wightman function}
    W_n:\mathcal{S}\times \mathcal{S}\times...\times \mathcal{S}\rightarrow \mathbb{C}
\end{equation}
defined explicitly as
\begin{equation}
    \label{Def Wightman function}
    W_n(f_1,...,f_n)\equiv \big<\Omega\big|\varphi(f_1)...\varphi(f_n)\big|\Omega\big>,
\end{equation}
where $\Omega$ is a vacuum state, thus, sometimes this correlation is called the vacuum expectation value. One can think about this quantity in the following sense: we have known that the test function $f$ spans the vector space $\mathcal{S}$, $W_n(f_1,...,f_n)$ can be realized as the complex-valued tensor of $(0,n)$-type defined on this vector space.\\
\indent There are two significant remarks. First of all, the Wightman distribution $W_n(f_1,...,f_n)$ defines a unique distribution on $\mathcal{S}(\mathbb{R}^{Dn})$, denoted again by $W_n$, as follows directly from the Schwartz's nuclear theorem \cite{Streater:1989vi}
\begin{theorem}[Schwartz's Nuclear Theorem]
Let $W$ be a separately continuous, linear functional on $(\mathcal{S}(\mathbb{R}^D))^n$. There exists a unique distribution $W'$ such that
\begin{equation}
    W(f_1,...,f_n)=W'(f_1...f_n).
\end{equation}
That is there is a unique correspondence between the dual space $\mathcal{S}^*((\mathbb{R}^D)^n)$ and $\mathcal{S}^*(\mathbb{R}^{Dn})$.
\end{theorem}
Another remark is that the correlation function appearing here is not the same as the correlation function in the previous chapter calculated from the variational derivative of the partition function. Observe that the Wightman distribution does not include the time-ordering operator into it but the chronological order of the operator-valued quantity must be seriously concerned. One might ask that can we modify the definition of the Wightman distribution by naively operating it with the time-ordering operator. This answer is we can not naively do that since the multiplication of the distribution with the discontinuous function (the definition of the time-ordering operator can be expressed through the discontinuous Heaviside step function) will make it loses its temperedness. However, it can be done eventually by approximating the Heaviside step function as a smooth function.\\
\indent Following from the Wightman axioms, one can prove the following theorem
\begin{theorem}
The Wightman function $W_n\in S^*(\mathbb{R}^{Dn})$ satisfies
\begin{itemize}
    \item \textbf{WD1} (\textbf{Covariance}) $W_n(f)=W_n(^{(\Lambda,a)}f),\;\forall(\Lambda,a)\in \mathcal{P}$. This comes from the Wightman's first axiom W1 and the fact that vacuum is Poincar\'e invariant.
    \item \textbf{WD2} (\textbf{Locality})
    \begin{equation}
        W_n(f_1,...,f_i,f_{i+1},...,f_n)=W_n(f_1,...,f_{i+1},f_i,...,f_n),
    \end{equation}
    if the supports of $f_i$ and $f_{i+1}$ are spacelike separated.
    \item \textbf{WD3} (\textbf{Spectral condition}) There exists a distribution $w_n\in \mathcal{S}^*(\mathbb{R}^{D(n-1)})$ supported in the product $(V^+)^{n-1}$ of forward cones such that 
    \begin{equation}
       \label{4.17} W_n(f_1,...,f_n)=\int\frac{d^{D(n-1)}p}{(2\pi)^{D(n-1)}}w_n(\tilde{f}(p))e^{ip_j\xi_j},
    \end{equation}
    where $\xi_j
    \equiv x_{j+1}-x_j$. This condition is the implicit consequence of the property WD1 telling that the Wightman function is translational invariant, i.e. $W_n(f_1(x_1-a),...,f_n(x_n-a))=W_n(f_1(x_1),...,f_n(x_n))$, implying that the Wightman function is dependent only on the difference of the spacetime coordinates. Thus, we can legitimately define
    \begin{equation}
      w_n(f_1(\xi_1),...,f_{n-1}(\xi_{n-1}))\equiv W_n(f_1(x_1),...,f_n(x_n)).
    \end{equation}
    Finally, one can show that the Fourier transform of $w_n(f)$, i.e. $w_n(\tilde{f})$ belongs to the closed future lightcone.
    \item \textbf{WD4} (\textbf{Positive definiteness}) For $f_n\in \mathcal{S}(\mathbb{R}^{Dn})$, one has
    \begin{equation}
        \sum_{m,n=0}^kW_{m+n}(f_m^*\otimes f_n)\geq 0,
    \end{equation}
    where $f\otimes g$ is defined as $f\otimes g(x_1,...,x_{m+n})\equiv f(x_1,...,x_m)g(x_{m+1},...,x_{m+n})$. This can be shown as follows: We first construct the operator $\sum^k_{m=1}\varphi(f_m)\Omega\in\mathcal{H}$. Since this operator belongs to the Hilbert space, thus we can compute the norm of this operator which is, in principle, positive-definite. Explicitly, we obtain
    \begin{equation}
        \begin{split}
            \bigg|\bigg|\sum^k_{m=1}\varphi(f_m)\Omega\bigg|\bigg|^2&\geq 0\\
            \bigg<\sum^k_{m=1}\varphi(f_m)\Omega\bigg|\sum^k_{n=1}\varphi(f_n)\Omega\bigg>&\geq 0\\
            \bigg<\Omega\bigg|\sum^k_{m=1}\varphi^\dagger(f_m)\sum^k_{n=1}\varphi(f_n)\bigg|\Omega\bigg>&\geq 0\\
            \sum_{m,n}W_{m+n}(f^*_m\otimes f_n)&\geq 0,
        \end{split}
    \end{equation}
    where we have used the definition of the self-adjoint operator $\varphi^\dagger(f)=\varphi(f^*)$ in the sub-sequential step.
\end{itemize}
\end{theorem}
In some place in the literature, after introducing the Wightman function, one may assign one more axiom into the definition of the Wightman QFT known as the cluster decomposition axiom \cite{Streater:1989vi} which states as follows: 
\begin{equation}
    \lim_{\lambda\rightarrow \infty}W_n(f_1,...,f_i,^{(\mathbb{1},\lambda a)}f_{i+1},...,^{(\mathbb{1},\lambda a)} f_n)
    -W_i(f_1,...,f_i)W_{n-i}(f_{i+1},...f_n)=0,
\end{equation}
for any arbitrary spacelike vector $a$.
The physical meaning of this axiom is that two experiments performing at the far spacelike-separated events in the spacetime can be decomposed into two separated clusters of the experiment apparatus. In the other words, the two experiments, that are far enough (in the spacetime) so that the information from one apparatus can not reach the other at all, will not correlate with each other at all. This axiom might be very obvious to see but, in particular, this axiom plays a very significant role in the perturbative QFT in the sense that it ensures that the disconnected Feynman diagram will not contribute to the theory \cite{Schwartz:2013pla}.\\
\indent Now we have seen precisely how the Wightman axioms constrain the properties of the Wightman functional. One might ask for the converse argument which is the most marvelous result of the Wightman QFT.
\begin{theorem}[Wightman Reconstruction Theorem]
For any sequence of the tempered distributions $W_n\in \mathcal{S}^*(\mathbb{R}^{Dn})$ satisfying the properties WD1-WD4 (along with the cluster decomposition principle), there exists a Wightman QFT of which sequence of the Wightman functions coincides with the given sequence of the tempered distributions above.
\end{theorem}
This theorem is extraordinary, it tells us that if we know the vacuum correlation function of any $n$-points, we can reconstruct back the whole QFT. Speaking novelly, only one sequence of the Wightman functional can rule them (QFT) all. This, in fact, is used to give the mathematical definition of the integrable or exactly solvable QFT. For example, a massless Thirring model, a Dirac field with the self-interaction of the Fermi-type in (1+1)-dimensions, is said to be exactly solvable since all $n$-points functions are known. We will not go through it in details because it does not relate much to this review article.\\
\indent So far we have discussed the axiomatic QFT in Minkowski spacetime, however, several things defined on Minkowski QFT is slightly ill-defined. The reader might have noticed that everything in all three previous chapters is constructed through the path integral quantization. Moreover, the reader might have also noticed that our path integration used so far has been performed in the manifold endowed with the Euclidean metric. The main reason is that the path integral measure in the Minkowski space is ill-defined in the sense that we even have no idea about the underlying mathematical structure of the measure space. Additionally, even if the structure is well-defined, there is no guarantee that the path integral will converge. Conversely, the path integral working on the Euclidean space is well-defined, at least in free theory, in the sense that its measure can be realized as the Borel probability measure (see also the Bochner-Minlos theorem, for example, theorem 2.1 of \cite{Dybalski2018}). Therefore, after axiomatizing the Minkowski QFT, we will then study how can we use the information from the Wightman QFT to construct the Euclidean field theory. To study this, we essentially discuss the notion of analytic continuation. The idea begins within the context of the complex analysis by studying the analytic function. Recall that, if $U$ is an open set in complex plane $\mathbb{C}$, the complex-valued function $F:U\rightarrow \mathbb{C}$ is said to be holomorphic or (complex) analytic if, for any point $z_0\in U$, the power series expansion of $F$ exists and converges in a open disk $B(z_0,\rho)\subset U$, i.e.
\begin{equation}
    F(z)=\sum^\infty_{k=0}a_k(z-z_0)^k.
\end{equation}
Alternatively, by using the fact that $a_k$ can be matched with $\frac{1}{k!}\frac{\partial^k F}{\partial z^k}\big|_{z=z_0}$ in the Taylor's series, we say that the holomorphic function is the complex-valued function of which partial derivative exists at any order. In more abstract language of the branch of mathematics known as the sheaf theory, $F$ is found to be just one representative of an equivalence class called germ of $F$. All analytic functions inside the same germ will have equal value of their restriction to $U$, namely, $F$ and $G$ will be said to belong to the same germ if $F\big|_U=G\big|_U$ for all points in $U$. Thus, this leads to the context of analytic continuation, that is, $F$ can be analytically continued to $G$ when $G\big|_U=F$, we will sometimes call $F$ the boundary value of $G$.\\
\indent In particular, we can repeat the same thing in the case of the distribution. Let's start with finding the appropriate domain of analytic continuation. Recall first that our distribution obeys the causality implying that our domain is nothing more and nothing less but the closed future lightcone $V^+$. Then, we define a so-called induced backward tube to be $T\equiv \mathbb{R}^D\times(-V^O)$ where $V^O$ is an open future lightcone $\equiv \{(x_0,\mathbf{x})\in \mathbb{R}^{1,d}|x_0>0,x^2_0-\mathbf{x}^2>0\}$.
\begin{theorem} \cite{Dybalski2018}
Suppose $\varphi\in \mathcal{S}^*(\mathbb{R}^D)$ whose Fourier transform is supported in $V^+$. There exists an analytic function $F$ on the tube $T$ satisfying
\begin{enumerate}
    \item $F(z)$ is polynomially bounded.
    \item $\varphi$ is the boundary value of the holomorphic function $F$ in the sense that
    \begin{equation}
        \lim_{t\rightarrow 0}\int d^Dx\; f(x)F(x+ity)=\varphi(f).
    \end{equation}
\end{enumerate}
\end{theorem}
Repeating again in the Wightman function's case. Recalling the Fourier transform \eqref{4.17}, instead of writing the test function over the coordinate chart of a real manifold, we perform the complexification and write the test function as a complex-valued function, i.e. Replacing $\xi\rightarrow \zeta\equiv \xi+i\eta\in \mathbb{C}^D$. We have the Wightman function's counterpart of the above theorem. Mathematically speaking, the distribution
\begin{equation}
    w_n(f(\zeta))=\int \frac{d^{D(n-1)}p}{(2\pi)^{D(n-1)}}w_n(\tilde{f}(p))e^{-ip_i\zeta_i},
\end{equation}
where $\zeta\in T_{n-1}\equiv (\mathbb{R}^D\times(-V^O))^{n-1}$, defines an analytic function in $T_{n-1}$ such that
\begin{equation}
    \lim_{t\rightarrow 0}w_n(\xi+it\eta)=w_n(\xi).
\end{equation}
This means that the Wightman distributions can be analytically continued into an associated holomorphic function on its complexified tube. One down but many to go. This is not yet the complete analytic continuation since the Wightman function is constrained by the properties WD1-WD4. The next thing to do is to impose the invariance under the action of the Lorentz group into the complexified tube. Since the complexification has been already done, the Lorentz group must be complexified accordingly.
\begin{theorem}[Bargmann-Hall-Wightman Theorem]
\cite{math-ph/9811002} If a function $W_n$ is holomorphic on $T_{n-1}$ and invariant under the action of the (orthochronous proper) Lorentz group $\mathcal{L}(\mathbb{R})$, then $W_n$ can be analytically continued to a single-valued function on a so-called extended tube
\begin{equation}
    T'_{n-1}\equiv \bigcup_{\Lambda\in\mathcal{L}(\mathbb{C})}\Lambda T_{n-1}.
\end{equation}
$W_n$ on this extended tube will be invariant under the action of the complexified Lorentz group denoted by $\mathcal{L}(\mathbb{C})$.
\end{theorem}
$T'_{n-1}$ contains real points also is called a Jost point. We say that $(\xi_1,...,\xi_{n-1})\in \mathbb{R}^{D(n-1)}$ is called a Jost point if its linear combination $\lambda_1\xi_1+...+\lambda_{n-1}\xi_{n-1}$ is spacelike for $\lambda_i\geq 0$. There is a lemma (Jost's lemma \cite{math-ph/9811002}) stating that $T'_{n-1}\cap \mathbb{R}^{Dn}$ coincides with the collection of Jost points. Thus, $W_n$ can be treated as real analytic functions on the set of Jost points,
since Jost points are altogether spacelike. Due to the property WD2 constrained on the Wightman functional, the argument of the Wightman function in the set of Jost points is, therefore, expected to be permutable. Finally, the tube will be further extended into the so-called permuted extended tube
\begin{equation}
    T'^\sigma_{n-1}\equiv\{\sigma\xi|\xi\in T'_{n-1},\;\sigma\in S_{n-1}\},
\end{equation}
where $S_n$ is discrete symmetric group or permutation group. $T'^\sigma_{n-1}$, in particular, contains the so-called non-coincident Euclidean points. Namely, it contains the point $(\xi_1,...,\xi_{n-1})\in\mathbb{C}^{D(n-1)}$ that can be written as $(\hat{w}\xi'_1,...,\hat{w}\xi'_{n-1})$ where $(\xi'_1,...,\xi'_{n-1})\in \mathbb{R}^{D(n-1)}$ and $\hat{w}$ denotes the operation known as the Wick rotation which rotate the real time into the purely imaginary time. Clearly speaking, $\hat{w}$ is concisely defined as $\hat{w}x^\mu=\hat{w}(x^0,x^1,...,x^d)\equiv (ix^0,x^1,...,x^d)$. In the end, we completed the analytic continuation of the Wightman function, and we define its boundary value as follows
\begin{equation}
    S_n\equiv W_n\big|_{E^{Dn}\backslash \Delta},
\end{equation}
$E^{Dn}\equiv (i\mathbb{R}_+\times \mathbb{R}^d)^n$ denotes the collection of the Euclidean points and $\Delta\equiv\{(\xi'_1,...,\xi'_{n-1})\in E^{Dn}|\xi_j=\xi_k\}$ is the coincide point. Henceforth, we will call the analytically continued Wightman function as the Schwinger function.
\begin{theorem}
The Schwinger function $S_n$ is a holomorphic function, $S_n:E^{Dn}\backslash\Delta\rightarrow \mathbb{C}$, satisfying the so-called Osterwalder-Schrader axioms \cite{Osterwalder:1973dx} (OS axioms for short)
\begin{itemize}
\item \textbf{OS21} (\textbf{Regularity}) $S_n$ is a tempered distribution on its domain $E^{Dn}\backslash \Delta$. To require that the Schwinger function will not grow fast enough at infinity. 
    \item \textbf{OS2} (\textbf{Covariance}) $S_n$ is invariant under the action of complexified Poincar\'e group or the Euclidean group $\mathcal{L}(\mathbb{C})\rtimes T$.
    \item \textbf{OS3} (\textbf{Locality}) $S_n(x_1,...,x_n)=S_n(x_{\sigma(1),...,\sigma(n)})$. Remember that the argument of the Schwinger function is the Jost points which are altogether spacelike at the beginning. Thus, permutation invariance always holds.
    \item \textbf{OS4} (\textbf{Reflection positivity}) 
    \begin{equation}
        \sum_{m,n}S_{m+n}(\Theta f_m\otimes f_n)\geq 0.
    \end{equation}
    This is the Euclidean counterpart of the WD4. The additional condition comes from the addition operator $\Theta$ which is the action of the so-called Euclidean time reflection mapping $ix_0\rightarrow -ix_0$. Intuitively, this axiom can be viewed as follows: we know that the Schwinger function or $n$-points correlation function has a probabilistic interpretation, and the negative probability is meaningless.
\end{itemize}
\end{theorem}
Again, the OS axioms are sufficient condition for the OS reconstruction theorem stating in the same sense as the Wightman reconstruction theorem above. In other words, if the Schwinger function satisfying the OS1-OS4 is known, we can then reconstruct the Euclidean field theory. To end this section, let us note that the OS axioms can be also written in the language of the path integration since, as mentioned previously, the path integral quantization of the QFT in the manifold equipped with the Euclidean metric is well-defined measure theory (the reader is recommended to see the reference, e.g. \cite{Dybalski2018}.)

\section{Violation of Reflection Positivity}
We have seen that the most powerful tool in the Euclidean QFT is the Schwinger function. Since, in the QYM theory, it is not too easy to evaluate the Schwinger function of any value of $n$, we will, therefore, restrict ourselves only with the 2-points correlation function or propagator $G(p)$. To see the reflection positivity in the explicit circumstance, we perform the Fourier transformation of the propagator in the momentum space (as computed out in the previous chapter) back to the position space. We have
\begin{equation}
    S_2(x)=\int\frac{d^Dp}{(2\pi)^D}\;e^{ip\cdot x}G(p).
\end{equation}
However, the multi-dimensional integral is still difficult to calculate out exactly. Fortunately, following the statement of \cite{Guerra:2005an}, the Euclidean QFT relies on only the hyperplane subspace of the Hilbert space $\mathcal{H}$. One might ask immediately that can we choose the hyperplane freely? In \cite{Guerra:2005an}, Guerra showed that we can project, through the projection operator denoted by $P_U$, the whole $\mathcal{H}$ into the subspace $\mathcal{H}_U$ defined by the distribution in $\mathcal{H}$ supported in the region $U$. The operation $P_U$ is proven to be Markov or have no memory ensuring that the choice of the hyperplane can be chosen almost freely. Let us choose the hyperplane in such a way that all spatial momenta $\mathbf{p}=0$. The problem is thus reduced into the evaluation of 
\begin{equation}
\label{4.31}
    S_2(t,\mathbf{x}=0)=\int \frac{dp^0}{2\pi}\;e^{ip^0t}G(p^0)\big|_{\mathbf{p}=0}.
\end{equation}
(We will denote the Euclidean time $ix^0\equiv t$.) Substitute the propagator of the Gribov type \eqref{Gribov propagator} into the projected Schwinger function \eqref{4.31}. So we obtain
\begin{equation}
\label{4.32}
    S_2(t)=\int^\infty_{-\infty}\frac{dp^0}{2\pi} e^{ip^0t}\frac{(p^0)^2}{(p^0)^4+\gamma^4}.
\end{equation}
First of all, notice that the integrand contains four distinguish poles at $p^0=\{\pm\frac{1}{\sqrt{2}}(1+i)\gamma,\pm\frac{1}{\sqrt{2}}(1-i)\}$. Since the integrand will be convergent when $p^0\rightarrow +i\infty$ (remember that the Schwinger function is supported in the forward lightcone, then we choose the value of $t$ such that $t\geq 0$), thus the coutour must be written as illustrated in the Figure 4.1 below.
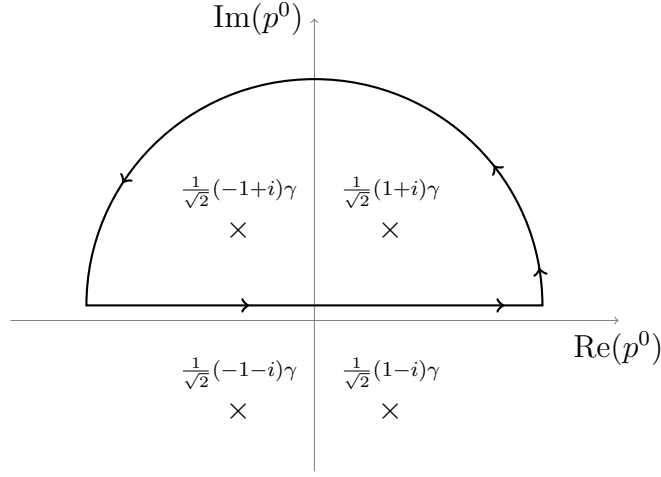
\begin{figure}[h!]
    \centering
  \begin{tikzpicture}
  [
    decoration={%
      markings,
      mark=at position 0.5cm with {\arrow[line width=1pt]{>}},
      mark=at position 2cm with {\arrow[line width=1pt]{>}},
      mark=at position 0.5 with {\arrow[line width=1pt]{>}},
      mark=at position 0.75 with {\arrow[line width=1pt]{>}},
      mark=at position -5mm with {\arrow[line width=1pt]{>}},
    }
  ]
  \draw [help lines,->] (-4,0) -- (4,0) coordinate (xaxis);
  \draw [help lines,->] (0,-2) -- (0,4) coordinate (yaxis);
  \node at (1,1.2) {$\times$};
  \node at (-1,1.2) {$\times$};
  \path [draw, line width=0.8pt, postaction=decorate] (3,0.2) arc (0:180:3)  -- cycle;
  \node at (1,-1.2) {$\times$};
  \node at (-1,-1.2) {$\times$};
  
  \node at (1,1.7) {$\scriptstyle{\frac{1}{\sqrt{2}}(1+i)\gamma}$};
  \node at (-1,1.7) {$\scriptstyle{\frac{1}{\sqrt{2}}(-1+i)\gamma}$};
  \node at (1,-0.7) {$\scriptstyle{\frac{1}{\sqrt{2}}(1-i)\gamma}$};
  \node at (-1,-0.7) {$\scriptstyle{\frac{1}{\sqrt{2}}(-1-i)\gamma}$};
  \node[below] at (xaxis) {$\mathrm{Re}(p^0)$};
 \node [left] at (yaxis) {$\mathrm{Im}(p^0)$};
\end{tikzpicture}
    \caption{The contour plot of the complex integration of \eqref{4.32}}.
\end{figure}
Thus, only two poles involve the complex integral. We can use the residue theorem to obtain the full expression of the Schwinger function
\begin{equation}
\label{4.33}
    \begin{split}
        S_2(t)&=\frac{2\pi i}{2\pi}\left(\frac{e^{ip^0t}}{4p^0}\bigg|_{\scriptstyle{p^0=\frac{1}{\sqrt{2}}(-1+i)\gamma}}+\frac{e^{ip^0t}}{4p^0}\bigg|_{\scriptstyle{p^0=\frac{1}{\sqrt{2}}(1+i)\gamma}}\right)\\
        &= \frac{ie^{-\frac{1}{\sqrt{2}}\gamma t}}{2\sqrt{2}}\left(\frac{e^{-\frac{i}{\sqrt{2}}\gamma t}}{(-1+i)\gamma}+\frac{e^{\frac{i}{\sqrt{2}}\gamma t}}{(1+i)\gamma}\right)\\
        &=\frac{ie^{-\frac{1}{\sqrt{2}}\gamma t}}{2\sqrt{2}\gamma}\left(-\frac{e^{-\frac{i}{\sqrt{2}}\gamma t}(1+i)}{2}+\frac{e^{\frac{i}{\sqrt{2}}\gamma t}(1-i)}{2}\right)\\
        &=\frac{ie^{-\frac{1}{\sqrt{2}}\gamma t}}{2\sqrt{2}\gamma}\left[i\sin\left(\frac{1}{\sqrt{2}}\gamma t\right)-i\cos\left(\frac{1}{\sqrt{2}}\gamma t\right)\right]\\
        &=\frac{e^{-\frac{1}{\sqrt{2}}\gamma t}}{2\gamma}\left[\cos\left(\frac{1}{\sqrt{2}}\gamma t\right)\cos\left(\frac{\pi}{4}\right)-\sin\left(\frac{1}{\sqrt{2}}\gamma t\right)\sin\left(\frac{\pi}{4}\right)\right]\\
        &=\frac{1}{2\gamma}e^{-\frac{1}{\sqrt{2}}\gamma t}\cos\left(\frac{\gamma t}{\sqrt{2}}+\frac{\pi}{4}\right),
    \end{split}
\end{equation}
which can be shown in the figure 4.2 below.
\begin{figure}[h!]
    \centering
    \includegraphics[width=6cm]{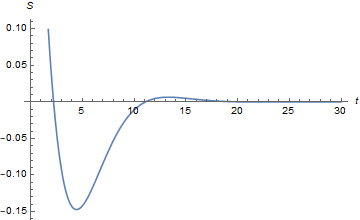}
    \caption{Plot of the Schwinger function of the Landau Gribov propagator \eqref{4.34} where the value of the Gribov mass parameter is chosen to be $\gamma=0.5$.}
    \label{fig:my_label}
\end{figure}
Note also that this plot considerably agrees with the Lattice result illustrated in the figure 4.3.
\begin{figure}[h!]
    \centering
    \includegraphics[width=6cm]{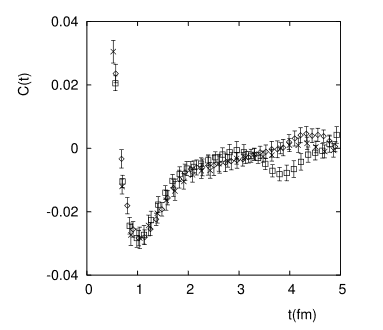}
    \caption{Shape of the Schwinger function plotted by lattice simulation obtained from \cite{cucchieri2004positivity}.}
    \label{fig:my_label}
\end{figure}\\
As shown explicit from the plot, this result shows that the Schwinger function is not positive-definite as the time parameter changes. On the other hand, this can be seen easily by recalling that the cosine function is a periodic function. In the other words, speaking strictly, there exists $t$ such that $S_2(t)<0$. In summary, the 2-points Schwinger function of the gluon propagator restricted in the Gribov region violates the reflection positivity axiom (OS3) of the Euclidean QFT. Thus, the modified gluon field cannot appear as the asymptotic state. Technically speaking, it is confined. Intuitive picture is quite precise since, in the previous chapter, we restrict the space of the gauge orbit into the bounded region. In the other words, our gluon field is generally confined inside the Gribov region at the first glance. Eventually, after a long journey, we have ended the analytical proof of the violation of the reflection positivity axiom for the restricted YM theory. This shows that by restriction into the Gribov region the gluon field behaves confined. However, this is not the unique mechanism to understand the gluon confinement. Strictly speaking, the violation of the reflection positivity is regarded as the sufficient condition for gluon confinement but not necessary condition. Note that the usage of the reflection positivity violation is studied further by Baulieu et al, they showed that such an unphysical gluon field can be used to constructed the physical bounded state of that fields (we will not go through it in details, however, we recommend the reader to read further, e.g. \cite{0912.5153}).\\
\indent We will give a final remark about this behavior. One might be curious that so what about other massive gauge fields in the presence of the spontaneous symmetry breaking and Higgs mechanism such as $W^\pm$ and $Z$ bosons in the electroweak theory? We have known evidently from the experiment at CERN that such a gauge field can appear as an asymptotic state. Let's compute the Schwinger function of the massive gauge field gained through the Higgs mechanism. We therefore have
\begin{equation}
\label{4.34}
    S_2(t)=\int \frac{dp^0}{2\pi}\;e^{ip^0t}\frac{1}{(p^0)^2+m^2},
\end{equation}
with the coutour of the shape
\begin{figure}[h!]
    \centering
    \begin{tikzpicture}
  [
    decoration={%
      markings,
      mark=at position 0.5cm with {\arrow[line width=1pt]{>}},
      mark=at position 2cm with {\arrow[line width=1pt]{>}},
      mark=at position 0.5 with {\arrow[line width=1pt]{>}},
      mark=at position 0.75 with {\arrow[line width=1pt]{>}},
      mark=at position -5mm with {\arrow[line width=1pt]{>}},
    }
  ]
  \draw [help lines,->] (-4,0) -- (4,0) coordinate (xaxis);
  \draw [help lines,->] (0,-2) -- (0,4) coordinate (yaxis);
  \node at (0,1.2) {$\times$};
  \node at (0,-1.2) {$\times$};
  \path [draw, line width=0.8pt, postaction=decorate] (3,0.2) arc (0:180:3)  -- cycle;

  \node at (0.5,1.2) {$\scriptstyle{im}$};
  \node at (0.5,-1.2) {$\scriptstyle{-im}$};
  \node[below] at (xaxis) {$\mathrm{Re}(p^0)$};
 \node [left] at (yaxis) {$\mathrm{Im}(p^0)$};
\end{tikzpicture}
    \caption{The coutour plot of the complex integration of \eqref{4.34}.}
\end{figure}\\
We end up with the 2-points Schwinger function of the form
\begin{equation}
\label{4.35}
    \begin{split}
      S_2(t)&=\frac{2\pi i}{2\pi}e^{ip^0t}\frac{1}{2p^0}\bigg|_{p^0=im}\\
      &=\frac{1}{2m}e^{-mt}.
    \end{split}
\end{equation}
Here, the Schwinger function of the conventional massive gauge propagator is positive definite. Therefore, it is allowed to appear in our experimental apparatus. This Schwinger function gives us more physical interpretation. By comparing the Schwinger function \eqref{4.35} with \eqref{4.33}. We see the Schwinger function for the Gribov propagator can be realized as the Schwinger function of the massive gauge field with the complex mass (in some literature, we thus call this kind of particle as an $i$-particle). The complex mass yields the tachyonic state that makes the vacuum of the state unstable. Thus, the Schwinger function of the Gribov gluon is consequently unphysical in this sense.
\newpage

\begin{appendix}
     \addcontentsline{toc}{chapter}{Appendices}
     \chapter{Mathematical Identities}
\section{Path Integration Identities}
When we are dealing with the path integration, we will focus on the integration over the weight factor of form $e^{-S}$ where $S$ is an action functional. Generally, for free theory, lack of interaction term, an action functional will take only the quadratic order in fields. Thus, the path integration of the kinetic part is normally the Gaussian integration as we have seen already in, for example, \eqref{2.5}. To evaluate it, let us recall the normal form of the Gaussian integration, we have
\begin{equation}
\label{A1}
    I=\int^\infty_{-\infty} dx\;e^{-\frac{1}{2}ax^2+px}=\sqrt{\frac{2\pi}{a}}\;e^{\frac{p^2}{2a}}.
\end{equation}
Now if we extend our consideration from the integration over ordinary function into the path integration over operator-valued case. Explicitly, we will calculate the integration of the kind
\begin{equation}
    I=\int \mathcal{D}\varphi \; e^{-\frac{1}{2}\varphi^TA\varphi+J\varphi}.
\end{equation}
Observe that we can choose the basis of $\varphi$ such that the matrix $A$ can be expressed as the diagonal matrix. That means we can actually diagonalize $A$ by performing similarity transformation by some orthogonal matrix $O^T=O^{-1}$ such that $A=OA^{diag}O^{-1}$ and then changing the basis of $\varphi\rightarrow O\varphi$. Since the path integration is nothing but an infinite-dimensional generalization of the ordinary integration, the particular factor $a$ in the expression \eqref{A1} will be replaced by-product of the eigenvalue (remember that we have already diagonalize the matrix $A$) of $A$ which is actually a determinant of that matrix. Moreover, $\frac{1}{a}$ in the exponent of \eqref{A1} will be represented by the inverse matrix (assume that it really exists) $A^{-1}$. We, therefore, obtain the result of the path integration counterpart of the Gaussian integration as (up to some constant which will be of course dropped out from the explicit calculation)
\begin{equation}
    \label{A.1}
    \int \mathcal{D}\varphi \; e^{-\frac{1}{2}\varphi^TA\varphi+J\varphi}=\frac{1}{\det A}\exp{\left[\frac{1}{2}J^TA^{-1}J\right]}.
\end{equation}
So far, we have considered the bosonic path integration, let's move to the fermionic one which is useful while dealing with ghost in the section 2.2. Fermions in the path integral can be described by using a so-called Grassmann variable which is the $\mathbb{Z}_2$-graded extension of the ordinary complex number. Suppose that the set of Grassmann variable spanned by the basis $\{\theta_i\}$ satisfying the anti-commutation relation
\begin{equation}
    \{\theta_i,\theta_j\}=0\;\;\mathrm{or}\;\;(\theta_i)^2=0.
\end{equation}
The integration over such a variable is called the Berezin integral, which is defined such that
\begin{equation}
\label{A5}
    \int d\theta =0,\;\;\;\;\;\int d\theta\; \theta=1.
\end{equation}
Another way to look at this integral, which will be helpful in the explicit calculation, is it is the integral that satisfies \say{integration=differentiation}, i.e.
\begin{equation}
    \int d\theta\;K(\theta)=\frac{\partial}{\partial\theta}\;K(\theta).
\end{equation}
Consider a Gaussian integral
\begin{equation}
    I=\int \mathcal{D}\bar\theta\int \mathcal{D}\theta\;e^{-\bar{\theta}A\theta}=\prod_i^n\prod_j^n\int d\bar\theta_i\int d\theta_j\; e^{-\frac{1}{2}\bar{\theta}A\theta},
\end{equation}
while we have supposed that there are only $n$ numbers of $\theta$ configuration and $n$ or $\bar\theta$. To evaluate this integral, we recall that such an exponential can be defined through the Taylor series as shown below
\begin{equation}
    I=\int d\bar{\theta}_1... d\bar\theta_n\int d\theta_1...d\theta_n\left(1-\bar\theta_i A_{ij}\theta_j+\frac{1}{2}\bar\theta_i A_{ij}\theta_j\bar{\theta}_kA_{kl}\theta_l+...\right).
\end{equation}
Due to the attractive property of the Berezin integration \eqref{A5}, there will be only one contribution left non-vanishing which is the contribution containing $n$ of $\theta$ and $n$ of $\bar\theta$. Thus,
\begin{equation}
    I=\frac{1}{n!}\sum_{\sigma}\;\pm A_{\sigma(1)\sigma(2)}A_{\sigma(3)\sigma(4)}...,
\end{equation}
where $\sigma$ denotes the permutation operator. Finally, we can deduce that the Gaussian path integration over two Grassmann variables yields
\begin{equation}
\label{A.2}
    \int \mathcal{D}\bar\theta\int\mathcal{D}\theta\;e^{-\bar{\theta}A\theta}=\det A.
\end{equation}
\section{BRST Transformation and Identities Among Super Lie brackets}
As being used in \eqref{2.69} and \eqref{2.70}. We will verify the following BRST transformation rules
\begin{equation}
\label{A.3}
    \begin{split}
        \delta(D_\mu c)^a&=\delta(\partial_\mu c^a-gf^{abc}A^b_\mu c^c)\\
        &=\partial_\mu\left(\frac{g}{2}f^{abc}c^bc^c\right)-gf^{abc}(D_\mu c)^bc^c-gf^{abc}A^b_\mu \left(\frac{g}{2}f^{cde}c^dc^e\right)\\
        &=D_\mu\left(\frac{g}{2}f^{abc}c^bc^c\right)-gf^{abc}(D_\mu c)^bc^c\\
        &=\frac{g}{2}f^{abc}(D_\mu c)^bc^c+\frac{g}{2}f^{abc}c^b(D_\mu c)^c-gf^{abc}(D_\mu c)^b c^c\\
        &=\frac{g}{2}f^{abc}(D_\mu c)^bc^c+\frac{g}{2}f^{abc}(D_\mu c)^bc^c-gf^{abc}(D_\mu c)^b c^c\\
        &=0,
    \end{split}
\end{equation}
where the Leibniz's rule of the covariant derivative has been carried out in the third step and in the fourth step, we have used the totally anti-symmetric property of $f^{abc}=-f^{acb}$ and anti-commutation relation between two ghost fields.
\\
\indent Consider the anti-commutator between $A,\;B$ and $C$, as appeared in \eqref{2.69}, which are fermionic, bosonic and fermionic, respectively. We can verify that
\begin{equation}
    \label{A.4}
    \begin{split}
        \{A,BC\}&=ABC+BCA\\
        &=[A,B]C+BAC+BCA\\
        &=[A,B]C+B\{A,C\}.
    \end{split}
\end{equation}
Additionally, we will also verify the relation that is used in \eqref{2.70}. Suppose $A,\; B$ and $C$ are all fermionic
\begin{equation}
    \label{A.5}
    \begin{split}
        [A,BC]&=ABC-BCA\\
        &=\{A,B\}C-BAC-BCA\\
        &=\{A,B\}-B\{A,C\}.
    \end{split}
\end{equation}
Sometime identities of the commutator between fermionic operator $Q$ and three operators $A,\;B$ and $C$ are useful. For (B and F stands for boson and fermion, respectively) $(A,B,C)$ be (B,B,B), we have
\begin{equation}
    \label{A.14}
    \begin{split}
        [Q,ABC]&=QABC-ABCQ\\
        &=[Q,A]BC+AQBC-AB[C,Q]-ABQC\\
        &=[Q,A]BC+A[Q,B]C+AB[Q,C].
    \end{split}
\end{equation}
Whereas, for other case, if $(A,B,C)$ be (F,B,F), we end up with the identity
\begin{equation}
    \label{A.15}
    \begin{split}
        [Q,ABC]&=QABC-ABCQ\\
        &=\{Q,A\}BC-AQBC-AB\{C,Q\}+ABQC\\
        &=\{Q,A\}BC-A[Q,B]C-AB\{Q,C\}.
    \end{split}
\end{equation}
Moreover, anti-commutation relations of (B,B,F) and (F,B,B) are also useful. They are expressed as
\begin{equation}
    \label{A.16}
        \{Q,ABC\}=[Q,A]BC+A[Q,B]C+AB\{Q,C\},
\end{equation}
and
\begin{equation}
    \label{A.17}
     \{Q,ABC\}=\{[Q,A\}BC-A[Q,B]C-AB[Q,C],
\end{equation}
respectively.
\section{Fourier Transformation}
Our world is a very interesting world in the sense that it seems like it is constructed to have non-trivial duality between positive-space and momentum-space. That is, it is mysteriously suitable with the Fourier transformation. In particular, the greatest advantage at all of the Fourier transformation is that, since the QFT is the theory of the operator-valued distribution (as discussed in chapter 4), in many situations, we will deal with the distribution function which is difficult to evaluate since it is the operator-valued quantity. The Fourier transformation solves it by just performing the transformation from the space of the operator-valued distribution into its Fourier space that turns out to the space of the $c$-number-valued distribution. Thus, in the Fourier space, the algebraic manipulation is allowed to do more conveniently. Moreover, by using the help of the Fourier transformation, several functions can be written in integral representation which is, once again, more convenient to deal with. Let's see the relevant examples. Firstly, the most used distribution is the Dirac delta function $\delta(x)$ which is defined to be $\displaystyle{\int dy\; f(x)\delta(x-y)=f(y)}$ where $f$ is any function and $x$ belongs to the domain of the integration. For simplicity, we will start with the 1-dimensional problem and then generalize to the general dimensional case. We can perform the Fourier transformation of the Dirac delta function as follows
\begin{equation}
    \delta(p)=\int^\infty_{-\infty} dx\;\delta(x)e^{ipx}=e^{0}=1.
\end{equation}
After that, we perform the Fourier transformation back to the position space. We thus have
\begin{equation}
    \delta(x)=\int^\infty_{-\infty} \frac{dp}{2\pi}\;\delta(p)e^{-ipx}=\int^\infty_{-\infty} \frac{dp}{2\pi}\;e^{-ipx}.
\end{equation}
By writing the delta function in the so-called integral representation. We can easily check many properties of the delta function, for example, $\delta(x)=\delta(-x)$ due to the fact that the measure $\displaystyle{\int}dp$ is integrated over the entire momentum-space, hence, the re-definition from $p\rightarrow -p$ is acceptable. In general dimensional case, $d$ said, we can write
\begin{equation}
\label{A.20}
    \delta^d(x-y)=\int\frac{d^dp}{(2\pi)^d}\;e^{-ip(x-y)}.
\end{equation}
Another useful function that can be written in the integral form is the Heaviside step function $\theta(x)$ defined as in \eqref{2.93}. We therefore obtain
\begin{equation}
    \theta(p)=\int^\infty_{-\infty} dx\;\theta(x)e^{ipx}=\int^\infty_0 dx\; e^{ipx}=\frac{i}{p}
\end{equation}
and then
\begin{equation}
    \theta(x)=\int^\infty_{-\infty}\frac{dp}{2\pi}\;\theta(p)e^{-ipx}=\int^\infty_{-\infty}\frac{dp}{2\pi}\;\frac{ie^{-ipx}}{p}.
\end{equation}
Let's redefine further that $\beta\equiv -ip$. We have
\begin{equation}
\label{A.23}
    \theta(x)=\int^{\infty}_{-\infty}\frac{d\beta}{2\pi i}\frac{e^{\beta x}}{\beta}.
\end{equation}
Last but not least, we will work mainly in the momentum space when we are trying to compute the propagator or derive the Feynman rule, thus we will show how to change the action from the position-space to the momentum-space. To be specific, let's Fourier transform the source's action
\begin{equation}
\begin{split}
    S_{source}&=\int d^Dx\; A^a_\mu(x)J^a_\mu(x)\\
   &=\int d^Dx\; \int \frac{d^Dp}{(2\pi)^D}\int \frac{d^Dq}{(2\pi)^D}\;A^a_\mu(p)J^a_\mu(q)e^{-i(p+q)\cdot x}\\
   &=\int\frac{d^Dp}{(2\pi)^D}\int \frac{d^Dq}{(2\pi)^D}\;A^a_\mu(p)J^a_\mu(q)(2\pi)^D\delta^D(p+q)\\
   &=\int\frac{d^Dp}{(2\pi)^D}\;A^a_\mu(p)J^a_\mu(-p)
\end{split}
\end{equation}
as appeared in \eqref{3.78}. This particular example might not that good to demonstrate how Fourier transformation can change the space of the operator-valued distribution to the c-number-valued one. Let's consider another example
\begin{equation}
    \begin{split}
        S_{YM}&=\int d^Dx\;\left(\frac{1}{2}A^a_\nu(x)(\partial_\mu \partial_\nu -g_{\mu\nu}\partial_\rho \partial_\rho )A^a_\mu(x)\right)\\
        &=\frac{1}{2}\int d^Dx\int \frac{d^Dp}{(2\pi)^D}A^a_\nu(p)e^{-ip\cdot x}(\partial_\mu \partial_\nu -g_{\mu\nu}\partial_\rho \partial_\rho )\int \frac{d^Dq}{(2\pi)^D}A^a_\mu(q)e^{-iq\cdot x}\\
        &=\frac{1}{2}\int d^Dx\int \frac{d^Dp}{(2\pi)^D}A^a_\nu(p)\int \frac{d^Dq}{(2\pi)^D}\;A^a_\mu(q)(-q_\mu q_\nu -g_{\mu\nu}(-q^2) )e^{-i(p+q)\cdot x}\\
        &=\frac{1}{2}\int \frac{d^Dp}{(2\pi)^D}A^a_\nu(p)\int \frac{d^Dq}{(2\pi)^D}\;A^a_\mu(q)(q^2g_{\mu\nu} -q_\mu q_\nu)(2\pi)^D\delta^D(p+q)\\
        &=\frac{1}{2}\int \frac{d^Dp}{(2\pi)^D}\;A^a_\nu(p)(p^2g_{\mu\nu} -p_\mu p_\nu)A^a_\nu(-p).
    \end{split}
\end{equation}
Here, in addition to change from $x$ to $p$, we have effectively changed from the operator $\partial_\mu=\frac{\partial}{\partial x^\mu}$ into just some complex number $-ip_\mu$ as have seen frequently in the QM101 and QFT101.Here, in addition to change from $x$ to $p$, we have effectively changed from the operator $\partial_\mu=\frac{\partial}{\partial x^\mu}$ into just some complex number $-ip_\mu$ as have seen frequently in the QM101 and QFT101.

\chapter{Elements of Differential Topology}
To be honest, the main reference used in this topic is actually the very handy book by Nakahara \cite{Nakahara:2003nw}. Caution! In this chapter, we will deal with the general case of spacetime, so the position of the subscripts and superscripts will be carefully studying.
\section{Differential Form and Cohomology}
The mathematical main tool mostly used in the language of differential topology for describing the gauge theory is a differential form. Loosely speaking, a differential form of degree $p$, alternatively called a $p$-form, is a tensor of type $(0,p)$ where all indices are totally antisymmetric under odd permutation. Let's denote the set of all differential $p$-forms defined at the specific point of the spacetime $\mathcal{M}$ as $\Omega^p(\mathcal{M})$ forming a vector space whose basis can be expressed by the help of the wedge product defined as following
\begin{equation}
\label{B.1}
    dx^\mu\wedge dx^\nu\equiv -dx^\nu\wedge dx^\mu.
\end{equation}
For any element $\omega\in\Omega^p(\mathcal{M})$ can be written explicitly as
\begin{equation}
\label{b.2}
    \omega=\frac{1}{p!}\omega_{\mu_1\mu_2...\mu_p}dx^{\mu_1}\wedge dx^{\mu_2}\wedge...\wedge dx^{\mu_p}.
\end{equation}
The diffential $(p+q)$-form can be constructed by using the differential $p$-form and $q$-form through the operation called exterior product $\wedge:\Omega^q(\mathcal{M})\times \Omega^p(\mathcal{M})\rightarrow \Omega^{p+q}(\mathcal{M})$.\\
\indent It is quite useful to define a so-called exterior derivative $d_p$ mapping from $\Omega^p(\mathcal{M})\rightarrow \Omega^{p+1}(\mathcal{M})$. To be specific, we define such an operation such that
\begin{equation}
    d_p\omega=\frac{1}{p!}\left(\frac{\partial}{\partial x^\nu}\omega_{\mu_1...\mu_p}\right)dx^\nu\wedge dx^{\mu_1}\wedge...\wedge dx^{\mu_p}
\end{equation}
By keeping track which degree of the differential form we are dealing with, we will usually drop the subscript on $d_p$ and writing only as $d$ for convenience. The attractive properties of the exterior derivative is it is nilpotent by its construction, i.e. $d^2=0$. We can check it directly by observing from 
\begin{equation}
    d^2\omega=\frac{1}{p!(p+1)!}\frac{\partial^2\omega_{\mu_1...\mu_p}}{\partial x^\nu\partial x^\rho}dx^\nu\wedge dx^\rho\wedge...\wedge dx^{\mu_p}.
\end{equation}
Since the Hessian matrix $\frac{\partial^2\omega_{\mu_1...\mu_p}}{\partial x^\nu\partial x^\rho}$ is symmetric under the change between $\nu$ and $\rho$ while the basis of the differential form is constructed in such a way that the indices $\nu$ and $\rho$ is anti-symmetric. Hence, the only way it can satisfy both conditions is to be zero at the beginning. As we have discussed in the main topic, the nilpotency of the exterior derivative can deduce the wonderful consequence that there exists the differential form that can be constructed by the action of the exterior derivative on the one lower differential form called an exact form. However, the set of all exact forms is just a subset of the whole set whose element, $\omega$ said, satisfies $d^2\omega=0$ known as a closed-form. The non-trivial class of differential $p$-forms is thus the form which is closed but not exact. Such a type of class is known as the $p$th de Rham cohomology $H^p(\mathcal{M})$. There are two remarkable structures of the de Rham cohomology group summarized below
\begin{enumerate}
    \item de Rham cohomologies satisfies the so-called Poincar\'e duality
    \begin{equation}
        \label{Poincare duality}
        H^p(\mathcal{M})\cong H^{D-p}(\mathcal{M}).
    \end{equation}
    \item Moreover, they satisfy the K\"unneth formula
    \begin{equation}
        \label{Kunneth formula}
        H^p(\mathcal{M}_1\times\mathcal{M}_2)=\bigoplus_{q+r=p}(H^q(\mathcal{M}_1)\otimes H^r(\mathcal{M}_2)).
    \end{equation}
\end{enumerate}
 One more attractive fact about the closed form lies on the very powerful theorem which people in the past called it as a lemma! The \say{Poincar\'e's lemma} stating that 
 
\begin{theorem}[Poincar\'e's lemma]
If a coordinate neighbourhood $U$ of a manifold $\mathcal{M}$ is contractible to a point $p_0\in \mathcal{M}$, any closed p-form on U will be exact.
\end{theorem}

(quoted from \cite{Nakahara:2003nw}). For the people who are in love with mathematics, we are so sorry about that we are not planning to prove anything inside this review article. However, we still can give a remark about the considering theorem. What Poincar\'e's lemma really tells us is any closed form defined on a manifold is \say{locally} exact since any local coordinate neighbourhood on any kind of manifold is typically isomorphic to a flat space $\mathbb{R}^{d+1}$ which is generally contractible to a point. We will use this fact later in the final discussion of the appendix B.3. \\
\indent It is very important to remark that since the differential form is a totally antisymmetric tensor, thus, the differential form of degree greater than the dimensions of the space where it defined on cannot be constructed. The highest degree which any differential form can reach is, of course, a degree $D$. Suppose that the manifold is orientable, i.e. the Jacobian $J\equiv \det(\partial x^\mu/\partial y^\alpha)>0$ where $x^\mu$ and $y^\alpha$ are local coordinates of the open regions $U_i$ and $U_j$ (under the assumption that $U_i\cup U_j\not=\O$), respectively. There will always exist a $D$-form which can be non-vanishing everywhere in the spacetime. Such a differential form is called a volume form playing a significant role to define a making sense integration of differential forms.\\
\indent If our manifold admits a metric structure where we will denote the metric with the notation $g\equiv g_{\mu\nu}dx^\mu\wedge dx^\nu$, the invariant volume form can be expressed as
\begin{equation}
    \Omega_\mathcal{M}\equiv \sqrt{|g|}\;dx^1\wedge...\wedge dx^{D}.
\end{equation}
In our review, we will focus on the flat spacetime, thus, we will usually neglect $|g|$ which is nothing more and nothing less but one. At most, we will perform 1-point compactification to effectively transform the flat $\mathbb{R}^{D}$ into the compact version $S^{D}$. The value of $|g|$, in this case, is $r^2\sin\theta$ which, in a practical situation, it will be cancelled out together with the invariant factor of the Levi-Civita tensor. In the light of such a form, we can accordingly define the integration of the function $f$ over the manifold by
\begin{equation}
    \int_\mathcal{M}f\Omega_\mathcal{M}\equiv \int_\mathcal{M} dx^1... dx^{D} \sqrt{|g|} f\equiv \int d^{D}x\;\sqrt{|g|}f.
\end{equation}
Last but not least, we will define another linear map called the Hodge star operator or sometimes known as the duality transformation denoted by $*$ which do a map from $\Omega^p(\mathcal{M})$ to $\Omega^{D-p}(\mathcal{M})$. Explicitly, on the basis of the differential p-form, we have
\begin{equation}
    *(dx^{\mu_1}\wedge...\wedge dx^{\mu_p})\equiv \frac{\sqrt{|g|}}{(D-p)!}{\epsilon^{\mu_1...\mu_p}}_{\nu_{p+1}...\nu_{d+1}}dx^{\nu_{p+1}}\wedge...\wedge dx^{\nu_{D}}.
\end{equation}
Consequently, we can perform the duality transformation of p-form $\omega$ \eqref{b.2} as follows
\begin{equation}
    *\omega=\frac{\sqrt{|g|}}{p!(D-p)!}\omega_{\mu_1...\mu_p}{\epsilon^{\mu_1...\mu_p}}_{\nu_{p+1}...\nu_{D}}dx^{\nu_{p+1}}\wedge...\wedge dx^{\nu_{D}}.
\end{equation}
One can show that the exterior product between two $p$-forms $\omega$ and $\eta$ is an $D$-form, expressed as
\begin{equation}
\label{b.11}
    \omega\wedge*\eta=\frac{1}{p!}\omega_{\mu_1...\mu_p}\eta^{\mu_1...\mu_p}\Omega_\mathcal{M}.
\end{equation}
The example of the expression \eqref{b.11} is what we are meant to always meet in this field. Let's consider the differential 2-form $F\equiv F_{\mu\nu}\;dx^\mu\wedge dx^\nu$ defined on the 4-dimensional Euclidean spacetime. We can accordingly define its dual 2-form $\Tilde{F}\equiv \Tilde{F}_{\mu\nu}dx^\mu\wedge dx^\nu$ as
\begin{equation}
\label{B.12}
    \Tilde{F}\equiv *F=\frac{1}{2}F_{\mu\nu}\epsilon_{\mu\nu\alpha\beta}\;dx^{\alpha}\wedge dx^\beta.
\end{equation}
Yes! The tensor we are referring to is the field strength tensor as mentioned at the beginning of chapter 2 and will be discussed again in section B.3. Therefore, we can write the differential form version of the YM action \eqref{2.3} as
\begin{equation}
    S_{YM}=\int_\mathcal{M}\mathrm{Tr}(F\wedge \Tilde{F})=\frac{1}{2}\int d^{d+1}x\;\mathrm{Tr} (F_{\mu\nu}F_{\mu\nu}),
\end{equation}
where Tr is a trace over the gauge group indices which will be discussed briefly in the appendix B.3.
\section{Homology, Euler Characteristic and Homotopy}
Sometimes we do not deal with the integration over a whole manifold $\mathcal{M}$ or topological space $X$ which are both non-visualizable in many sense, instead, we might integrate over more general shape, the polyhedron. To study the polyhedrons more systematically, we introduce the notion of the so-called simplex. A $p$-simplex is defined to be $p$-dimensional object demanded to be geometrical independent (of course, since topology does not care much about geometrical structures). In fact, a set of all oriented $p$-simplex generates an Abelian group called a $p$-chain group $C_p$. Parallelly to the case of differential $p$-forms, we can introduce the operation which, in this case, is regarded to a map from a $p$-chain group to one lower chain, the boundary operator $\partial_r$ said. Accordingly, we can define the set containing the elements of the $p$-chain group that satisfies $\partial_r c=0$, where $c\in C_p$ as a subset known as a $p$-cycle group $Z_p$. Once again, due to the nilpotency property, which can be shown, of the boundary operation, we can define the smaller subset called $p$-boundary group $B_p$. Finally, we can define the class of the $p$-simplexes which is a $p$-cycle (closed) but not $p$-boundary (exact) to be an equivalence class called the $p$th degree homology group $H_p$.\\
\indent Amazingly, the homology group plays a very powerful role as a tool to calculate the topological invariance, specifically, the Euler characteristic $\chi(\mathcal{M})$. By defining the Betti number $b_p$ as a dimension of the homology group, i.e. $b_p\equiv \dim H_p(\mathcal{M})$, we can compute the Euler characteristic through the light of the Euler-Poincar\'e theorem
\begin{theorem}[Euler-Poincar\'e Theorem]
\begin{equation}
   \chi(\mathcal{M})=\sum_p(-1)^pb_p.
\end{equation}
\end{theorem}
Note that the name of the homology group is very suggestive by leading us to think that the homology group of simplexes is a dual vector space of the cohomology group of differential forms. To see it relation more explicitly, we might define the inner product between an element of the homology group $c\in H_p$ and cohomology $\omega\in H^p$ group as
\begin{equation}
    <c,\omega>\equiv \int_c\;\omega.
\end{equation}
Strictly speaking, the inner product between two vector spaces defined to be the integration of the differential form over the simplex. One can check that such a form of the inner product is well-defined in the sense that the definition is independent on the representation of both equivalence classes $H_p$ and $H^p$. However, to verify this proposition, we need to know one more information that there is a traditional theorem by Stoke in the language of the differential forms. The generalized Stoke's theorem can be expressed as follows
\begin{equation}
\label{B.16}
    \int_c d\omega=\int_{\partial c}\omega.
\end{equation}
In the inner product language, we have
\begin{equation}
\label{B.17}
    <c,d\omega>=<\partial c,\omega>.
\end{equation}
Due to this suggestive form, we can, therefore, observe that the exterior derivative $d$ can be realized to be the adjoint operator of boundary operator $\partial$. Now we can check the independence of the representative by, for example, picking another representative of the cohomology such as $\omega+d\omega'$. We have
\begin{equation}
    <c,\omega+d\omega'>=<c,\omega>+<c,d\omega'>=<c,\omega>+<\partial c,\omega'>=<c,\omega>,
\end{equation}
where the generalized Stoke's theorem \eqref{B.17} and the cycle property $\partial c=0$ have been used. All we have done so far also suggests that the homology and cohomology are dual to each other. Honestly, there is the theorem that supports this idea
\begin{theorem}[de Rham's Theorem]
\label{de Rham's theorem}
If $\mathcal{M}$ is a compact manifold, $H_p(\mathcal{M})$ and $H^p(\mathcal{M})$ are finite-dimensional and the inner product between them is bilinear and non-degenerate.
\end{theorem}
In English, this implies really that $H^p(\mathcal{M})$ is the dual vector space of $H_p(\mathcal{M})$, thus $\dim H^p=\dim H_p$ implying that the Betti number can be also evaluated through the cohomology group instead of using the homology group. Moreover, by recalling the K\"unneth formula \eqref{Kunneth formula}, we can eventually derive the very powerful formula to compute the Euler characteristic of the Cartesian product space as
\begin{equation}
\label{b.19}
    \chi(\mathcal{M}_1\times \mathcal{M}_2)=\chi(\mathcal{M}_1)\chi(\mathcal{M}_2),
\end{equation}
following from $b^p(\mathcal{M}_1\times \mathcal{M}_2)=\sum_{q+r=p}b^q(\mathcal{M}_1)b^r(\mathcal{M}_2)$ which is, in fact, a direct consequence of the K\"unneth formula.\\
\indent Let's consider the example of using the homology group. Consider the homology group of $n$-sphere. Since physicist is more familiar with the differential form, we therefore rather evaluate the cohomology of the sphere. Firstly, consider the zeroth-order cohomology group $H^0(S^n)$, let us remind the reader that the zeroth-order cohomology group is just an equivalence class of zeroth degree differential form which is nothing but an ordinary function. The $(-1)$-form is not defined in principle, thus the set of corresponding exact form is an empty set consequently. The closed condition $d\omega=0$ implies that the considering 0-form or function must be a constant function over space if the topological space where such a function defined on is connected (remember that a sphere is a connected space). A set of constant real-valued function is isomorphic to a real line $\mathbb{R}$. Finally, we end up with $H^0(S^n)\cong \mathbb{R}$. According to the Poincar\'e duality \eqref{Poincare duality}, we get the result
\begin{equation}
    H^0(S^{n})\cong H^{n}(S^{n})\cong \mathbb{R}.
\end{equation}
One can also show that otherwise, all cohomology group are trivial. Namely,
\begin{equation}
    H^p(S^{n})\cong\{e\},
\end{equation}
for $p\in[1,d]$. As a result, the Euler characteristic of the sphere of $n$ dimensions can be expressed as
\begin{equation}
    \label{B.22}
    \chi(S^{n})=(-1)^0b_0+(-1)^nb_n=1+(-1)^n,
\end{equation}
where the fact that real line is 1-dimensional space implying $b_0=b_n=\dim\mathbb{R}=1$ has been used.\\
\indent To end this section of appendix, we will discuss one more important group used mostly in differential topology. Since differential topology does study the structure of topological space without concerning the geometrical shape of such a space which means that any topological space must be \say{continuously} deformable into other topological space of the same topological structure. The deformation property encodes inside the language of the homotopy group.\\
\indent Let's begin with defining the unit k-cube $I^k$ as the product of $I\equiv[0,1]$ k times. After that, we then define the k-loop $\alpha$ of the topological space $X$ as a map $\alpha:I^k\rightarrow X$. The loop $\alpha$ is said to be homotopic to another loop $\beta$, $\alpha\sim\beta$, if there exists a continuous map $F:I^k\times I\rightarrow X$ such that 
\begin{equation}
    \begin{split}
        F(s_1,...,s_k;0)&=\alpha(s_1,...,s_k),\\
        F(s_1,...,s_k;1)&=\beta(s_1,...,s_k).
    \end{split}
\end{equation}
Along with the condition that points of the boundary of the $n$-cube hold fixed for any parameter $t\in I$. Intuitively, this map can be realized as the map that shows the deformation from the loops $\alpha$ into $\beta$ parametrized by the real parameter $t\in I$. One can think $t$ as a time variable, initially, the loop begins with of the same form as the loop $\alpha$ when $t=0$ and then it starts to deform. Eventually, the time stopped at $t=1$, the loop becomes $\beta$ at the end of the time. One can show that the relation defines the associated equivalence class. Additionally, the set of all equivalence class on the topological space $X$ also forms a group called the kth order homotopy group, denoted by $\pi_k(X)$.\\
\indent Since, the $n$-cube is topologically equivalent to an $n$-sphere, thus, the homotopy group can be thought as a group of equivalence classes of the map from $n$-sphere to the considering topological space. The general properties of the homotopy group are listed down below
\begin{enumerate}
    \item \begin{theorem}[Bott Periodicity Theorem]
    \label{Bott theorem}
     For the case that $N\geq (k+1)/2$, we have
    \begin{equation}
    \label{Bott unitary}
        \pi_k(U(N))\cong\pi_k(SU(N))\cong\begin{cases}
        \{e\}&k\;\mathrm{is}\;\mathrm{even}\\
        \mathbb{Z}&k\;\mathrm{is}\;\mathrm{odd}
        \end{cases}
    \end{equation}
    and for $N\geq k+2$,
    \begin{equation}
    \label{Bott orthogonal}
        \begin{split}
            \pi_k(O(N))&\cong \pi_k(SO(N))\\
            &\cong \begin{cases}
          \{e\}&k=2,4,5,6 \mod8\\
          \mathbb{Z}_2&k=0,1\mod8\\
          \mathbb{Z}&k=3,7\mod8
            \end{cases}
        \end{split}
    \end{equation}
    \end{theorem}
    (obtained from, of course, \cite{Nakahara:2003nw}).
    \item There exists a so-called J-homomorphism that maps from $\pi_k(SO(N))$ to $\pi_{k+N}(S^N)$. However, in the case that $k=1$, such a homomorphism is instead an isomorphism meaning that
    \begin{equation}
        \label{B.26}
        \pi_1(SO(N))\cong \pi_{N+1}(S^N).
    \end{equation}
\end{enumerate}
\section{Fiber Bundle and Chern Character}
In differential topologist's perspective, what we refer to \say{gauge theory} is a theory described by a fibre bundle. What is it then? Let's begin with the terminology, suppose (and we always suppose) that we live in a so-called base space $\mathcal{M}$ which is nothing but our spacetime background. Moreover, we suppose further that there exists some set that defined on each point on the base space, we will henceforth call it a fibre denoted by $F$. Finally, the fibre bundle is a disjoint union of all fibres $F$ over the whole region (or sometimes over subregion) of the base space $\mathcal{M}$. We will denote the fibre bundle by $E$. To visualize this structure, let's think of a male hair comb. The base of the comb is regarded as a base space, each tooth of such comb is referred to the fibre that defined on each point on base space (base of comb). Thus, all teeth comb are meant to be a fibre bundle in this perspective. To have well-defined fibre bundle, there need to have a surjection (onto map) from the total space $E$ onto $\mathcal{M}$, known as a projection map $\pi$ said, whose pre-image of this projection will be fibre at that point. The true inverse operation of the projection map is a section $s$ defined conversely as a map from $\mathcal{M}$ to $E$ such that $\pi s=\mathbb{1}_{\mathcal{M}}$. Note that the section map can be realized as picking a point on the fibre bundle associated to each point on the base point. The realization has been used once in section 3.1 when discussing Singer's proof of the existence of the Gribov's ambiguity.\\
\indent There are two more important maps used when working with the fibre bundle. The first one is a diffeomorphism (differentiable invertible map) $\phi_i:U_i\times F\rightarrow \pi^{-1}(U_i)$ where $\{U_i\}$ is an open covering of $\mathcal{M}$. Such a map is sometimes called a local trivialization (we will understand why soon). The local trivialization behaves almost the same way as a chart map on the manifold. On the intersected area on the base space, we can define the transition function $t_{ij}(p)\equiv \phi^{-1}_i(p)\phi_j(p):F\rightarrow F$ where $p$ is a point on the intersected region. We demand that the transition function need to form a group since the sequential action of the transition function must be essentially well-behaved in the sense that the transition must be closed. That group will be from now on called a structure group $G$. If the transition function is an identity globally, that is the transition function is typically trivial, the fibre bundle is said to be a trivial bundle also. A trivial bundle can be generally written in the form of the direct product $\mathcal{M}\times F$. This fact is the particular reason why the local trivialization serves that name since the inverse of the local trivialization do map from the fibre $\pi^{-1}(U_i)$ on the local region $U_i$ to its trivial version $F\times U_i$. Note that there is, in particular, important theorem (in fact corollary of another theorem) about the fibre bundle stating that
\begin{theorem}
A fibre bundle $E$ will be trivial if the base space $\mathcal{M}$ is contractible to a point.
\end{theorem}
This corollary is useful to explore the triviality of any bundle so it is important to be remarked.\\
\indent Let's move to the discussion on the structure group $G$. In a very general situation, $G$ will be very general matrix group $GL(N,\mathbb{R})$ (for the real vector space). At most, we can introduce the local coordinate with the metric structure into the fibre bundle to change from the general linear one to the orthogonal group $O(N)$ and then imposing the orientability to change from $O(N)$ to $SO(N)$. This kind of situation is what we found in Einstein's theory of general relativity. However, the standard model's gauge theory is mainly dealing with the unitary group $U(N)$ rather than the $SO(N)$. Fortunately, by using the structure group, we can re-construct back the fibre bundle. Thus, we can always introduce the fibre bundle which have its own non-trivial structure group known as the principal fibre bundle denoted by $P(\mathcal{M},G)$. For example, the real world (3+1-dimensional) QCD is amount to study the principal fibre bundle $P(\mathbb{R}^{3,1},SU(3))$. Triviality of the principal fibre bundle generally related to the global property of the section supported by the following theorem
\begin{theorem}
A principal fibre bundle is trivial if and only if it admits a global section.
\end{theorem}
\end{appendix}
The principal fibre bundle seems too much abstract to physicists. However, when we introduce a little bit more structure, this will become more familiar. Suppose that the principal fibre bundle admits the connection, we can then introduce a Lie algebra-valued one form $A\equiv A^a_\mu(x)T^a dx^\mu$, which is known to be a gauge field, to represent the connection of this fibre bundle. This gauge field can be used to further construct the curvature tensor which is a Lie algebra-valued 2-form $F=\frac{1}{2}F_{\mu\nu}^a(x)T^a dx^\mu\wedge dx^\nu\equiv DA\equiv dA+A\wedge A$ (where the coupling and the imaginary basis is neglected for simplifying the computation). Observe first that, by construction, the component of such 2-form is antisymmetric under the interchanging $\mu\leftrightarrow\nu$ since the two form's basis must satisfy \eqref{B.1}.
\\
\indent As suggested by its name, the curvature 2-form (field strength tensor) determines how twist the fibre bundle is. On other words, it describes the topological structure of the fibre bundle. Unfortunately, since the curvature 2-form is not unique since they must satisfy the gauge transformation. Thus, even though field strength is very powerful to determine the non-triviality of the fibre bundle, but it is actually not the global information. Therefore, we need to find out other quantity that can be used to fulfil this blind spot of the field strength tensor. So let's try to think that what is that? After taking some time, we suddenly realize that no such a thing that can be used beside the curvature tensor. Thus, the global information must be constructed from the non-trivial combination between the field strength tensor itself. The simplest object we can think of is a polynomial. Suppose there is a polynomial in term of the field strength denoted as $P(F)$, if such a polynomial is invariant under the gauge transformation, it is said to be the invariant polynomial. The properties of invariant polynomial are stated inside the theorem given by Chern and Weil
\begin{theorem}[Chern-Weil Theorem]For an invariant polynomial $P(F)$
\begin{enumerate}
    \item It is closed in the sense of differential form, i.e.
    \begin{equation}
        dP(F)=0
    \end{equation}
    \item Suppose $F$ and $F'$ are two different field strength tensors. $P(F')-P(F)$ is an exact form.
\end{enumerate}
\end{theorem}
This theorem demonstrates that the invariant polynomial defines a cohomology class which is gauge-independent according to the second theorem of the Chern-Weil theorem. This particular class is known as a characteristic class.\\
\indent The characteristic class formally used here is the one called the Chern character defined such that
\begin{equation}
    ch(F)\equiv \mathrm{Tr}\exp\left(\frac{F}{2\pi}\right)=\sum_{n=0}\frac{1}{n!}\mathrm{Tr}\left(\frac{F}{2\pi}\right)^n,
\end{equation}
where Tr is trace over the basis of Lie algebra $\mathfrak{g}$ associated to the Lie group $G$ conventionally normalized to be $\mathrm{Tr}(T^aT^b)=\frac{\delta^{ab}}{2}$. In our spacetime, i.e. 4-dimensional spacetime, the series of the Chern character stops at $n=2$ since the differential $p$-forms of $p>4$ is meaningless to be defined. The first term of the series can be also neglected since it is a merely trivial constant. The remaining interesting terms are only $n=1$ and $n=2$. However, for principal fiber bundle with the group strucutre $G=SU(N)$, the trace over single $F=F^aT^a$ vanishes because of the tracelessness of $SU(N)$ generator $T^a$. Thus, in the $SU(2)$ YM in 4-dimensional spacetime case, the Chern character boils down into only one left non-trivial term. The full expression of this particular term is 
\begin{equation}
\label{B.29}
    ch(F)=\frac{1}{8\pi^2}\mathrm{Tr}(F\wedge F).
\end{equation}
To measure the topological charge, we consider the integration of this characteristic class over an entire space, i.e. the topological charge can be obtained from
\begin{equation}
\label{B.30} 
    \nu\equiv \int_{\mathcal{M}} ch(F)=\frac{1}{16\pi^2}\int_{\mathcal{M}} d^4x\;\mathrm{Tr}(F_{\mu\nu}\Tilde{F}_{\mu\nu})\;
\end{equation} (here we take $\mathcal{M}$, in this case, to be a 4-dimensional manifold of any kind) where an extra $\frac{1}{2}$ factor coming from $(1/2)^2$ of the definition $F\wedge F=(\frac{1}{2}F_{\mu\nu}dx^\mu\wedge dx^\nu)^2$ multiplies by $2$ from the definition $\epsilon_{\mu\nu\alpha\beta}F_{\alpha\beta}=2\Tilde{F}_{\mu\nu}$.\\
\indent One thing must be emphasized is, due to the Chern-Weil theorem, the Chern character which is a characteristic class is a closed form in its construction. Recalling also the Poincar\'e lemma tells us that any closed form defined on a manifold will be locally exact meaning that we can always write any characteristic class (Chern character in this case) as a exterior derivative of some lower degree differential form, $K$ said. Namely,
\begin{equation}
\label{B.31}
    ch(F)=dK,
\end{equation}
where $K$ is usually called the Chern-Simon form associated to the Chern character. The explicit expression of this Chern-Simon form is written as follows
\begin{equation}
\label{B.32}
    K=\frac{1}{8\pi^2}\mathrm{Tr}\left(A\wedge dA+\frac{2}{3}A\wedge A\wedge A\right)=\frac{1}{8\pi^2}\mathrm{Tr}\left(A\wedge F-\frac{1}{3}A\wedge A\wedge A\right),
\end{equation}
where the definition $F=dA+A\wedge A$ has been used. One can check by straightforward computation that the Chern-Simon current \eqref{B.32} truly satisfies \eqref{B.31}.
\begin{equation}
\label{B.33}
    \begin{split}
        dK&=\frac{1}{8\pi^2}\mathrm{Tr}\left(dA^2+\frac{2}{3}(dA A^2-AdA A+A^2dA)\right)\\
        &=\frac{1}{8\pi^2}\mathrm{Tr}\left((F-A^2)^2+\frac{2}{3}((F-A^2)A^2-A(F-A^2)A+A^2(F-A^2)\right).
    \end{split}
\end{equation}
To evaluate this, we need to know two important identities. The first one is \begin{equation}
    \label{B.34}
    \begin{split}
        \mathrm{Tr}(A^4)&=\mathrm{Tr}(\epsilon_{\mu\nu\alpha\beta}A_\mu A_\nu A_\alpha A_\beta)\\
        &=\mathrm{Tr}(\epsilon_{\mu\nu\alpha\beta}A_\beta A_\mu A_\nu A_\alpha)\\
        &=-\mathrm{Tr}(\epsilon_{\beta\mu\nu\alpha}A_\beta A_\mu A_\nu A_\alpha)\\
        &=-\mathrm{Tr}(A^4)=0.
    \end{split}
\end{equation}
To explain what actually happened in \eqref{B.34}, we firstly use the cyclic permutation property of trace operation. After that, we perform three times permutations on the Levi-Civita indices yielding the extra minus sign. Finally, we merely substitute back the origin of the expression in the third line of \eqref{B.34}. Logically, anything that equals to its negative value is nothing but zero. So now we can get rid several term inside the relation \eqref{B.33}. The field strength dependence term can be evaluated easily by using another identity
\begin{equation}
\label{B.35}
\begin{split}
    \mathrm{Tr}(AFA)&=\frac{1}{2}\mathrm{Tr}(\epsilon_{\mu\nu\alpha\beta}A_\mu F_{\nu\alpha} A_\beta)\\
    &=-\frac{1}{2}\mathrm{Tr}(\epsilon_{\beta\mu\nu\alpha}A_\beta A_\mu F_{\nu\alpha})=-Tr(A^2F),\\
    &=-\frac{1}{2}\mathrm{Tr}(\epsilon_{\nu\alpha\beta\mu}F_{\nu\alpha}A_\beta A_\mu)=-\mathrm{Tr}(FA^2).
\end{split}
\end{equation}
By the help of both identities \eqref{B.34} and \eqref{B.35}. The expression of $dK$ in the right hand side of \eqref{B.33} clearly yields the right side of \eqref{B.29}. Then, we are done. Let's plug in the Chern-Simon current \eqref{B.31} into the topological charge \eqref{B.30}. We obtain the alternative form of the topological charge as follows
\begin{equation}
    \nu=\int_{\mathcal{M}}dK=\int_{\partial \mathcal{M}}K,
\end{equation}
where the generalized Stoke's theorem \eqref{B.16} has been used in the last step. Remarkably, the evaluation of the integration of the Chern-Simon current $K$ on the boundary of the considering manifold must be done carefully in the sense that one must remind oneself that the evaluation implicitly subjected to the sufficient boundary condition chosen such that the field strength tensor must vanish there to maintain the finiteness of the YM energy (note also that even if $F=0$, $F\wedge F$ is not necessary). Thus, the true expression of such a topological charge, sometimes known as the Chern-Simon number, will be
\begin{equation}
    \label{CS number}
    \nu=-\frac{1}{24\pi^2}\int_{\partial \mathcal{M}}\mathrm{Tr}(A^3)=-\frac{1}{24\pi^2}\int_{\partial \mathcal{M}}d^3x\; n_\mu\epsilon_{\mu\nu\rho\sigma}\mathrm{Tr}(A_\nu A_\rho A_\sigma),
\end{equation}
where $n_\mu$ is a normal vector pointed out from the boundary surface. Nevertheless, the vanishing of the field strength tensor implies that the corresponding gauge field must be of the pure gauge form meaning that there exists an element of the gauge group $U\in G$ such that $A_\mu=U^{-1}\partial_\mu U$. Therefore, we can write \begin{equation}
    \label{CS number pure}
    \nu=-\frac{1}{24\pi^2}\int_{\partial\mathcal{M}}d^3x\; n_\mu\epsilon_{\mu\nu\rho\sigma}\mathrm{Tr}(U^{-1}\partial_\nu U U^{-1}\partial_\rho U U^{-1}\partial_\sigma U).
\end{equation}

	\renewcommand{\bibname}{Bibliography}
	\addcontentsline{toc}{chapter}{Bibliography}
	\bibliographystyle{plain}
	\bibliography{Cite.bib}
\nocite{*}

\end{document}